\documentclass[openacc]{rsproca_new}%%%%where rsproca is the template name

\usepackage{amsmath}
\usepackage{mathrsfs}
\usepackage{amssymb}
\graphicspath{{figures/}}
\usepackage{tikz}
\usepackage{pstricks}
\usepackage{xcolor}
\usepackage{soul}

\newcommand{\pd}[2]{\frac{\partial #1}{\partial #2}}
\newcommand{\be}{\begin{equation}}
\newcommand{\ee}{\end{equation}}
\newcommand{\norm}[1]{\left\lVert#1\right\rVert}

\begin{document}

%%%% Article title to be placed here
\title{Non-intrusive Balancing Transformation of Highly Stiff Systems with Lightly-damped Impulse Response}

\author{%%%% Author details
Elnaz Rezaian$^{1}$, Cheng Huang$^{1}$ and Karthik Duraisamy$^{1}$}

%%%%%%%%% Insert author address here
\address{$^{1}$University of Michigan, Ann Arbor, MI}
%$^{2}$Second author address\\
%$^{3}$Third author address}

%%%% Subject entries to be placed here %%%%
\subject{computational science, control, data-driven modeling}

%%%% Keyword entries to be placed here %%%%
\keywords{reduced-order modeling, balanced truncation, eigensystem realization algorithm, combustion}

%%%% Insert corresponding author and its email address}
\corres{Karthik Duraisamy\\
\email{kdur@umich.edu}}

%%%% Abstract text to be placed here %%%%%%%%%%%%
\begin{abstract}
Balanced truncation (BT) is a model reduction method that utilizes a coordinate transformation to retain eigen-directions that are highly observable and reachable. To address realizability and scalability of BT applied to highly stiff and lightly-damped systems, a non-intrusive data-driven method is developed for balancing discrete-time systems via the eigensystem realization algorithm (ERA). The advantage of ERA for balancing transformation makes full-state outputs tractable. Further, ERA enables balancing despite stiffness, by eliminating computation of balancing modes and adjoint simulations. As a demonstrative example, we create balanced ROMs for a one-dimensional reactive flow with pressure forcing, where the stiffness introduced by the chemical source term is extreme (condition number $10^{13}$), preventing analytical implementation of BT.  We investigate the performance of ROMs in prediction of dynamics with unseen forcing inputs and demonstrate stability and accuracy of balanced ROMs in {\em truly} predictive scenarios whereas without ERA, POD-Galerkin and Least-squares Petrov-Galerkin projections fail to represent the true dynamics. We show that after the initial transients under unit impulse forcing, the system undergoes lightly-damped oscillations, which magnifies the influence of sampling properties on predictive performance of the balanced ROMs.  We propose an output domain decomposition approach and couple it with tangential interpolation to resolve sharp gradients at reduced computational costs.
\end{abstract}
%%%%%%%%%%%%%%%%%%%%%%%%%%%

%%%%%%%%%% Insert the texts which can accomdate on firstpage in the tag "fmtext" %%%%%

%\begin{fmtext}
%\section{Insert A head here}

%\subsection{Insert B head here}
%%%% Insert B head here
%Subsection text here.

%\subsubsection{Insert C head here}
%%%% Insert C head here
%Subsubsection text here.

%\section{Equations}

%\end{fmtext}

%%%%%%%%%%%%%%% End of first page %%%%%%%%%%%%%%%%%%%%%

\maketitle

\section{Introduction}
High-fidelity simulations have pushed the boundaries of computational science for decades by showcasing detailed views of complex dynamical systems underlying phenomena such as turbulence, chemical reactions, transitional effects, shock interactions, etc \cite{Aditya:19, Elghobashi:2019, zhang:18, Lee:2015}. High-fidelity simulations will continue to serve the realm of engineering applications with outstanding insight into the multi-scale character of nature. However, they remain too expensive to use in many-query settings such as design optimization, control systems or in uncertainty quantification. As an example, model-predictive control (MPC) employs control decisions based on real-time simulation of the quantities of interest (QoI), which precludes the use of expensive high-fidelity models (HFM). Repeating simulations with costly HFMs for optimization of design decisions also leads to backlogs that cannot be supported by manufacturing processes.

Complex dynamical systems - or quantities of interest thereof - typically evolve on a low-dimensional attractor, which suggests there is a potential to reduce the number of degrees of freedom by projecting the dynamics onto a lower-dimensional space \cite{noack2013, Wei_Rowley:2009, Wei_Qawa:2012}. For linearly reducible problems in the sense of rapidly decaying Kolmogorov n-widths \cite{Ohlberger2015}, proper-orthogonal decomposition (POD) is a computationally efficient tool for data compression that identifies coherent structures via singular value decomposition or the method of snapshots developed by Sirovich \cite{holmes:96bk, Berkooz:1993, sirovich:87qam, rowley:04pdnp}. In advection-dominated flows characterized by slowly-decaying Kolmogorov n-widths, POD is known to be less effective, as strong nonlinearities cannot be described by a low-dimensional linear representation \cite{Lee:2020}. POD modes are computed by minimizing projection error in the sense of $L^2$ norm, thus, the POD subspace is naturally comprised of the energetically dominant structures, while the high-frequency components that are critical in transport problems are removed from the system and the linear subspace is incapable of recovering the lost information. 

In projection-based model reduction, the governing equations are projected onto a low dimensional space (e.g. the POD subspace) using a projection method (typically Galerkin projection) \cite{Ghattas:2021, benner:2015}. However, Galerkin projection is only optimal when the HFM Jacobian is symmetric, which is not the case in non-normal systems and advection-dominated flows \cite{Serre:12, rezaian:2021IJNME}. In turn, the difficulty of POD in maintaining the high-frequency observable modes (a.k.a., POD closure problem) leaves POD-Galerkin ROMs vulnerable to instabilities. Stabilization methods and closure terms have improved the performance of POD-Galerkin ROMs in various applications \cite{noack2013, Pan2018, Iliescu:13, Wells:2017}. Eigenvalue reassignment methods \cite{kalashnikova:14cmame, rezaian:2020IJNME, rezaian:2021IJNME, Mojgani:2020}, subspace rotation \cite{amsallem:12ijnme, Balajewicz:16ijcp, rezaian:2020IJNME}, linear quadratic regulation and numerical dissipation \cite{lucia:03jcp} recover ROMs by post-processing non-intrusive stabilization. Leveraging energy stability in linear POD-Galerkin ROMs through symmetrization by an energy-based inner product \cite{barone:09jcp, rezaian:2020IJNME}, and transformation of nonlinear POD-Galerkin ROMs using entropy-conservative variables \cite{kalashnikova:11AIAA, Chan:2020} on the other hand, stabilize ROMs through an intrusive framework that modifies and reformulates ROM equations. Symmetrization of POD-Galerkin ROMs is inherently equivalent to the least-squares Petrov-Galerkin (LSPG) projection \cite{Carlberg:2011, BuiThanh:2008, benner:2015}, and its variants~\cite{Huang:2020}.  Closure modeling~\cite{wang2012proper,ahmed2021closures} via eddy-viscosity terms \cite{Rempfer:1994} and more recently using the Mori-Zwanzig formalism \cite{Gouasmi:2017,parish2020adjoint} have also  contributed to the utility of POD-based ROMs. 

When applied to unit impulse response data of a linear system, POD retains the most controllable directions while disregarding observability of the modes \cite{rowley:05}. However, it is possible to transform the system to coordinates in which the controllability and observability Gramians are equal and diagonal, such that the leading modes after balancing transformation are both highly controllable and observable. This is the idea behind balanced truncation (BT) to create ROMs that satisfy theoretical error bounds \cite{Moore:1981}. The balancing transformation is obtained by singular value decomposition of the Hankel matrix, which is originally constructed via the system Gramians computed by solving Lyapunov equations \cite{Moore:1981, Antoulas:05bk}. Solving the Lyapunov equations however, carries a few pitfalls that have encouraged data-driven approaches to compute the balancing transformation. 
The computational cost of solving Lyapunov equations prevents application of the analytical BT in large-scale problems. 
One way to alleviate this cost in high-dimensional systems is through low-rank solvers \cite{Vettermann:2021, behr:2021, benner:2013, shank:16}.
On the other hand, analytical evaluation of the Gramians requires the system to be stable, which is undesirable in control applications, where ROMs are expected to accurately predict instabilities to promote effective control decisions. These difficulties are addressed to some extent by resorting to data-driven methods for computing the system Gramians \cite{Moore:1981, Willcox:02,Flinois:2015}. Impulse response of the direct and adjoint systems are used to assemble matrices known as empirical Gramians, that are equivalent to analytical Gramians when the impulse response is properly sampled \cite{Antoulas:05bk}. Flinois et al. \cite{Flinois:2015} used empirical Gramians for balancing transformation of unstable systems. Computing the Gramians on the other hand, can be entirely bypassed using approximate balancing transformation (or balanced POD) \cite{Willcox:02, rowley:05}. However, balanced POD (BPOD) still requires adjoint system simulations, which is prohibitively expensive in systems with full-state output, and impossible to access in experimental studies.

Eigensystem realization algorithm (ERA) is originally a system identification method \cite{Juang:94bk}, that was later adopted by Ma et al. \cite{Ma:2011} as an adjoint-free solution for scalable balancing transformation. ERA operates in a data-driven non-intrusive setting, in the sense that it does not require access to the FOM operators \cite{Peherstorfer2016, McQuarrie:2021, Ghattas:2021}. Identifying balanced ROM matrices merely based on the direct system response is one of the advantages of this algorithm over the preceding versions of BT. Therefore, ERA is applicable to experiments and systems with full-state output. More recently, Gosea et al. \cite{Gosea:2021} extended this method for non-intrusive balancing transformation of continuous-time systems. Tu et al. \cite{Tu:14dmd} showed that the linear operator obtained by ERA is related to the linear operator in dynamic mode decomposition (DMD) via a similarity transform. As a result, eigenvalues of the two operators are the same and DMD modes can be obtained by eigenvectors of the linear operator identified by ERA. On the other hand, exact DMD modes and eigenvalues are related to the Koopman eigenvalues and modes for a linearly consistent snapshots matrix \cite{Tu:14dmd}. Therefore, the eigenvalues and eigenvectors of the dynamic matrix identified by ERA are in turn related to the Koopman modes and eigenvectors.

{\em This work is focused on balancing stiff systems with lightly-damped (slowly-decaying) impulse response: two factors that make BT more challenging by triggering instabilities and intensifying sensitivity to sampling properties, respectively.} We use the balanced ROM to predict one-dimensional reactive flow problems with species transport. The system is impulsively excited through pressure forcing to compute Markov parameters. The Cholesky factorizations at the heart of the analytical BT are prone to instabilities and numerical errors under extreme stiffness \cite{Antoulas:05bk, Hammarling:1982}. In the present work, stiffness is mainy a consequence of  the chemical source term, that results in a full-order model Jacobian with a condition number of $\sim 10^{13}$. On the other hand, we are interested in  the full-state output, thus, BPOD carries prohibitive computational costs. One way to address this issue is through output projection \cite{rowley:05}. However, with the slow convergence of the impulse response modal energy in our application, output projection also does not alleviate adjoint simulation costs. This motivates us to use ERA, which does not require the adjoint system impulse response for balancing transformation.

We show in this study that under adequate sampling, balancing transformation via ERA generates truly predictive ROMs that compensate for the training cost by replicating the full-order model (FOM) behavior in response to unseen forcing inputs. We also demonstrate that POD-Galerkin and LSPG ROMs are not predictive under the same conditions. The main contributions of this work are as follows:
\begin{itemize}
    \item The original balanced truncation method developed by Moore \cite{Moore:1981} has two challenges : (i) The computational expense of solving the Lyapunov equations to evaluate analytical Gramians, and (ii) Computing the balancing transformation using matrix inversions and Cholesky factorization introduces numerical errors in highly stiff systems. This issue is commonly addressed by approximate balancing \cite{Moore:1981, Willcox:02, rowley:05}, however, adjoint simulations append prohibitive costs in systems with full-state output. We propose the use of a data-driven balanced truncation (via the ERA) for balancing transformation of a stiff reactive flow problem with full-state output to generate stable ROMs while bypassing computation of the system Gramians, balancing modes, and adjoint simulations.
    
    \item The overarching goal for construction of ROMs is to predict the dynamics beyond the training regime. This is a non-trivial task for most projection-based ROMs in practical applications, including POD-Galerkin projection and least-squares Petrov-Galerkin projection. While major effort is spent in the literature to validate ROMs against high-fidelity simulations in the training regime, it is  critical to investigate predictive performance of ROMs. Here, we focus on ROM predictions and examine accuracy of balanced ROMs in a purely predictive scenario and compare the results against standard model reduction techniques.
    
    \item Data-driven methods are inherently sensitive to sampling properties. In non-intrusive balanced truncation methods similar to ERA in particular, lack of proper sampling compromises balancing, which in turn results in ROMs that do not satisfy the theoretical error bounds \cite{Glover:1984, Hinrichsen:1990}. It is critical to avoid treating sampling as a pre-processing step and ignoring the influence of sampling properties on the performance of ROMs. We highlight the importance of this factor by detailed sensitivity analysis of the balanced ROMs with respect to impulse response sampling properties.
    
    \item Resolving sharp gradients by the data-driven BT requires high sampling frequencies and introduces a computational bottleneck in large-scale systems. Here, we propose an output domain decomposition approach to divide the computational domain into subdomains and gradually change the sampling frequency between the subdomains to capture the gradients while reducing the cost of offline computations. The subdomains that contain sharp gradients are handled by balanced ROMs trained with higher sampling frequencies, while in the rest of the domain we use ROMs trained with lower sampling frequencies to minimize computation costs. To further alleviate the training costs, we use a tangential interpolation method proposed by Kramer and Gugercin \cite{Kramer:2016} to project the impulse response sequence onto the dominant tangential directions and reduce the number of outputs before balancing transformation.
\end{itemize}

The paper is organized as follows: in section~\ref{MD} proper orthogonal decomposition is explained as a dimensionality reduction tool. In section~\ref{MOR} we discuss projection-based model reduction methods including the standard POD-Galerkin method, LSPG, and balancing transformation. Eigensystem realization algorithm is then described in section~\ref{era} along with the output domain decomposition approach and tangential interpolation. In section~\ref{appl} we show the governing equations and numerical setup for the one-dimensional reactive flow problem of interest. Results and discussions are then demonstrated in section~\ref{results} and concluding remarks and research perspective are laid out in section~\ref{conclusion}.

The data and code for the numerical experiments in this paper can be found at \url{https://github.com/cwentland0/perform.git}.

\section{Modal Decomposition}\label{MD}

\subsection{Proper Orthogonal Decomposition}
Proper orthogonal decomposition (POD), also known as Karhunen-Loeve expansion and principal component analysis, is a data compression method for linear dimensionality reduction that learns the low-dimensional space by maximizing the projection energy in the sense of an energy inner product (e.g., $L^2$),
\be
\label{pod}
max_{\phi \in \mathscr{H}(\Omega)} \frac{(\langle q(\boldsymbol x, t), \boldsymbol \phi(\boldsymbol x) \rangle^2)}{\norm{\boldsymbol \phi (\boldsymbol x)}^2},
\ee
where $q(\boldsymbol x, t)$ is the FOM solution, centered about the time-averaged mean or steady-state solution, $\phi(\boldsymbol x)$ is the POD basis, $\boldsymbol x \in \Omega$, and $\mathscr{H}$ is the Hilbert space associated with an inner product. $(\cdot, \cdot)$ represents time average, $\langle \cdot, \cdot \rangle$ shows the inner product, and $\norm{\cdot}$ is the inner product norm. Optimization problem (\ref{pod}) is solved here by singular value decomposition of the FOM snapshots matrix $\mathbf{Q} \in \mathbb{R}^{n \times n_t}$,
$
\mathbf{Q} = \mathbf{V} \boldsymbol{\Sigma} \mathbf{U}^T.$ POD modes are the column vectors of $\mathbf{V}$ (the trial subspace), and $\boldsymbol{\Sigma}$ is the diagonal matrix of singular values, where the singular values are arranged in descending order and represent the amount of energy captured by the modes. The method of snapshots proposed by Sirovich \cite{sirovich:87qam} is an alternative way for computing the basis functions. We use scalar-valued POD \cite{rowley:04pdnp}, where the POD modes are computed separately for each variable. However, when the governing equations are dimensional, snapshots are normalized before computing the POD modes to remove bias in projection-based ROMs \cite{lumley:1997}. In this sense, the FOM solution is decomposed as,
\be
\label{exp}
\Tilde{\mathbf{q}}(t) = \bar{\mathbf{q}} + \mathbf{S}^{-1} \mathbf{V} \hat{\mathbf{q}}(t),
\ee
where, $\mathbf{V} \in \mathbb{R}^{n \times m}$, $m$ is the number of retained POD modes, $\bar{\mathbf{q}} \in \mathbb{R}^{n}$ is the steady-state solution, $\hat{\mathbf{q}}(t) \in \mathbb{R}^{m}$ is the vector of generalized coordinates (modal amplitudes), and $\mathbf{S} \in \mathbb{R}^{n \times n}$ is a diagonal scaling matrix, where the diagonal elements of this matrix are scaled by $L^2$ norm of the FOM solution as $\mathbf{S}_{i, i} = \frac{1}{\alpha_i}$ and,
\be
\alpha_i = \frac{1}{n_t} \sum_{j=1}^{n_t} \frac{1}{n} \sum_{k=1}^{n} (\mathbf{q}_{j, k})^2, \qquad i = 1, \dots, n.
\ee
Here, $n$ is the spatial degrees of freedom (i.e., number of cells in a control volume approach), and $n_t$ is the number of snapshots. 

\section{Projection-based Model Reduction}\label{MOR}
Projection-based model reduction consists of two components: a) dimensionality reduction to identify the coordinates in which the dynamics of the system evolves in a reduced-order setting, and b) projection of the governing equations onto the low-dimensional space \cite{benner:2015}. We use POD for dimensionality reduction in this work. Galerkin and LSPG projections are described in this section and compared against balancing transformation in application to the reactive flow problem in section~\ref{appl}. The original balanced truncation is also an intrusive model reduction approach that is explained here for completeness.

\subsection{The Standard Galerkin Projection}
Consider a linear time-invariant (LTI) system obtained by semi-discretization of the linearized governing partial differential equations (PDEs),
\be
\label{lti}
\frac{d \mathbf{q}(t)}{dt} = \mathbf{J} \mathbf{q}(t) + \mathbf{B} u(t), \qquad \mathbf{q}(0) = \mathbf{q}_0,
\ee
where, $\mathbf{q}(t) \in \mathbb{R}^{n}$ is the FOM state, $\mathbf{J} \in \mathbb{R}^{n \times n}$ is the FOM Jacobian matrix, $\mathbf{B} \in \mathbb{R}^{n}$ contains boundary flux information and $u(t)$ is the input. We normalize the LTI system equations using the scaling matrix $\mathbf{S}$,
\be
\label{scaledlti}
\mathbf{S}  \frac{d \mathbf{q}(t)}{dt} = \mathbf{S} \mathbf{J} \mathbf{q}(t) + \mathbf{S} \mathbf{B} u(t).
\ee
Substituting modal expansion (\ref{exp}), and projecting onto the test subspace $\mathbf{W} \in \mathbb{R}^{n \times m}$ yields,
\be
\label{rom}
\mathbf{W}^T \mathbf{V}  \frac{d \hat{\mathbf{q}} (t)}{dt} = \mathbf{W}^T \mathbf{S} \mathbf{J} \mathbf{S}^{-1} \mathbf{V} \hat{\mathbf{q}}(t) + \mathbf{W}^T \mathbf{S} \mathbf{B} u(t),
\ee
where, $\mathbf{W} = \mathbf{V}$ in the standard Galerkin projection. Therefore, $\mathbf{W}^T \mathbf{V} = \mathbf{I}$, $\mathbf{I} \in \mathbb{R}^{m \times m}$ is the identity matrix, and,
\be
\frac{d \hat{\mathbf{q}} (t)}{dt} = \mathbf{W}^T \mathbf{S} \mathbf{J} \mathbf{S}^{-1} \mathbf{V} \hat{\mathbf{q}}(t) + \mathbf{W}^T \mathbf{S} \mathbf{B} u(t), \qquad \hat{\mathbf{q}}(0) = \mathbf{W}^T \mathbf{q}_0,
\ee
which is solved here with the third-order Runge-Kutta scheme.

\subsection{Least-squares Petrov-Galerkin Projection}
Substituting approximation (\ref{exp}) in equation (\ref{scaledlti}) and discretizing  with an implicit time integration scheme (e.g. backward Euler) yields,
\be
\mathbf{S} \frac{\tilde{\mathbf{q}}^k - \tilde{\mathbf{q}}^{k-1}}{\Delta t} = \mathbf{S} \mathbf{J} \tilde{\mathbf{q}}^k + \mathbf{S} \mathbf{B} \mathbf{u}^k.
\ee
Least-squares Petrov-Galerkin (LSPG) seeks to minimize the residual $\mathbf{r}$,
\be
\label{minlspg}
\tilde{\mathbf{q}}^k = \textrm{argmin}_{\tilde{\mathbf{q}}^k \in \mathbb{R}^m} \norm{\mathbf{r}( \tilde{\mathbf{q}}^k)}_2^2
\ee
where, 
\be
\mathbf{r} = \mathbf{S} (\Delta \tilde{\mathbf{q}} - \Delta t \mathbf{J}  \tilde{\mathbf{q}}^k -  \Delta t  \mathbf{B} \mathbf{u}^k),
\ee
is the FOM residual and $\Delta \tilde{\mathbf{q}} = \tilde{\mathbf{q}}^k - \tilde{\mathbf{q}}^{k-1}$. Minimization problem (\ref{minlspg}) is equivalent to $
\mathbf{W}^T \mathbf{r} = 0,$
which yields $\mathbf{W} = \mathbf{S} (\mathbf{I} - \Delta t \beta_0 \mathbf{J}) \mathbf{S}^{-1} \mathbf{V}$ as the test subspace, where $\beta_0 = 0$ for explicit time integration schemes and $\beta_0 = 1$ for implicit schemes \cite{Huang:2020, BuiThanh:2008, benner:2015}. Therefore, 
\be
\mathbf{W}^T \mathbf{V} \frac{\hat{\mathbf{q}}^k - \hat{\mathbf{q}}^{k-1}}{\Delta t} = \mathbf{W}^T \mathbf{S} \mathbf{J} \mathbf{S}^{-1} \mathbf{V} \hat{\mathbf{q}}^k + \mathbf{W}^T \mathbf{S} \mathbf{B} \mathbf{u}^k,
\ee
and with further rearrangements we have,
\be
\mathbf{W}^T \mathbf{V} \Delta \hat{\mathbf{q}} = \mathbf{W}^T \mathbf{S} \Delta t \mathbf{J} \mathbf{S}^{-1} \mathbf{V} \hat{\mathbf{q}}^k + \mathbf{W}^T \mathbf{S} \Delta t  \mathbf{B} \mathbf{u}^k,
\ee
where, $ \Delta \hat{\mathbf{q}} = \hat{\mathbf{q}}^k - \hat{\mathbf{q}}^{k-1}$. It is easy to verify that LSPG is equivalent to Galerkin projection: a) when using an explicit time integration scheme, or b) when $\Delta t$ approaches zero. Note that LSPG is sensitive to the time step $\Delta t$, therefore, some fine-tuning is required for the LSPG ROM to achieve optimal performance \cite{Carlberg:17}. 

\subsection{Balanced Truncation}
Consider an LTI system in state-space form,
\be
\label{FOMss}
\begin{aligned}
& \dot{\mathbf{x}}(t) = \mathbf{A} \mathbf{x}(t) + \mathbf{B} u(t)\\
& \mathbf{y}(t) = \mathbf{C} \mathbf{x}(t) + \mathbf{D} u(t),
\end{aligned}
\ee
where, $\mathbf{x}(t) \in \mathbb{R}^{n}$ is the FOM state, $\dot{\mathbf{x}}(t)  = \frac{d \mathbf{x}(t)}{dt}$, $u(t) \in \mathbb{R}^{p}$ is the input, and $\mathbf{y}(t) \in \mathbb{R}^{q}$ is the system output ($q=n$ for full-state output). Without loss of generality, we remove the feedthrough term (i.e., $\mathbf{D} = \mathbf{0}$) in what follows. Here, $\mathbf{A} \in \mathbb{R}^{n \times n}$, $\mathbf{B} \in \mathbb{R}^{n \times p}$,  and $\mathbf{C} \in \mathbb{R}^{q \times n}$ are constant matrices. BT consists of identifying a transformation such that the reachability and observability Gramians are equal and diagonal \cite{Antoulas:05bk, Willcox:02}. Reachability (equivalent to controllability in continuous-time systems) and observability Gramians are Hermitian matrices that are positive definite if and only if the system is reachable (i.e., $\mathcal{R}(\mathscr{P}) = n$) and observable (i.e., $\mathcal{R}(\mathscr{O}) = n$), respectively, where, $\mathscr{P}$ is the reachability matrix,
\be
\mathscr{P} = \begin{bmatrix}
    \mathbf{B} & \mathbf{A} \mathbf{B} & \dots & \mathbf{A}^{n-1} \mathbf{B} 
    \end{bmatrix}
\ee
and $\mathscr{O}$ is the observability matrix,
\be
\mathscr{O} = \begin{bmatrix}
        \mathbf{C}  \\ \mathbf{C} \mathbf{A} \\ \vdots \\ \mathbf{C}  \mathbf{A}^{n-1}
    \end{bmatrix}.
\ee

The Reachability Gramian ($\mathcal{W}_p$) is a measure of the degree of reachability, that is, the amount of energy required to drive the system from zero to a desired state \cite{Antoulas:05bk}. Therefore,
\be
\label{CWp}
\mathcal{W}_p = \int_0^{\infty} e^{\mathbf{A}t} \mathbf{B} \mathbf{B}^{*} e^{\mathbf{A}^* t} dt,
\ee
in continuous-time systems and $\mathcal{W}_p = \mathscr{P} \mathscr{P}^{*}$ in discrete-time systems, where, ($^*$) denotes complex conjugate transpose. Reachability Gramian is the solution of the following Lyaounov equation,
\be
\label{ctrbLyap}
\mathbf{A}  \mathcal{W}_p + \mathcal{W}_p \mathbf{A} ^{*} + \mathbf{B} \mathbf{B}^{*} = 0.
\ee
Reachability and observability are known as dual concepts \cite{Antoulas:05bk}. Observability Gramian ($\mathcal{W}_o$) therefore, identifies the impact of the state on the system output, when $u(t) = 0$. For continuous-time systems we have,
\be
\label{CWo}
\mathcal{W}_o = \int_0^{\infty} e^{\mathbf{A}^* t} \mathbf{C}^* \mathbf{C} e^{\mathbf{A} t} dt,
\ee
which reduces to $\mathcal{W}_o = \mathscr{O}^{*} \mathscr{O}$ in discrete-time systems. Similar to reachability Gramian, observability Gramian is the solution of a Lyapunov equation,
\be
\label{bsvLyap}
\mathbf{A}^{*} \mathcal{W}_o + \mathcal{W}_o \mathbf{A} + \mathbf{C}^{*} \mathbf{C} = 0.
\ee
Equations (\ref{CWp}) and (\ref{CWo}) cannot be evaluated for unstable systems, therefore, the standard balanced truncation based on analytical Gramians is only applicable to stable systems. 

The intent is to identify a transformation that balances the reachability and observability Gramians in the following sense,
\be
\mathbf{T}^{-1} \mathcal{W}_c \mathbf{T}^{-*} = \mathbf{T}^{*} \mathcal{W}_o \mathbf{T} = \boldsymbol{\Sigma}, 
\ee
where, $\boldsymbol{\Sigma}$ is a diagonal matrix. Its diagonal components are Hankel singular values, that are invariant under coordinate transformation and represent the input-output energy of the system. Subsequently, the balancing transformation $\mathbf{T}_r$ is computed by $\mathbf{T}_r = \mathbf{U} \mathbf{W}_r \boldsymbol{\Sigma}_r^{-1/2}$, where, subscript $r$ shows that the directions corresponding to the smaller Hankel singular values are truncated. $\mathbf{U}$ is the Cholesky factor of the reachability Gramian $\mathcal{W}_p = \mathbf{U} \mathbf{U}^*$, and $\mathbf{W}$ is obtained by SVD of the product $
\mathbf{U}^* \mathbf{L} = \mathbf{W} \boldsymbol{\Sigma} \mathbf{V}^*$, where $\mathbf{L}$ is computed by Cholesky factorization of the observability Gramian $\mathcal{W}_o = \mathbf{L} \mathbf{L}^*$. Column vectors of $\mathbf{T}_r$ are called the direct modes and adjoint modes are row vectors of the inverse transformation $\mathbf{T}_r^{-1} = \boldsymbol{\Sigma}_r^{-1/2} \mathbf{V}_r^{*} \mathbf{L}^{*}$ \cite{Antoulas:05bk}. Balanced ROMs can then be constructed by transformation of the FOM (\ref{FOMss}) as,
\be
\begin{aligned}
\label{BROM}
& \dot{\mathbf{x}}_r(t) = \mathbf{A}_r \mathbf{x}_r(t) + \mathbf{B}_r u(t)\\
& \mathbf{y}(t) = \mathbf{C}_r \mathbf{x}_r(t),
\end{aligned}
\ee
where, 
\be
\label{BROMmatrices}
\begin{aligned}
& \mathbf{A}_r = \mathbf{T}_r^{-1} \mathbf{A} \mathbf{T}_r,\ \
& \mathbf{B}_r = \mathbf{T}_r^{-1} \mathbf{B},\ \
& \mathbf{C}_r = \mathbf{C} \mathbf{T}_r.
\end{aligned}
\ee
Regardless of the model reduction method, ROMs satisfy the following lower bound,
\be
\norm{\mathbf{G} - \mathbf{G}_r}_{\infty} > \sigma_{r+1}.
\ee
However, the error encountered with the balanced low-dimensional system in (\ref{BROM}) is also upper-bounded,
\be
\norm{\mathbf{G} - \mathbf{G}_r}_{\infty} < 2 \sum_{i=r+1}^n \sigma_i,
\ee
where, $\mathbf{G} = \mathbf{C} (s \mathbf{E} - \mathbf{A})^{-1} \mathbf{B}$ and $\mathbf{G}_r = \mathbf{C}_r (s \mathbf{E}_r - \mathbf{A}_r)^{-1} \mathbf{B}_r$ are the transfer functions of FOM and ROM, respectively \cite{Glover:1984, Hinrichsen:1990, Willcox:02}. Here, $\mathbf{E} = \mathbf{I} \in \mathbb{R}^{n \times n}$, $\mathbf{E}_r = \mathbf{I}_r \in \mathbb{R}^{r \times r}$, and $s \in \mathbb{C}$. It is important to note that BT is equivalent to the standard POD-Galerkin projection when the observability Gramian is used as the inner product, that is, when the test subspace is modified as $\mathbf{W} = \mathcal{W}_o \mathbf{V}$ \cite{rowley:05, Carlberg:15}.

\section{The Eigensystem Realization Algorithm}\label{era}
While intrusive model reduction processes FOM operators, non-intrusive ROMs are discovered based on the input-output behavior of the system, accompanied by some knowledge about the structure of the system (glass-box models), or in a purely data-driven fashion (black-box models) \cite{Peherstorfer2016, McQuarrie:2021, Ghattas:2021}. Therefore, non-intrusive methods are more suitable for applications with limited access to FOM operators and in stiff problems in which manipulation of the FOM Jacobian is susceptible to numerical errors. ERA is described in this section as a non-intrusive BT approach for discrete-time systems.

Computing system Gramians by solving the Lyapunov equations (\ref{ctrbLyap}) and (\ref{bsvLyap}) is expensive and requires access to the FOM operators. However, using the method of snapshots \cite{sirovich:87qam}, it is possible to bypass direct computation of the Gramians and evaluate the Hankel matrix as $ \mathbf{H} = \mathscr{O}^{*} \mathscr{P}$. This is the idea behind approximate balanced truncation (a.k.a, balanced POD) \cite{Willcox:02, rowley:05}. 

System identification via eigensystem realization algorithm (ERA) balances the system Gramians in a purely data-driven setting. For a discrete-time linear system,
\be
\label{dlti}
\begin{aligned}
& \mathbf{x}_{k+1} = \mathbf{A} \mathbf{x}_k +  \mathbf{B} u_k\\
& \mathbf{y}_k = \mathbf{C} \mathbf{x}_k,
\end{aligned}
\ee
the Hankel matrix takes the following form,
\be
\begin{aligned}
\mathbf{H} & = \begin{bmatrix}
                \mathbf{y}_1 & \mathbf{y}_2 & \dots & \mathbf{y}_{m_p} \\
                \mathbf{y}_2 & \mathbf{y}_3 & \dots  & \mathbf{y}_{m_p + 1}  \\
                 \vdots & \dots & \ddots & \vdots \\
                 \mathbf{y}_{m_o} & \mathbf{y}_{m_o + 1}  & \dots & \mathbf{y}_{m_o + m_p - 1}   \\
\end{bmatrix}  
 = \begin{bmatrix}
                \mathbf{C}  \mathbf{B} & \mathbf{C} \mathbf{A}  \mathbf{B} & \dots & \mathbf{C} \mathbf{A}^{m_p - 1}  \mathbf{B} \\
                \mathbf{C} \mathbf{A}  \mathbf{B} & \mathbf{C} \mathbf{A}^2  \mathbf{B} & \dots & \mathbf{C} \mathbf{A}^{m_p}  \mathbf{B} \\
               \vdots & \dots & \ddots & \vdots \\
                \mathbf{C} \mathbf{A} ^{m_o - 1} \mathbf{B} & \mathbf{C} \mathbf{A}^{m_o}  \mathbf{B} & \dots & \mathbf{C} \mathbf{A}^{m_p + m_o - 2}  \mathbf{B} \\
\end{bmatrix},
\end{aligned}
\ee
where, the terms of the sequence,
\be
\label{markov}
\mathbf{y}_i = \mathbf{C} \mathbf{A}^i \mathbf{B}, \qquad i = 0, \dots, m_p + m_o - 2,
\ee
are known as Markov parameters \cite{Ma:2011}. For a linear system, Markov parameters are equivalent to the unit impulse response \cite{Antoulas:05bk}. The SVD of the Hankel matrix gives $\mathbf{H} = \mathbf{U} \boldsymbol{\Sigma} \mathbf{V}^{*}$, where we retain only the singular values that capture most of the input-output energy. Advancing the sequence (\ref{markov}) one step in time yields the shifted Hankel matrix \cite{Ma:2011},
\be
\begin{aligned}
\mathbf{H}^{\prime} & = \begin{bmatrix}
                \mathbf{y}_2 & \mathbf{y}_3 & \dots & \mathbf{y}_{m_p + 1} \\
                \mathbf{y}_3 & \mathbf{y}_4 & \dots  & \mathbf{y}_{m_p + 2}  \\
                 \vdots & \dots & \ddots & \vdots \\
                 \mathbf{y}_{m_o + 1} & \mathbf{y}_{m_o + 2}  & \dots & \mathbf{y}_{m_o + m_p}   \\
\end{bmatrix} 
 = \begin{bmatrix}
                \mathbf{C}  \mathbf{A}  \mathbf{B} & \mathbf{C} \mathbf{A}^2  \mathbf{B} & \dots & \mathbf{C} \mathbf{A}^{m_p}  \mathbf{B} \\
                \mathbf{C} \mathbf{A}^2  \mathbf{B} & \mathbf{C} \mathbf{A}^3  \mathbf{B} & \dots & \mathbf{C} \mathbf{A}^{m_p + 1}  \mathbf{B} \\
               \vdots & \dots & \ddots & \vdots \\
                \mathbf{C} \mathbf{A} ^{m_o} \mathbf{B} & \mathbf{C} \mathbf{A}^{m_o + 1}  \mathbf{B} & \dots & \mathbf{C} \mathbf{A}^{m_p + m_o - 1}  \mathbf{B} \\
\end{bmatrix},
\end{aligned}
\ee
which is the key to the adjoint-free character of ERA. Given this matrix we no longer need to run adjoint simulations and explicitly compute the adjoint modes required to build ROM matrices. Note that the direct modes,
\be
\label{discdirmodes}
\mathbf{T}_r = \mathscr{P} \mathbf{V}_r \boldsymbol{\Sigma}_r^{-1/2},
\ee
and the adjoint modes,
\be
\label{discadjmodes}
\mathbf{T}_r^{-1} = \boldsymbol{\Sigma}_r^{-1/2} \mathbf{U}_r^{*} \mathscr{O}^{*},
\ee
can be computed by the reachability ($\mathscr{P}$) and observability ($\mathscr{O}$) matrices of the discrete-time system, where, $\mathbf{U}$ and $\mathbf{V}$ are obtained by SVD of the Hankel matrix and subscript $r$ denotes the retained singular values and singular vectors. Combining (\ref{BROMmatrices}), (\ref{discdirmodes}) and (\ref{discadjmodes}), balanced ROM matrices in ERA are obtained as,
\be
\begin{aligned}
& \mathbf{A}_r = \boldsymbol{\Sigma}_r^{-1/2} \mathbf{U}_r^{*} \mathbf{H}^{\prime}\mathbf{V}_r \boldsymbol{\Sigma}_r^{-1/2},\ \
& \mathbf{B}_r = \boldsymbol{\Sigma}_r^{1/2} \mathbf{V}_r^{*} \begin{bmatrix}
\mathbf{I}_p & \mathbf{0} \\
\mathbf{0} & \mathbf{0} 
\end{bmatrix},\ \
& \mathbf{C}_r = \begin{bmatrix}
\mathbf{I}_q & \mathbf{0} \\
\mathbf{0} & \mathbf{0} 
\end{bmatrix} \mathbf{U}_r  \boldsymbol{\Sigma}_r^{1/2},
\end{aligned}
\ee
without the need for adjoint simulations to compute the inverse transformation matrix (\ref{discadjmodes}). Here, $\mathbf{I}_p \in \mathbb{R}^{p \times p}$ and $\mathbf{I}_q \in \mathbb{R}^{q \times q}$ are identity matrices, $p$ is the number of inputs and $q$ is the number of outputs \cite{Ma:2011, Brunton:19bk}.

\subsection{Output Domain Decomposition}
Under-sampling the FOM solution through training reduces the offline computation costs associated with model reduction. However, when sharp gradients, discontinuities or shock waves are present, higher sampling rates are required to avoid numerical oscillations as a result of unresolved dynamics. We propose an output domain decomposition method to confine high sampling frequencies to subdomains with sharp gradients and reduce offline computation costs of model reduction. Considering the discrete-time linear system (\ref{dlti}), we proceed by partitioning the output matrix $\mathbf{C}$ into $q_m$ blocks that represent $q_m$ subdomains in the output space:
\begin{equation}
\label{dd}
    \mathbf{C} = [
\mathbf{C}_1^T \ \
\mathbf{C}_2^T \ \
\vdots \ \
\mathbf{C}_{q_m}^T]^T,
\end{equation}

where each block is used in a ROM trained with a different sampling frequency that is adjusted according to the smoothness of the solution in the subdomain. Sampling frequency is increased in subdomains that contain sharp gradients to resolve the gradients, but the high sampling frequency is confined only to individual subdomains to maintain computational feasibility. As a result, we avoid training one full-state output ROM with high sampling frequencies that entail prohibitive computation costs. The balanced ROM with output domain decomposition takes the following structure:
\be
\begin{aligned}
\label{BROMDD}
& \dot{\mathbf{x}}_{r}(t) = \mathbf{A}_{rdd} \mathbf{x}_{r}(t) + \mathbf{B}_{rdd} u(t)\ \ ; \ \
& \mathbf{y}(t) = \mathbf{C}_{rdd} \mathbf{x}_{r}(t),
\end{aligned}
\ee
where,
\be
\begin{aligned}
& \mathbf{A}_{rdd} = diag(\mathbf{A}_{r_i}),\ \
& \mathbf{B}_{rdd} = \begin{bmatrix}
\mathbf{B}_{r_1} & \dots & \mathbf{B}_{r_{q_m}} \\
\end{bmatrix}^T,\ \
& \mathbf{C}_{rdd} = diag(\mathbf{C}_{r_i}),
\end{aligned}
\ee
and for $i = 1, \dots, q_m$, we have,
\be
\begin{aligned}
& \mathbf{A}_{r_i} = \boldsymbol{\Sigma}_{r_i}^{-1/2} \mathbf{U}_{r_i}^{*} \mathbf{H}^{\prime}\mathbf{V}_{r_i} \boldsymbol{\Sigma}_{r_i}^{-1/2},\ \
& \mathbf{B}_{r_i} = \boldsymbol{\Sigma}_{r_i}^{1/2} \mathbf{V}_{r_i}^{*} \begin{bmatrix}
\mathbf{I}_p & \mathbf{0} \\
\mathbf{0} & \mathbf{0} 
\end{bmatrix},\ \
& \mathbf{C}_{r_i} = \begin{bmatrix}
\mathbf{I}_{q_i} & \mathbf{0} \\
\mathbf{0} & \mathbf{0} 
\end{bmatrix} \mathbf{U}_{r_i}  \boldsymbol{\Sigma}_{r_i}^{1/2}.
\end{aligned}
\ee

\subsection{Tangential Interpolation}
Systems with lightly-damped impulse response - and hence a longer Markov sequence - require excessive storage and a high computational cost for SVD of a high-dimensional Hankel matrix. Tangential interpolation is employed as a way to reduce the offline cost of ERA in multi-input multi-output (MIMO) systems by projecting the impulse response onto the leading left and right tangential directions \cite{Kramer:2016}. Two orthogonal projection matrices are chosen for the entire sequence of Markov parameters to retain the structure of the Hankel matrix \cite{Kramer:2016}. Therefore, the impulse response snapshots are first assembled in matrix $\mathbf{Q}_L$ as,
\be
\mathbf{Q}_L = \begin{bmatrix}
\mathbf{y}_1 & \mathbf{y}_2 & \dots & \mathbf{y}_{m_o + m_p -1}
\end{bmatrix}.
\ee
We seek the solution to the following optimization problem,
\be
\mathbf{P}_1 = argmin_{rank(\tilde{\mathbf{P}}_1) = l_1} \norm{\tilde{\mathbf{P}}_1 \mathbf{Q}_L - \mathbf{Q}_L}^2_F, 
\ee
where, $\mathbf{P}_1 = \mathbf{W}_1 \mathbf{W}_1^*$, and $\mathbf{W}_1$ contains the first $l_1$ columns of $\mathbf{U}_L$ given by SVD of the impulse response snapshots $\mathbf{Q}_L = \mathbf{U}_L \boldsymbol{\Sigma}_L \mathbf{V}_L^*$. Similarly, to obtain the right tangential directions we form,
$
\mathbf{Q}_R = [
\mathbf{y}_1^T \ \
\mathbf{y}_2^T \ \
\vdots \ \
\mathbf{y}_{m_o + m_p -1}^T
]^T,
$
and solve the optimization problem,
\be
\mathbf{P}_2 = argmin_{rank(\tilde{\mathbf{P}}_2) = l_2} \norm{\mathbf{Q}_R \tilde{\mathbf{P}}_2 - \mathbf{Q}_R}^2_F,
\ee
that is equivalent to SVD of the sequence of Markov parameters $\mathbf{Q}_R = \mathbf{U}_R \boldsymbol{\Sigma}_R \mathbf{V}_R^*$, where, $\mathbf{P}_2 = \mathbf{W}_2 \mathbf{W}_2^*$ and $\mathbf{W}_2 = \mathbf{V}_R (:, 1:l_2)$ \cite{Kramer:2016}. The new sequence of Markov parameters for ERA is obtained by projection of the original impulse response sequence $\hat{\mathbf{y}}_i = \mathbf{W}_1^* \mathbf{y}_i \mathbf{W}_2$. However, the balanced ROM trained with this sequence has $l_1$ outputs and $l_2$ inputs. Therefore, the original input-output dimensions are recovered by $\mathbf{A}_r = \hat{\mathbf{A}}$, $\mathbf{B}_r = \hat{\mathbf{B}} \mathbf{W}_2^*$, and $\mathbf{C}_r = \mathbf{W}_1 \hat{\mathbf{C}}$, where, $\hat{\mathbf{A}}$, $\hat{\mathbf{B}}$, and $\hat{\mathbf{C}}$ are the balanced ROM matrices obtained by applying ERA to the projected impulse response sequence $\hat{\mathbf{y}}_i$ \cite{Kramer:2016}.

\section{Application}\label{appl}
The one-dimensional Navier-Stokes equations with species transport and reaction are given by,
\be
\label{nonlinearfom}
\pd{\mathbf{q}_c}{t} + \pd{\mathbf{f}}{x} - \pd{\mathbf{f}_v}{x} = \mathbf{s},
\ee
and, 
\be
\mathbf{q}_c =  \begin{bmatrix} \rho \\ \rho u \\ \rho h^0 - p \\ \rho Y_l  \end{bmatrix}, \quad
\mathbf{f} =  \begin{bmatrix} \rho u \\ \rho u^2 + p \\ \rho h^0 u \\ \rho Y_l  \end{bmatrix}, \quad
\mathbf{f}_v =  \begin{bmatrix} 0 \\ \tau \\ u \tau - q \\ -\rho V_l Y_l  \end{bmatrix}, \quad
\mathbf{s} =  \begin{bmatrix} 0 \\ 0 \\ 0 \\ \dot{\omega}_l  \end{bmatrix}, 
\ee
where, $\mathbf{q}_c$ is the vector of conservative variables, $\mathbf{f}$ and $\mathbf{f}_v$ are vectors of inviscid and viscous fluxes, respectively, and $\mathbf{s}$ denotes the source term. $\rho$, $u$, $h^0$, and $p$ are density, velocity, stagnation enthalpy, and pressure respectively. $\tau$ is the shear stress, $q$ is the heat flux, $Y_l$, and $V_l$ are the species mass fraction and diffusion velocity of the $l^{th}$ chemical species. Here, $\dot{\omega}_l$ is the production rate of the $l^{th}$ chemical species,
\be
\dot{\omega}_l = W_l \sum_{m=1}^M (\nu_{lm}'' - \nu_{lm}') w_m,
\ee
where, $W_l$ is the molecular weight of the $l^{th}$ species. $\nu_{lm}'$ and $\nu_{lm}''$ are the stoichiometric coefficients of the $l^{th}$ reactant and product species in the $m^{th}$ reaction, respectively. Considering the $m^{th}$ reaction between $n_s$ species,
\be
\sum_{l=1}^{n_s} \nu_{l,m}' X_l \mathrel{\mathop{\rightleftharpoons}^{\mathrm{k_{f,m}}}_{\mathrm{k_{b,m}}}} \sum_{l=1}^{n_s} \nu_{l,m}'' X_l,
\ee
$w_m$ represents the rate of progress for the $m^{th}$ reaction,
\be
w_m = k_{f,m} \prod_{l=1}^{n_s} [X_l]^{\nu_{l,m}'} - k_{b,m} \prod_{l=1}^{n_s} [X_l]^{\nu_{l,m}''},
\ee
where, $X_l$ is the molar concentration, and $k_{f,m}$ is the forward reaction rate of the $m^{th}$ reaction that takes the following Arrhenius form \cite{Turns:96bk},
\be
k_{f} = D T^b exp(\frac{-E_a}{RT}).
\ee
$D=2.12 \times 10^{10}$ is the pre-exponential factor, and $T^b$ identifies the temperature dependence. Here, $b=1$, $\frac{E_a}{R}=24358$, $E_a$ is the activation energy threshold for the reaction, and $R=8.31446 \frac{J}{K \cdot mol}$ is the universal gas constant. The backward reaction rate is obtained using the forward rate as $k_{b,m} = \frac{k_{f,m}}{K_{C,m}}$, where, $K_{C,m}$ is the equilibrium constant. 

We use an open access solver called Prototyping Environment for Reacting Flow Order Reduction Methods (PERFORM) \cite{PERFORM:2021} to solve the nonlinear equations with a single-step two-species reaction using a finite volume approach and the second-order Roe scheme. The nonlinear equations are then linearized about the steady-state solution to form the LTI system in equation (\ref{lti}). The linearized system Jacobian is,
\be
\label{linearJ}
\mathbf{J} = \left [ \boldsymbol{\Gamma}^{-1} (\mathbf{D} - \mathbf{L}_{R} + \mathbf{L}_{L}) \right ]_{\bar{\mathbf{q}}},
\ee
where, $\boldsymbol{\Gamma} = \pd{\mathbf{q}_c}{\mathbf{q}}$, $\mathbf{D} = \pd{\mathbf{s}}{\mathbf{q}}$, $\mathbf{L} = \pd{\mathbf{f}}{\mathbf{q}} - \pd{\mathbf{f_v}}{\mathbf{q}}$ with R and L subscripts denoting the right and left neighboring cells, and $\mathbf{q} = \left [p \quad u \quad T \quad Y_k \right ]^T$ is the vector of primitive variables. The steady-state solution ($\bar{\mathbf{q}}$) is obtained by initializing the velocity, temperature, and species mass fraction with step functions and reducing the bulk velocity until the flame is stationary. The semi-discretized equations are advanced in time with the third-order Runge-Kutta scheme. Two-species reaction is considered here and a calorically perfect gas is assumed. Non-reflective characteristic boundary conditions are applied through ghost cells at the left and right boundaries and input forcing is applied through back pressure perturbation. 

ERA is used for balancing transformation of the linearized equations. The linearized system is impulsively excited by the following step function to generate training snapshots for balancing transformation,
\be
p_{back} = \delta (t) = \begin{cases}
            1 \quad Pa \qquad t=0 \\
            0 \quad Pa \qquad t>0
            \end{cases}.
\ee
The performance of the balanced ROMs is then evaluated in reconstruction of the impulse response, and prediction of the dynamics for unseen sinusoidal pressure forcing applied through, 
\be
\label{sine}
p_{back} = \alpha \sin(2 \pi \omega t),
\ee
where, $\alpha$ and $\omega$ are perturbation amplitude and frequency, respectively. Predictive performance of the balanced ROM is compared against the standard Galerkin projection and LSPG. Accuracy of the linearized ROMs is measured by the relative error,
\be
e^k = \frac{\norm{\mathbf{q}^k - \tilde{\mathbf{q}}^k}_2}{\norm{\mathbf{q}^k}_2}, \qquad k=1, \dots, n_t
\ee
where, $\mathbf{q}^k$ and $\tilde{\mathbf{q}}^k$ are the FOM and ROM solutions at the $k^{th}$ time step, respectively,  and $n_t$ is the total number of time steps.

\section{Results}\label{results}
\subsection{Full Order Model}
\begin{figure}[htbp!]
  \centering
  \begin{minipage}[a]{0.59\textwidth}
    \includegraphics[trim=0.3cm 4 0.5cm 4, clip, width=\textwidth]{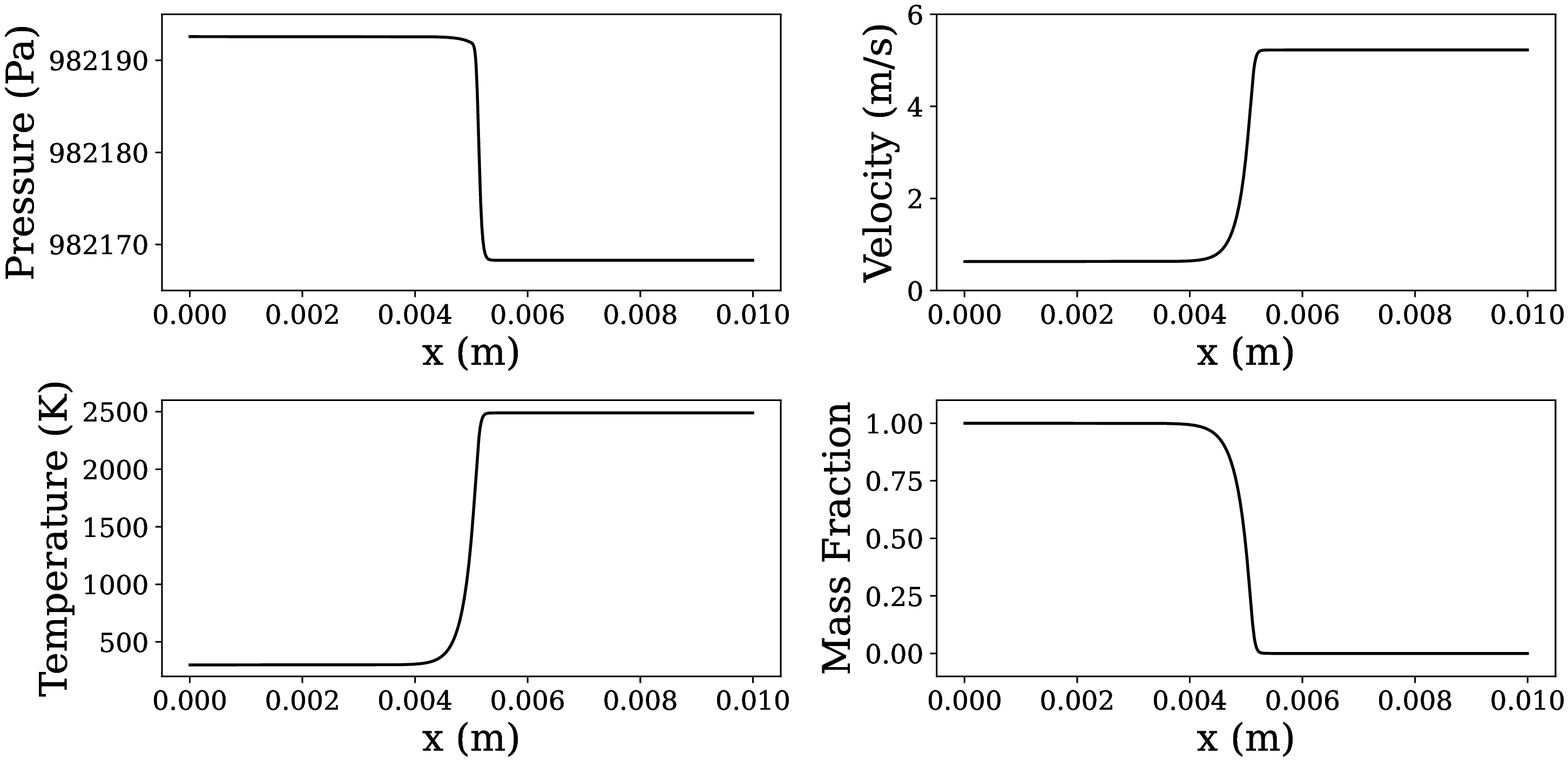}
  \end{minipage}
  \centering
  \begin{minipage}[a]{0.40\textwidth}
    \begin{tikzpicture}
    \node[above left] (img) at (0,0) {\includegraphics[trim=3.1cm 4.05cm 2.5cm 6cm, clip, width=0.93\textwidth]{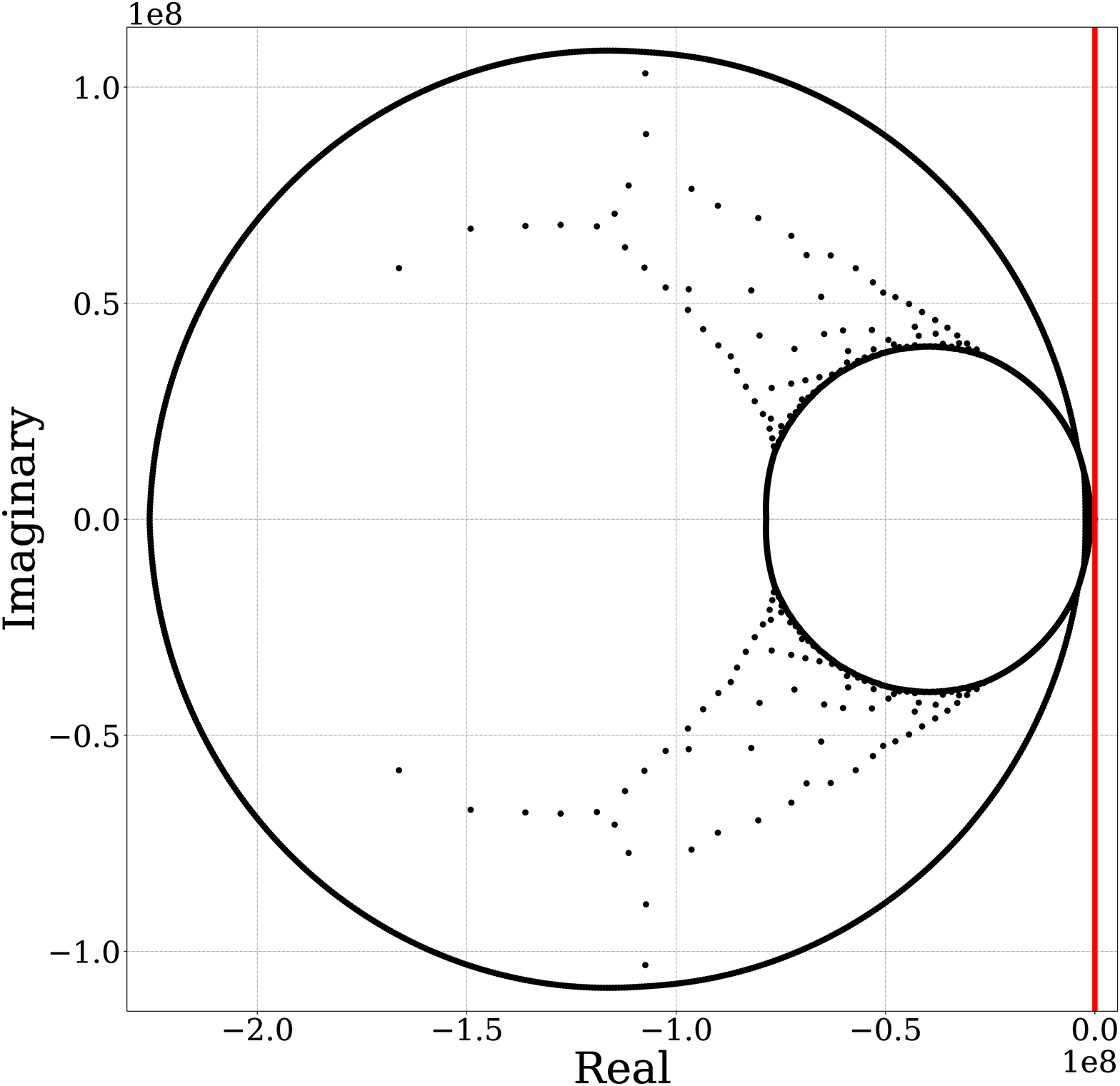}};
    \node at (-5.2,2.5){\rotatebox{90}{Imaginary}};
    \node at (-4.3,4.9){$1 \times 10^{8}$};
    \node at (-0.5,0.0){$1 \times 10^{8}$};
    \node at (-2.5,-0.1){Real};
    \end{tikzpicture}
  \end{minipage}
  \centering
  \caption{Steady-state solution (left) obtained by numerical solution of equation (\ref{nonlinearfom}). Eigenvalues of the linearized FOM (right). The red vertical line shows the continuous-time stability margin.}
   \label{f:ss_FOMEigs}
\end{figure}

The steady-state solution shown in Figure~\ref{f:ss_FOMEigs} is obtained by solving the nonlinear Navier-Stokes equations (\ref{nonlinearfom}) with dual time-stepping in a computational domain consisting of 1000 cells with $\Delta x = 1 \times 10^{-5} m$ and a time step of $\Delta t = 1 \times 10^{-8} sec$.
For subsonic inlet boundary conditions, we specify upstream pressure, temperature, and the reactant and product species mass fraction, whereas, back pressure is specified for the subsonic outlet conditions. Table~\ref{t:bc} lists the prescribed upstream and back pressure values at the boundaries.
\begin{table}[h!]
 \begin{center}
  \caption{Prescribed inlet and outlet boundary conditions.}
  \label{t:bc}
  \begin{tabular}{lllll}\hline
        & Upstream pressure & 984.284 kPa \\\hline
        & Upstream temperature & 300.16 K  \\\hline
        & Upstream species mass fraction & $[1.0, 0.0]$  \\\hline
        & Back pressure & 976.139 kPa  \\\hline
  \end{tabular}
 \end{center}
\end{table}

The eigenvalues of the linearized system Jacobian in equation (\ref{linearJ}) are shown in Figure~\ref{f:ss_FOMEigs}, which shows that all of the eigenvalues reside in the left half of the complex plane (the maximum real component is -180.44) and the continuous-time system is stable.

\begin{figure}[h!]
  \centering
  \begin{minipage}[a]{0.45\textwidth}
    \includegraphics[trim=4 4 4 2cm, clip, width=\textwidth]{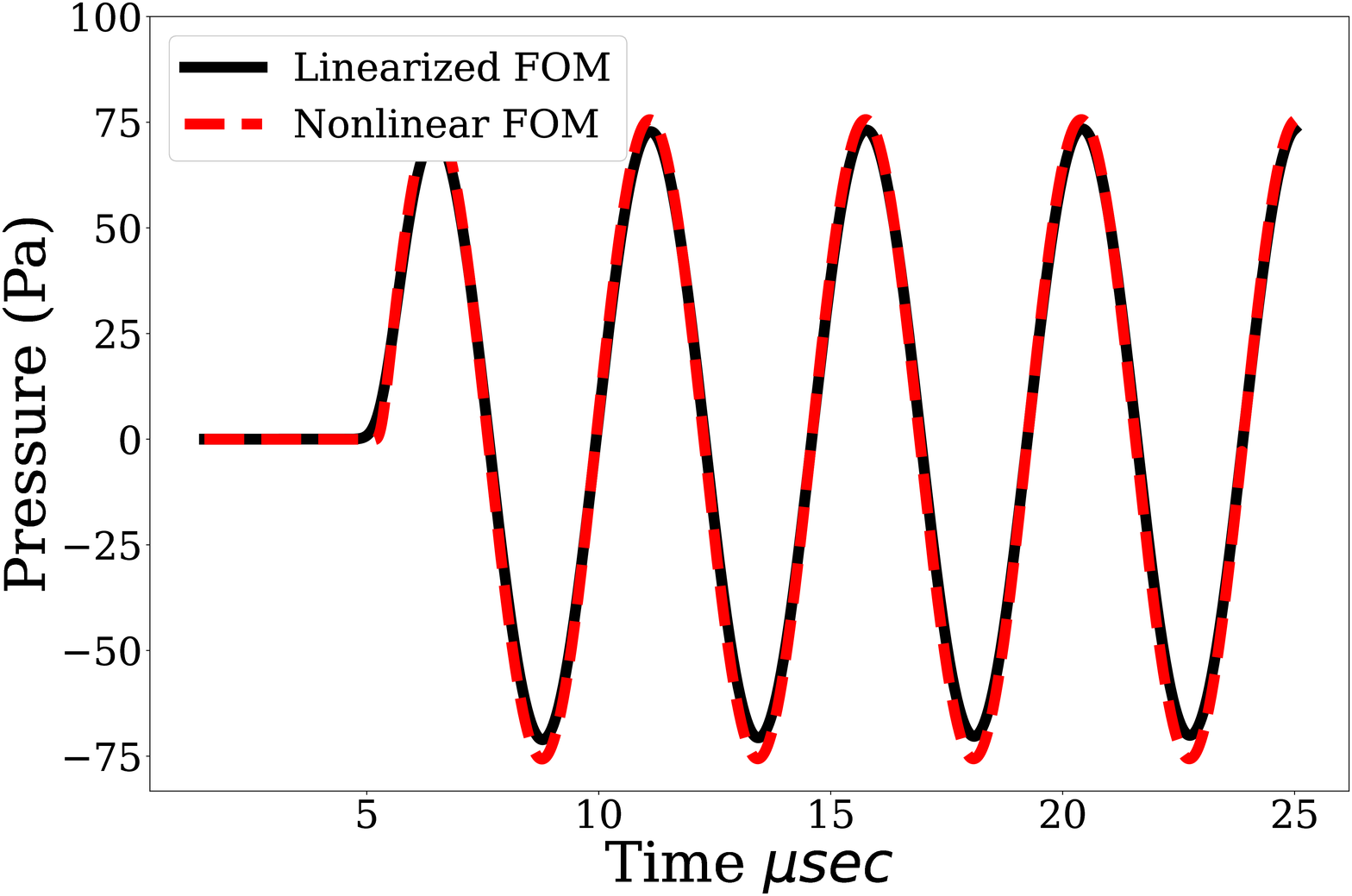}
  \end{minipage}
  \centering
  \begin{minipage}[a]{0.45\textwidth}
    \includegraphics[trim=4 4 4 2cm, clip, width=\textwidth]{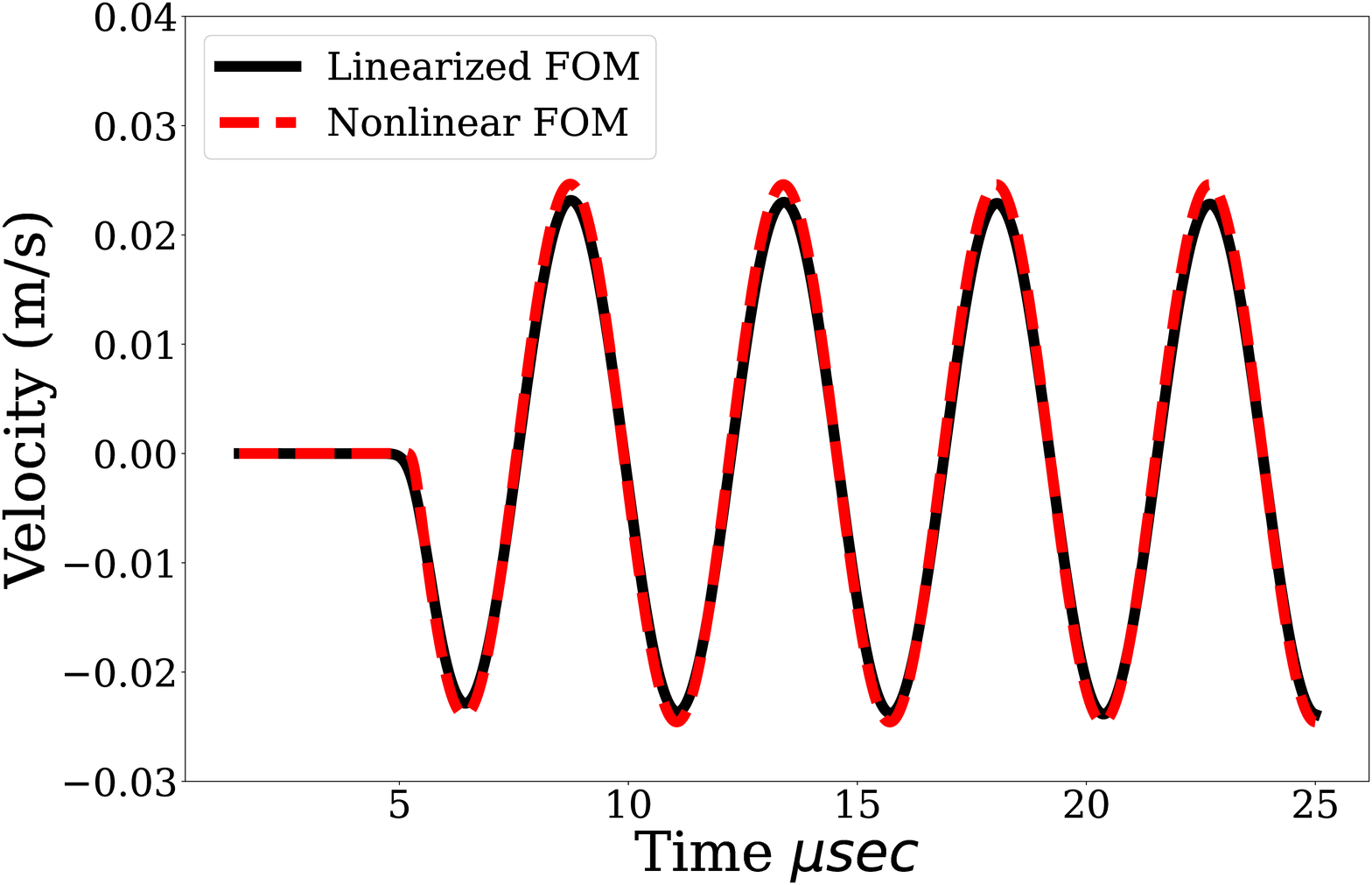}
  \end{minipage}
   \centering
  \begin{minipage}[a]{0.45\textwidth}
    \includegraphics[trim=4 4 4 2cm, clip, width=\textwidth]{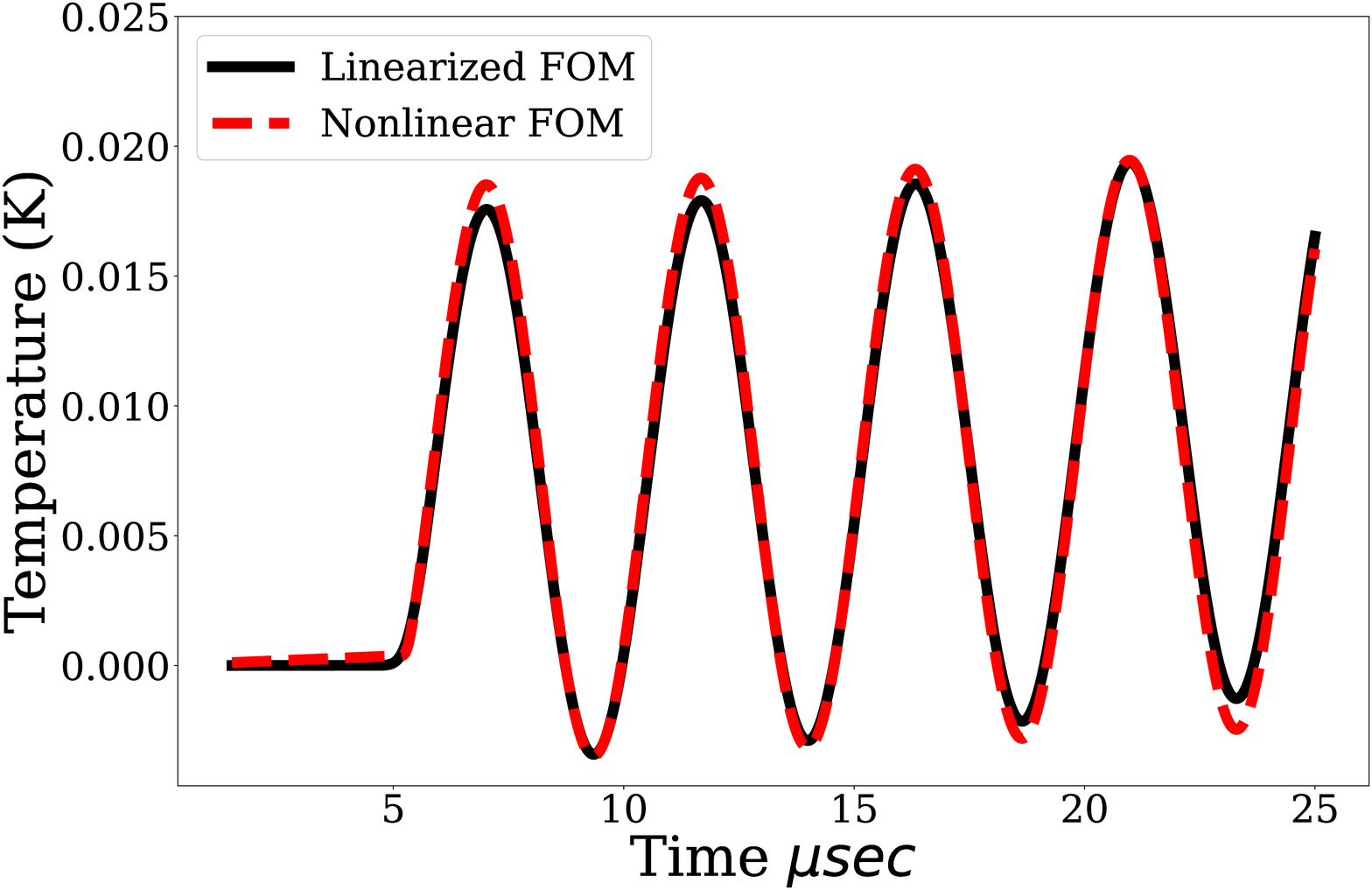}
  \end{minipage}
   \centering
  \begin{minipage}[a]{0.45\textwidth}
    \includegraphics[trim=4 4 4 2cm, clip, width=\textwidth]{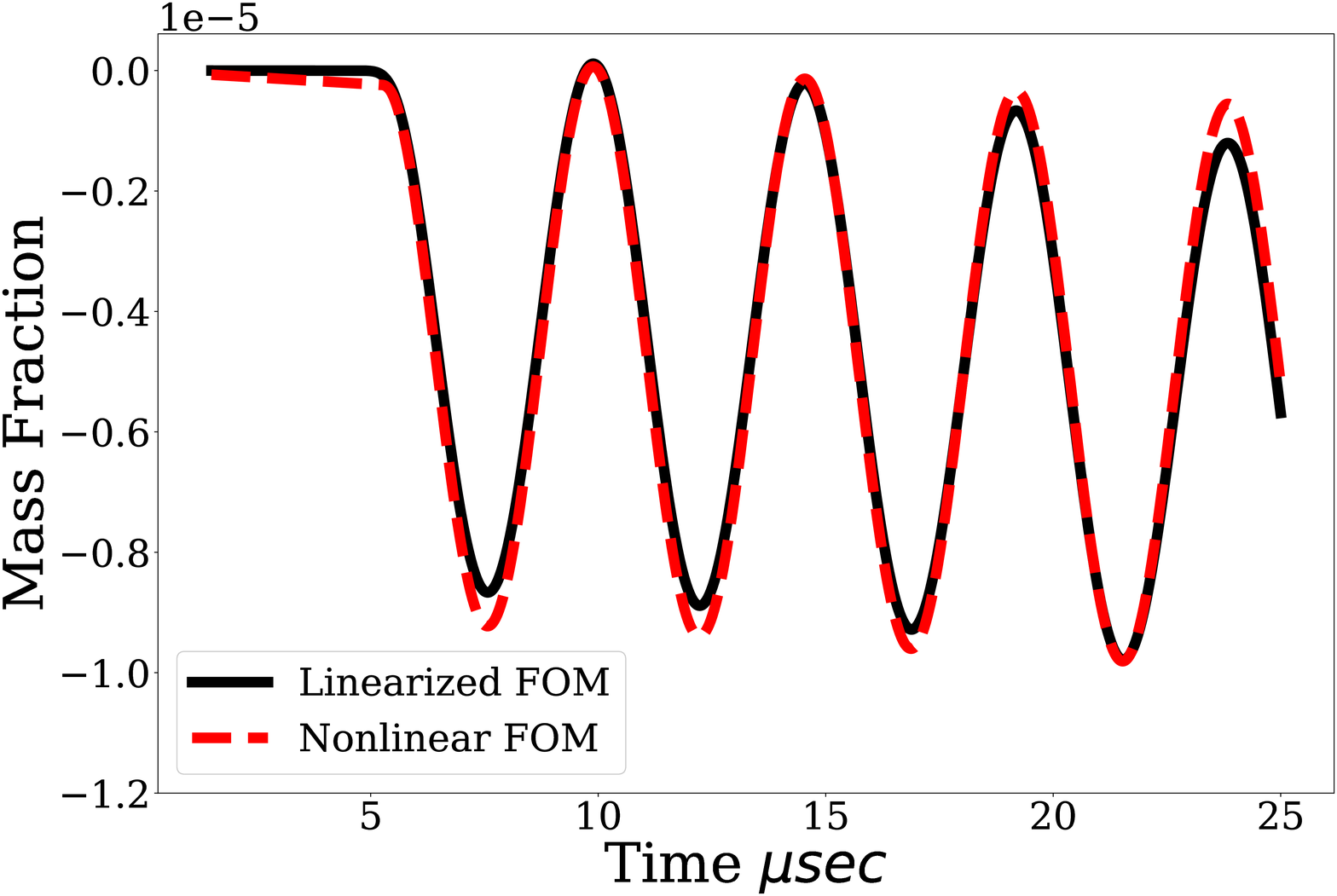}
  \end{minipage}
  \centering
  \caption{Perturbation variables computed by the linearized and nonlinear FOMs at $x=0.0045 m$.}
   \label{linearvsnonlinear}
\end{figure}

Figure~\ref{linearvsnonlinear} compares the time evolution of the state perturbation at $x = 0.0045 m$ computed by the linearized and nonlinear FOMs. The back pressure is perturbed by the sinusoidal input shown in equation (\ref{sine}) with a perturbation amplitude of $0.01 \% p_{back}$ and a frequency of 215 kHz. The third-order Runge-Kutta scheme is then used to solve the linearized and nonlinear equations with a time step of $\Delta t = 1 \times 10^{-9} sec$. 

\subsection{ROM Construction}

The stiffness introduced by the chemical source term results in an ill-conditioned Jacobian matrix, where the condition number of the Jacobian of the linear system of equations (\ref{linearJ}) is as high as $4.82 \times 10^{13}$. An ill-conditioned transformation matrix is a consequence of the stiffness of the linearized system, which presents challenges in  transforming the system via the analytical BT approach. ERA on the other hand, uses the empirical shifted Hankel matrix to bypass computation of the Gramians and the inverse transformation matrix without additional costs associated with adjoint system simulations.

\begin{figure}[h!]
  \centering
  \begin{minipage}[a]{0.40\textwidth}
    \includegraphics[trim=4 0.3cm 1cm 2cm, clip, width=\textwidth]{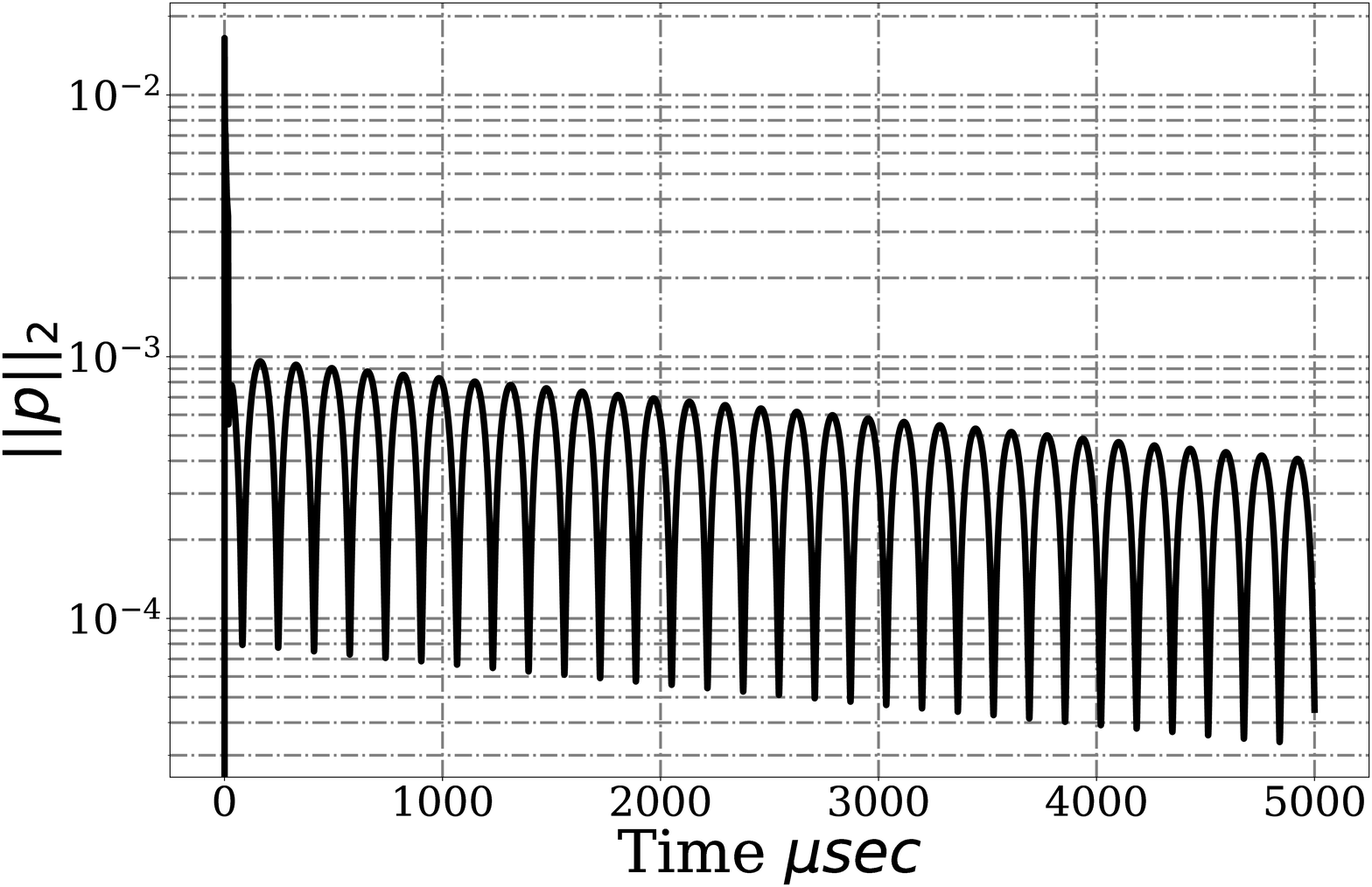}
     \rput(0.9,0.45){\psscalebox{0.9}{\color{black} \textbf{a)}}}
     \vspace{0.1cm}
  \end{minipage}
  \centering
  \begin{minipage}[a]{0.40\textwidth}
    \includegraphics[trim=4 0.3cm 1cm 2cm, clip, width=\textwidth]{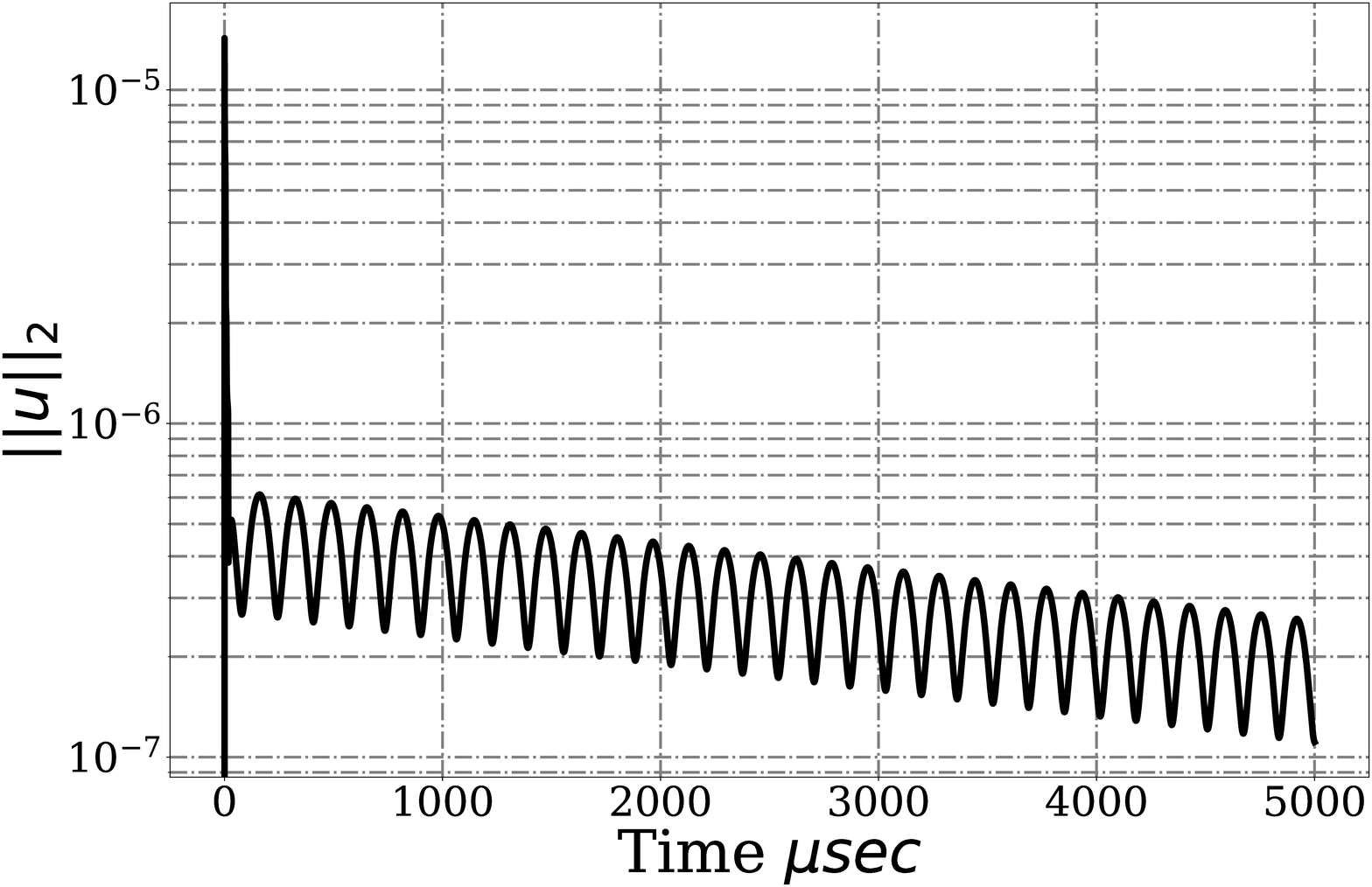}
     \rput(0.9,0.45){\psscalebox{0.9}{\color{black} \textbf{b)}}}
     \vspace{0.1cm}
  \end{minipage}
   \centering
  \begin{minipage}[a]{0.40\textwidth}
    \includegraphics[trim=4 0.3cm 1cm 2cm, clip, width=\textwidth]{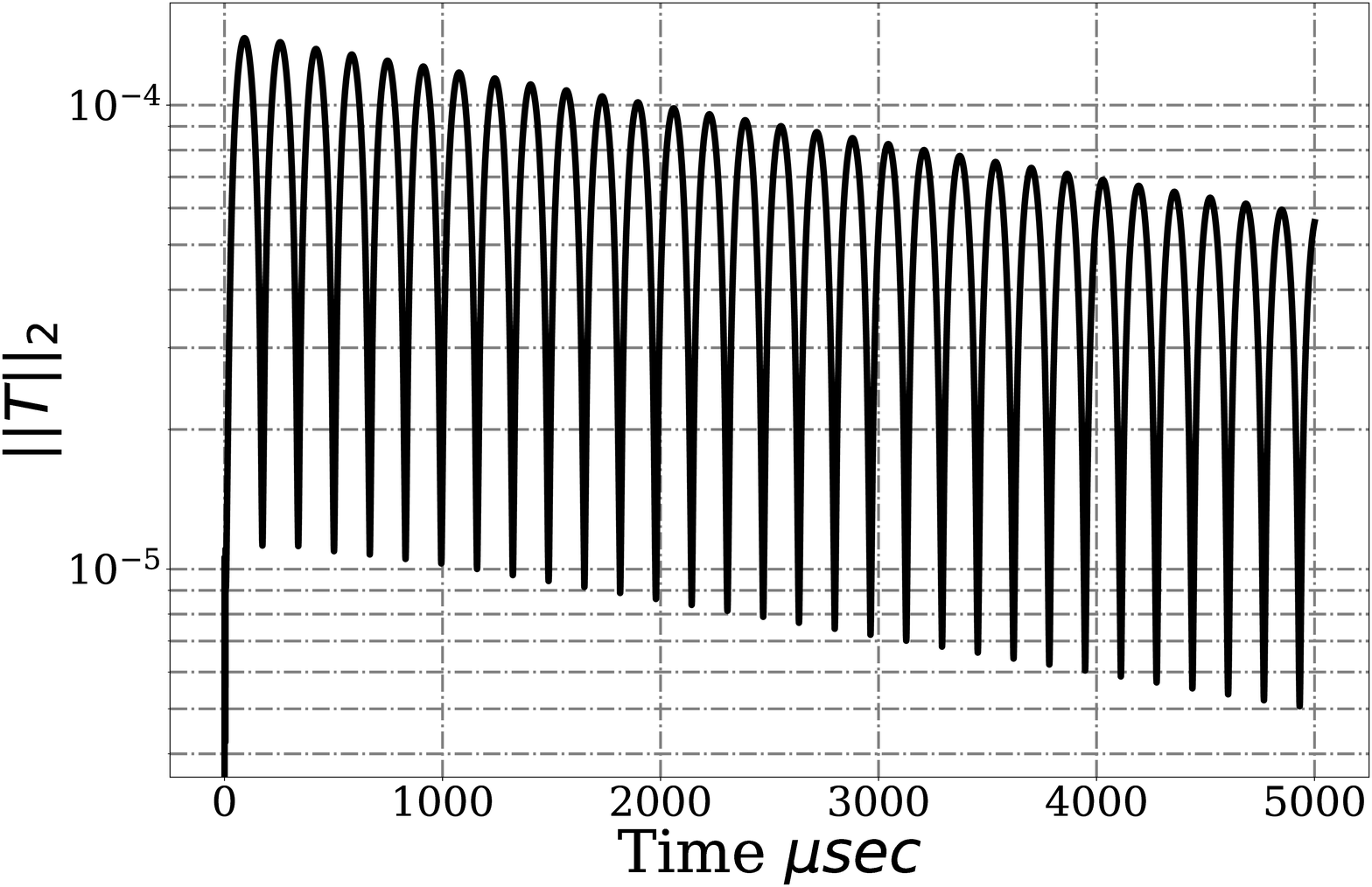}
     \rput(0.9,0.45){\psscalebox{0.9}{\color{black} \textbf{c)}}}
  \end{minipage}
   \centering
  \begin{minipage}[a]{0.40\textwidth}
    \includegraphics[trim=4 0.3cm 1cm 2cm, clip, width=\textwidth]{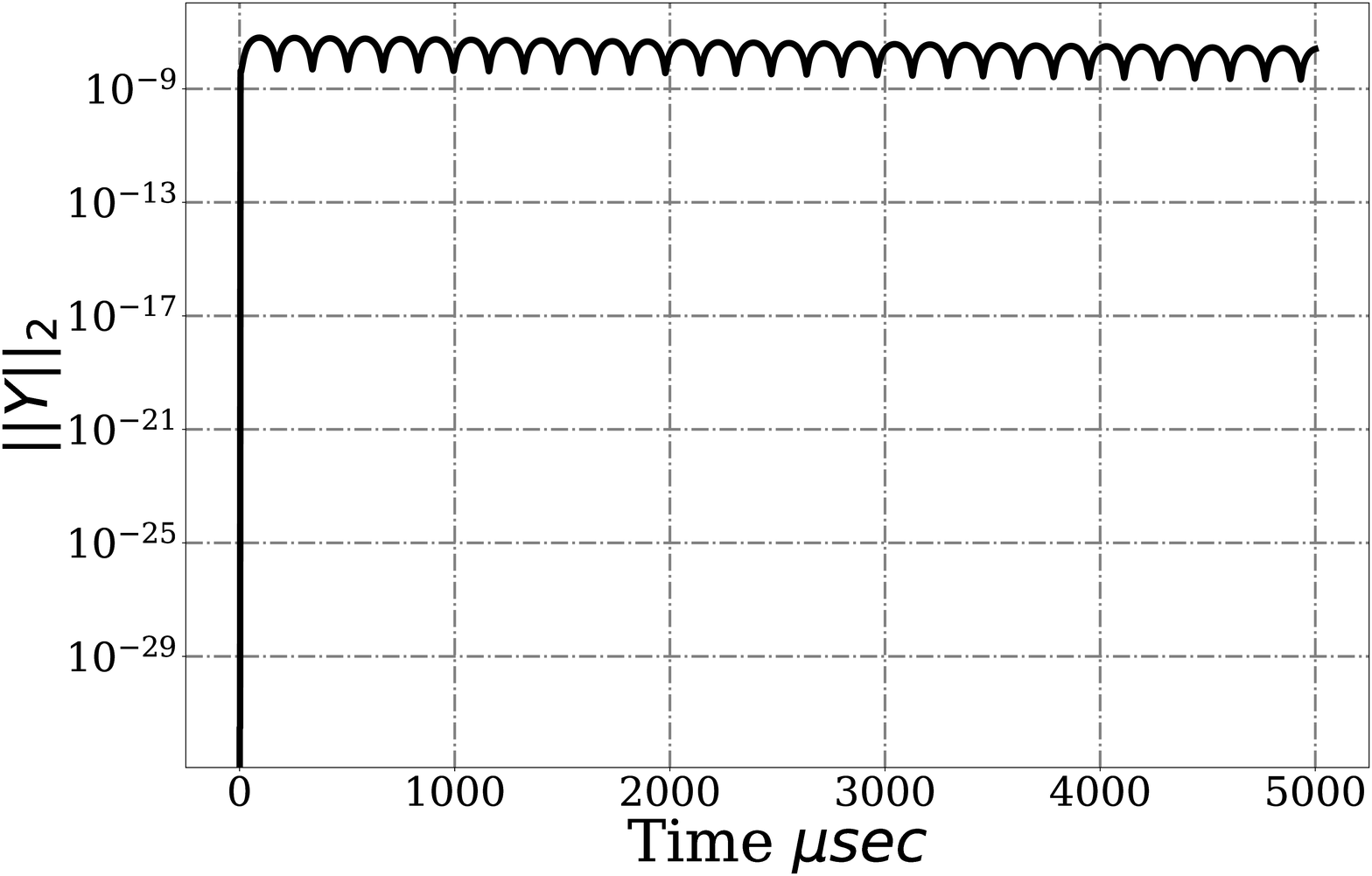}
     \rput(0.9,0.45){\psscalebox{0.9}{\color{black} \textbf{d)}}}
  \end{minipage}
  \centering
  \caption{Norm of impulse response computed for a) pressure, b) velocity, c) temperature, and d) species mass fraction.}
   \label{f:statenorm}
\end{figure}

The performance of data-driven models is driven by sampling properties. In non-intrusive balancing transformation, the frequency and sampling time of the impulse response impacts robustness and accuracy of the balanced ROMs. Figure~\ref{f:statenorm} shows the norm of the state when snapshots are collected every 100 time steps. It is clear that after the initial transients, the system undergoes lightly-damped oscillations that take a long time to decay. Tu and Rowley \cite{Tu:12JCP} classified this behavior as a simple impulse response tail, where following the initial non-normal growth, only a few eigenvectors remain active in the flow, and these eigenvectors persist for a long interval of time. Therefore, it is critical to sample the impulse response for a long enough time to capture the decay of the remaining lightly-damped structures.

\begin{figure}[h!]
  \centering
  \begin{minipage}[a]{0.32\textwidth}
    \includegraphics[trim=4 0.3cm 4 1.2cm, clip, width=\textwidth]{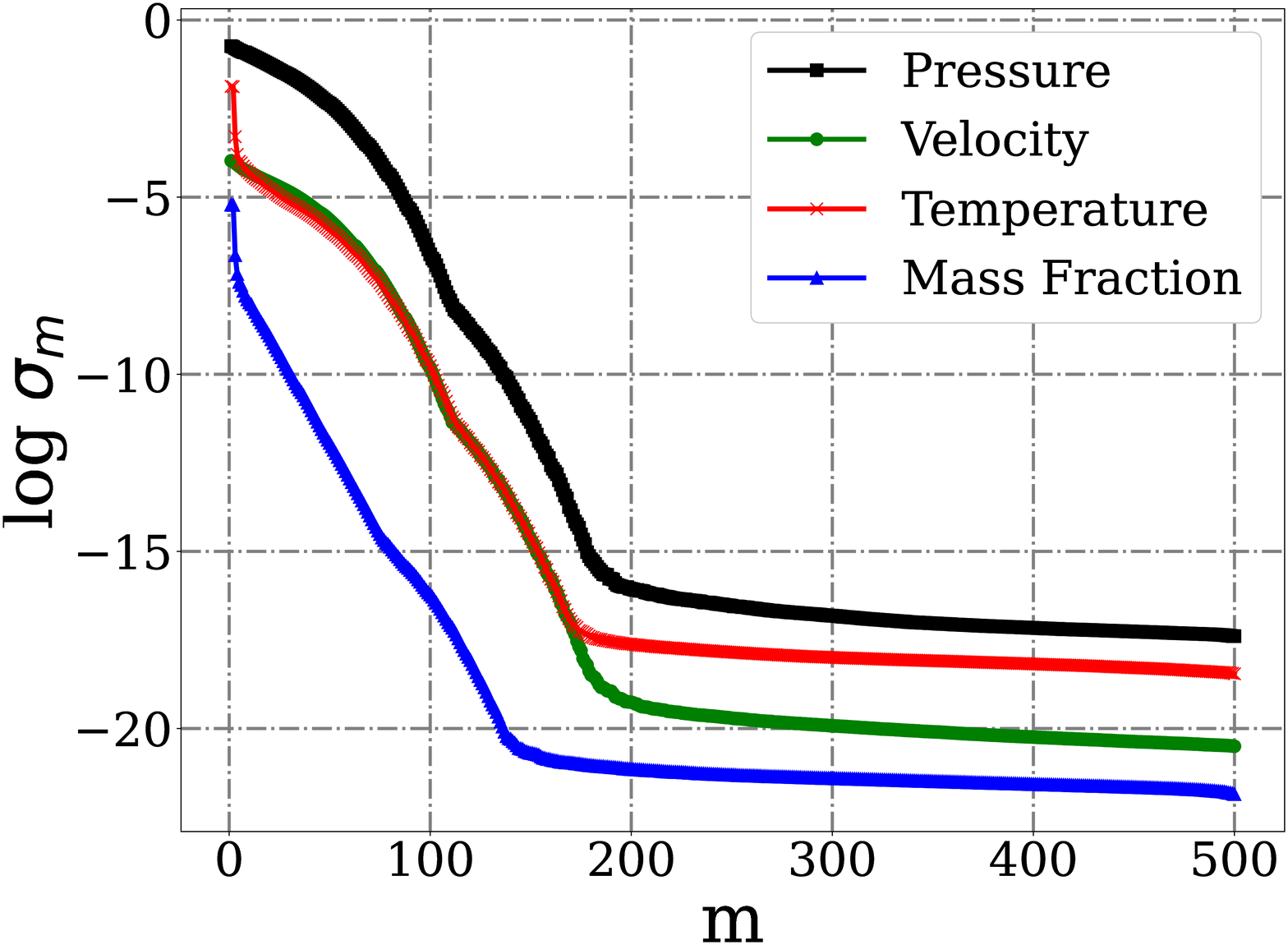}
    %\rput(0.9,0.0){\psscalebox{0.9}{\color{black} \textbf{a)}}}
     \includegraphics[trim=4 0.2cm 4 0.2cm, clip, width=\textwidth]{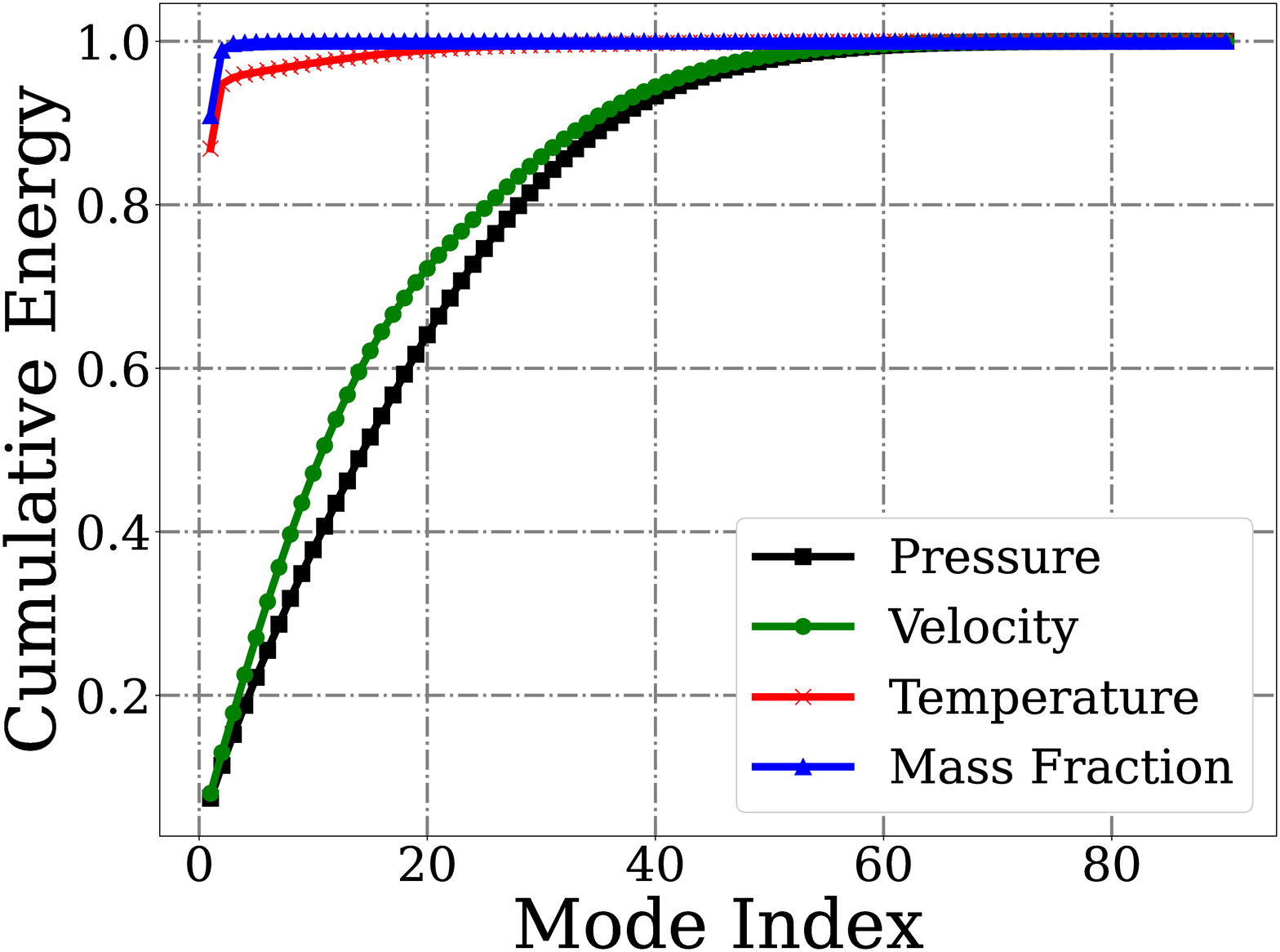}
  \end{minipage}
  \centering
  \begin{minipage}[a]{0.55\textwidth}
    \includegraphics[trim=4 1cm 4 1.2cm, clip, width=\textwidth]{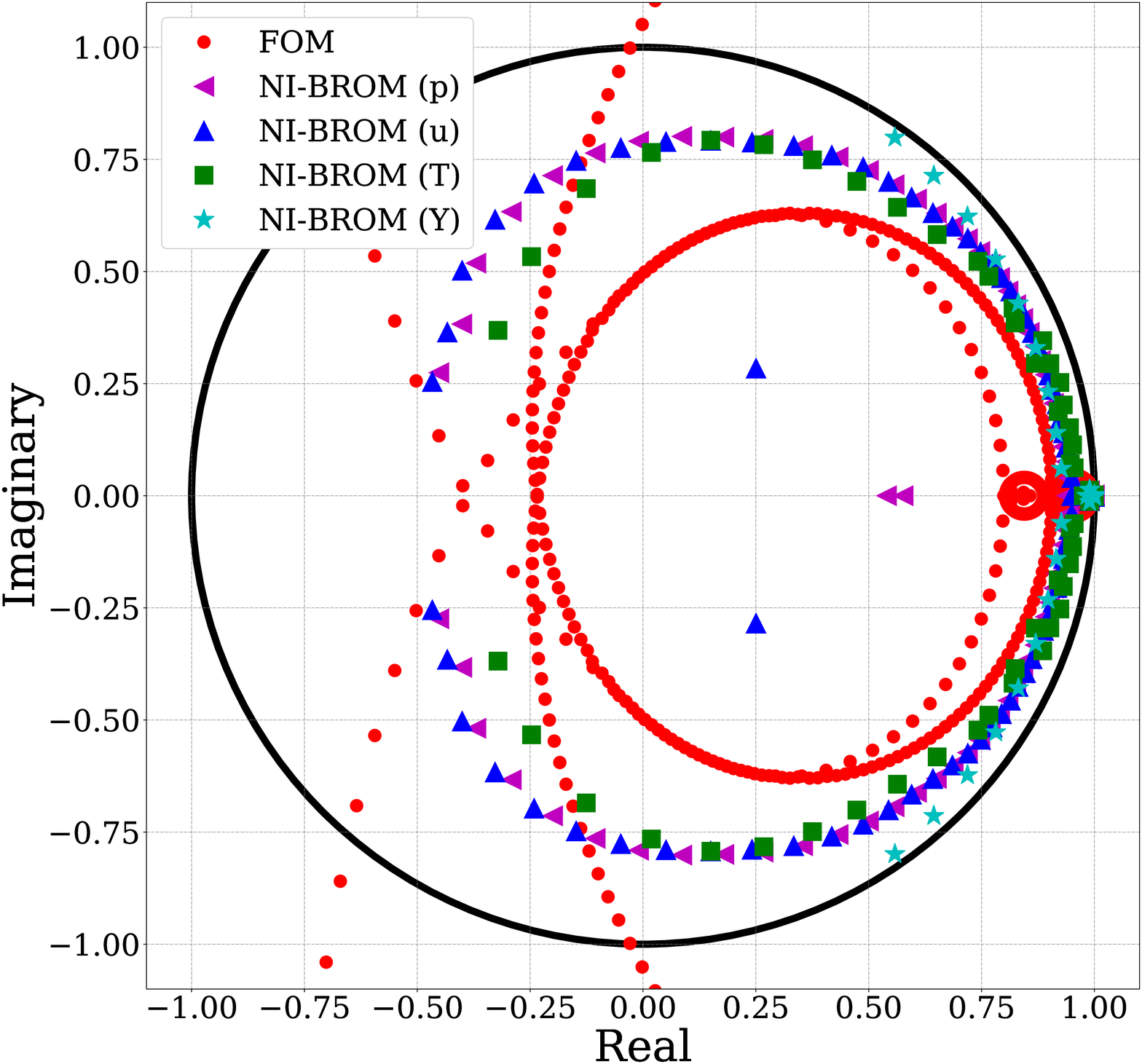}
    %\rput(0.9,0.35){\psscalebox{0.9}{\color{black} \textbf{b)}}}
  \end{minipage}
   \centering
  \caption{Hankel singular values (top left). Cumulative energy of POD modes based on the snapshots of unit impulse response (bottom left). Eigenvalues of balanced ROMs (NI-BROM) and the linearized FOM (right). FOM is discretized with the third-order Runge-Kutta scheme and time step is increased to $\Delta t = 1 \times 10^{-7} sec$ to match the time step of balanced ROMs. The black unit circle marks the discrete-time margin of stability.}
   \label{f:HSV_POD_ERAEigs}
\end{figure}

We collect a total of 1000 snapshots of the impulse response, which covers $100 \mu sec$ of the dynamics, every 100 time steps (equivalent to a sampling period of $T_s=0.1 \mu sec$) to train ROMs. Shorter sampling times were found to miss the evolution of the lightly-damped low-frequency eigenvector. As a result, the underlying empirical Gramians are not balanced, and theoretical error bounds are not satisfied, which leads to inaccurate or unstable ROMs. On the other hand, lower sampling frequencies prevent capturing some of the critical high-frequency structures and affects the predictive performance of the balanced ROMs. Figure~\ref{f:HSV_POD_ERAEigs} shows the Hankel singular values for pressure, velocity, temperature, and species mass fraction. Capturing $99.99 \%$ of the input-output energy requires 77 modes for pressure and velocity, 50 modes for temperature, and 19 modes for species mass fraction.

Balanced ROMs are then constructed with ERA. Eigenvalues of the ROMs are shown in Figure~\ref{f:HSV_POD_ERAEigs}. All eigenvalues are located inside the unit circle (with maximum absolute value of $|\lambda|=0.9999976$), which demonstrates discrete-time stability of the balanced ROMs, despite the fact that the time step of balanced ROMs is 100 times larger than the original time step of the FOM ($\Delta t = 1 \times 10^{-9} sec$). Clearly, the linearized FOM is unstable with the larger time step of $\Delta t = 1 \times 10^{-7} sec$. 

\begin{figure}[h!]  
  \centering
  \begin{minipage}[a]{0.42\textwidth}
    \includegraphics[trim=4 4 4 2cm, clip, width=\textwidth]{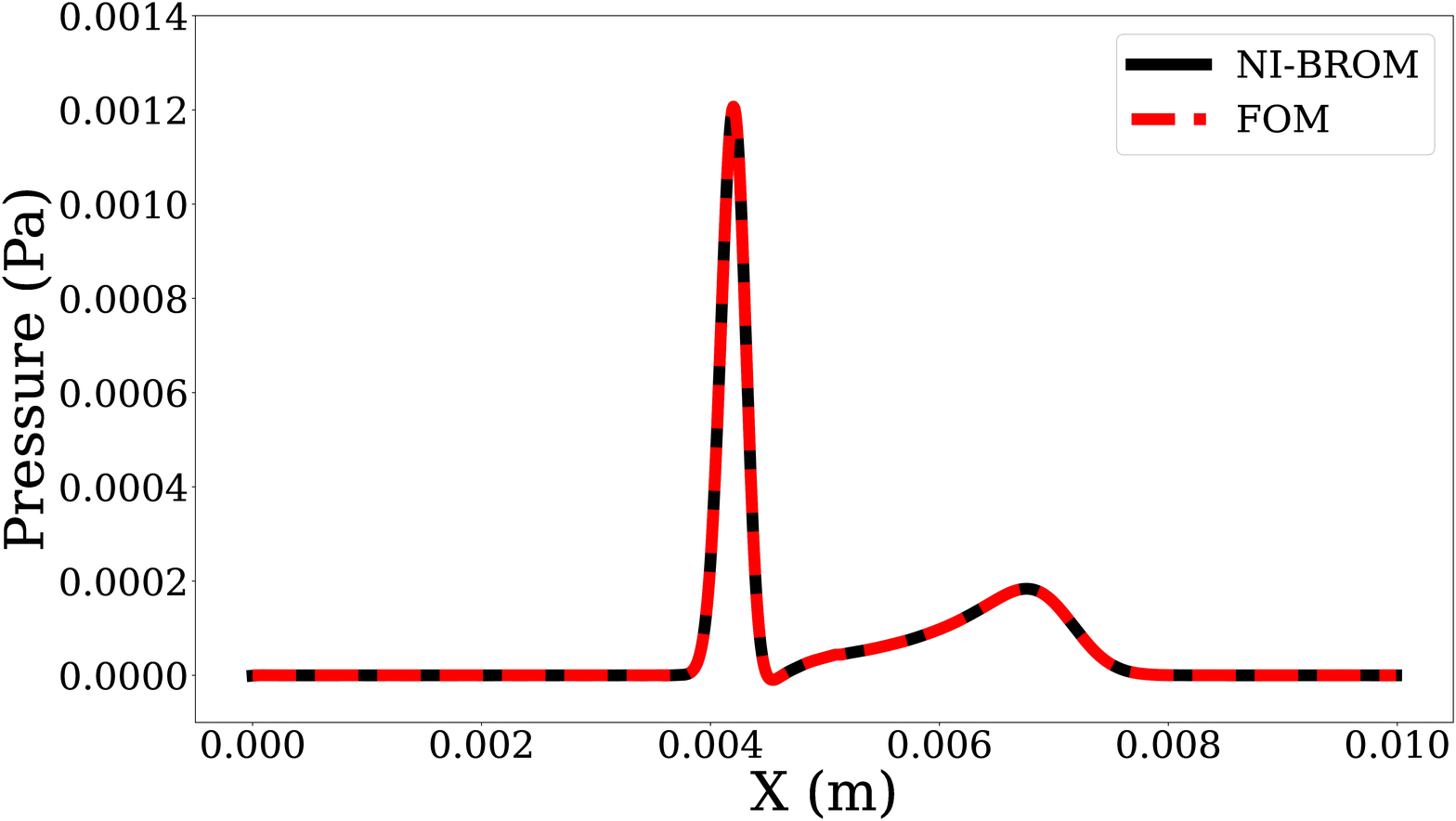}
  \end{minipage}
  \centering
  \begin{minipage}[a]{0.42\textwidth}
    \includegraphics[trim=4 4 4 2cm, clip, width=\textwidth]{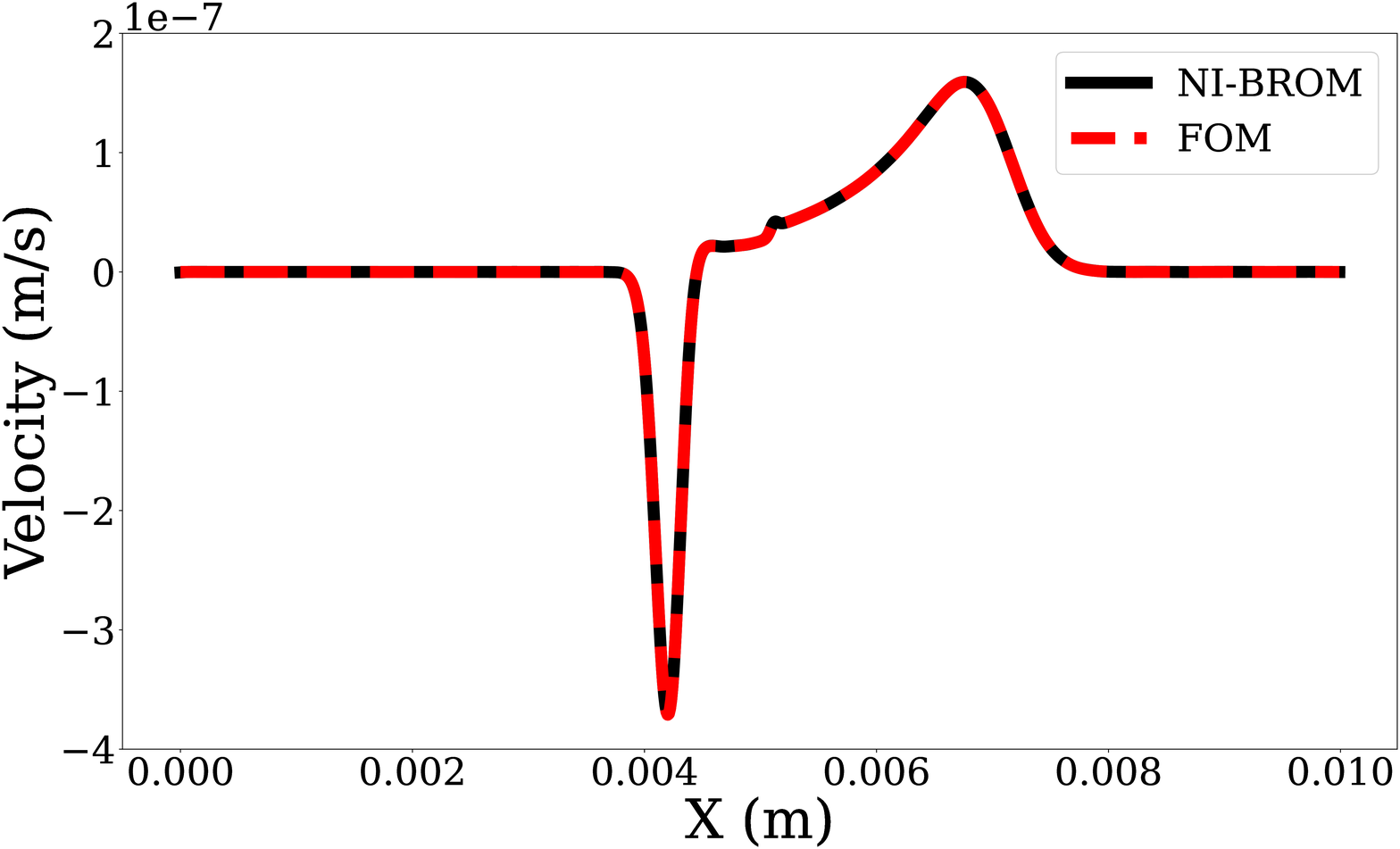}
  \end{minipage}
   \centering
  \begin{minipage}[a]{0.42\textwidth}
    \includegraphics[trim=4 4 4 2cm, clip, width=\textwidth]{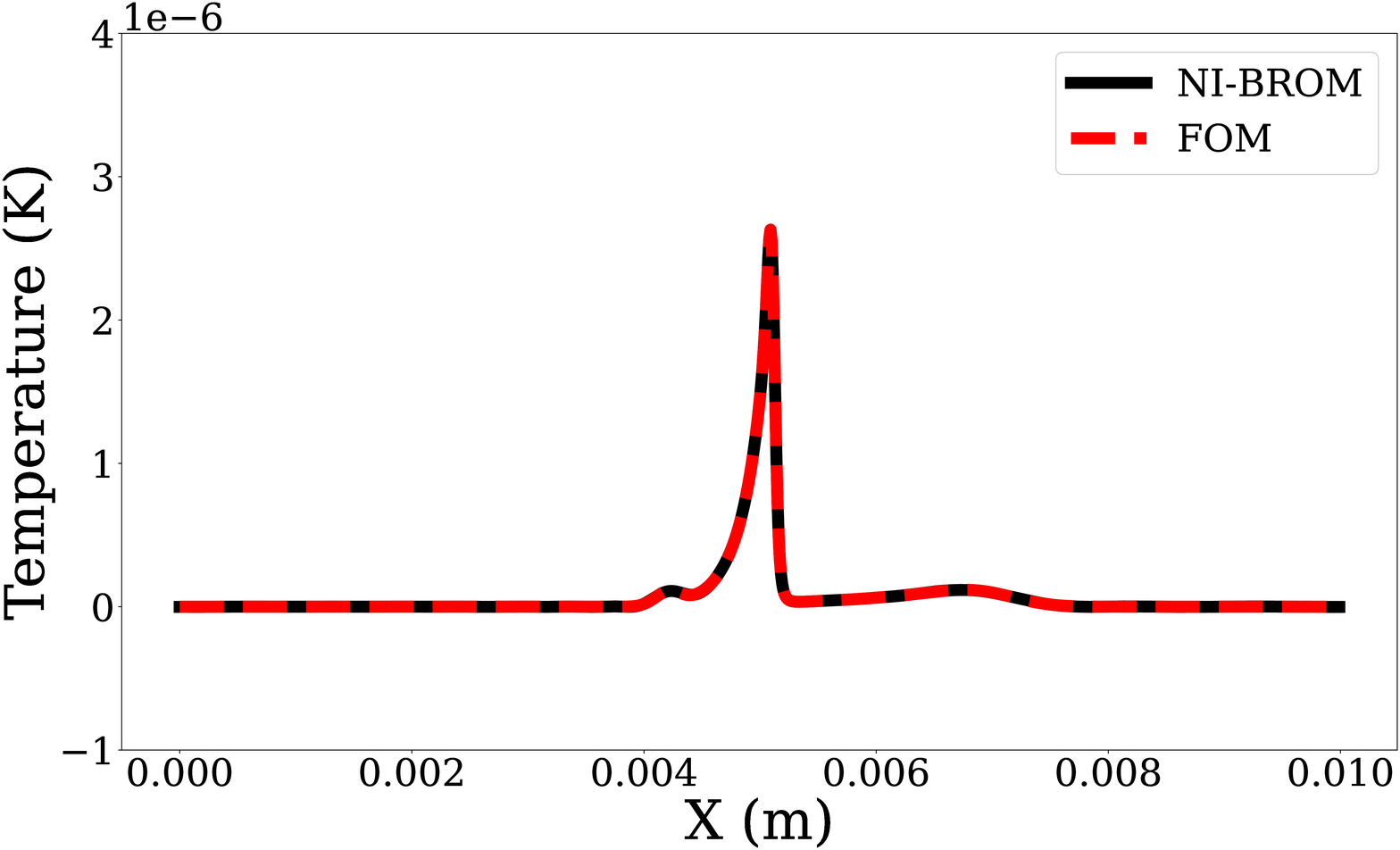}
  \end{minipage}
   \centering
  \begin{minipage}[a]{0.42\textwidth}
    \includegraphics[trim=4 4 4 2cm, clip, width=\textwidth]{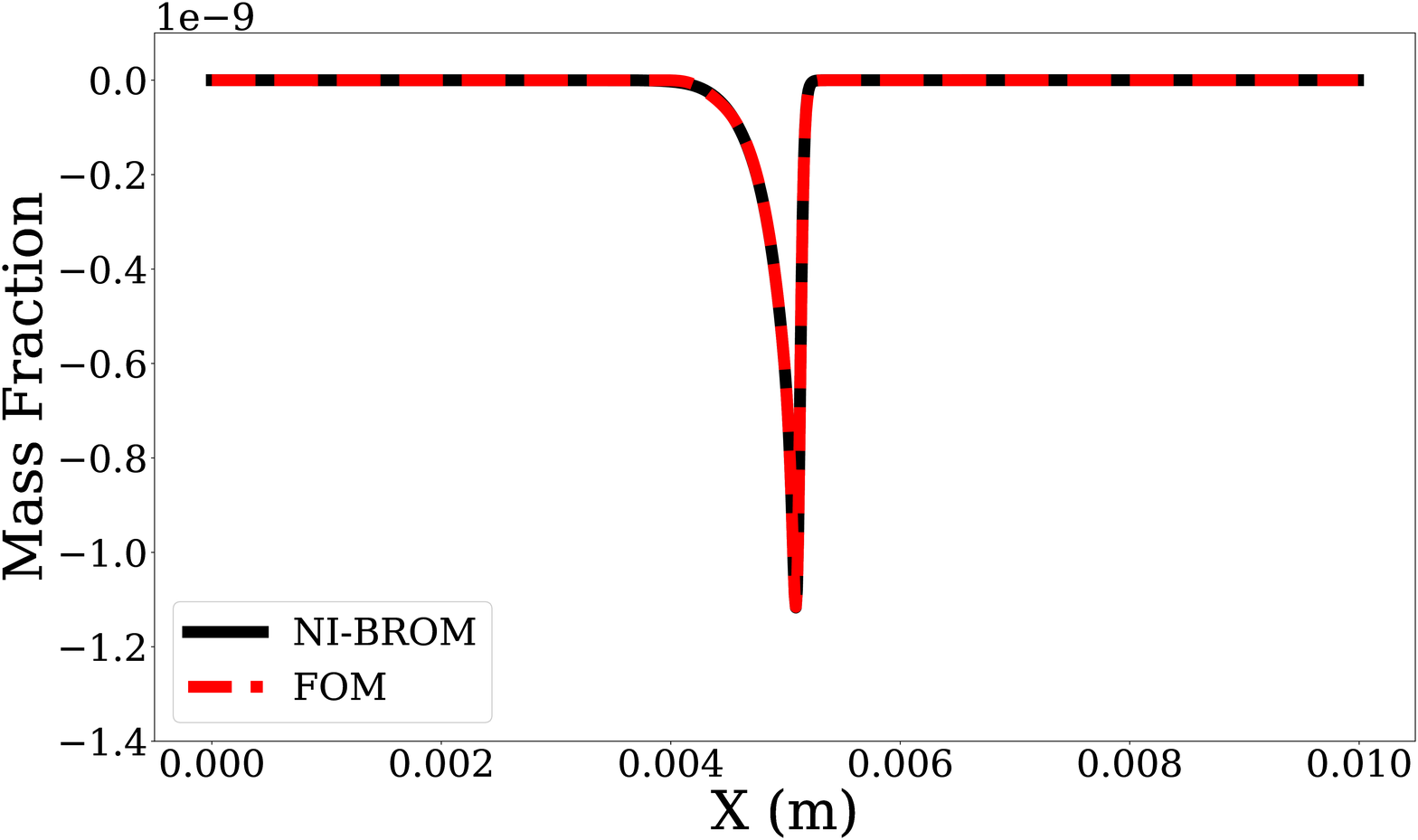}
  \end{minipage}
   \centering
  \caption{Impulse response reconstruction by balanced ROMs (NI-BROM) compared against FOM at $t = 6 \mu sec$.}
   \label{f:IRespReconst}
\end{figure}

 The snapshot of the perturbation variables at time $t=6 \mu sec$ in Figure~\ref{f:IRespReconst} demonstrates good agreement of the impulse response reconstruction by ERA with the linearized FOM.

In order to compare the predictive performance of Galerkin and LSPG ROMs with the balanced ROMs, POD modes are computed based on the same impulse response snapshots that are used for training the balanced ROMs. Figure~\ref{f:HSV_POD_ERAEigs} shows cumulative energy of the POD modes,
\be
E_{cum}^k = \frac{\sum_{j=1}^k \sigma_j}{\sum_{j=1}^n \sigma_j}, \qquad k = 1, \dots, n,
\ee
where, $\sigma_j$ is the $j^{th}$ singular value of the snapshots matrix. The first 80 modes of pressure and velocity, 70 modes of temperature, and 10 modes of species mass fraction capture $99.99 \%$ of the impulse response energy. It is clear here that with the slow decay of modal energies adjoint simulation via output projection \cite{rowley:05} is still expensive. Thus, ERA enables feasible implementation of BT by avoiding adjoint simulations.

\subsection{ROM Performance}

\begin{figure}[htbp!]   
  \centering
  \begin{minipage}[a]{0.42\textwidth}
    \includegraphics[trim=4 4 4 2cm, clip, width=\textwidth]{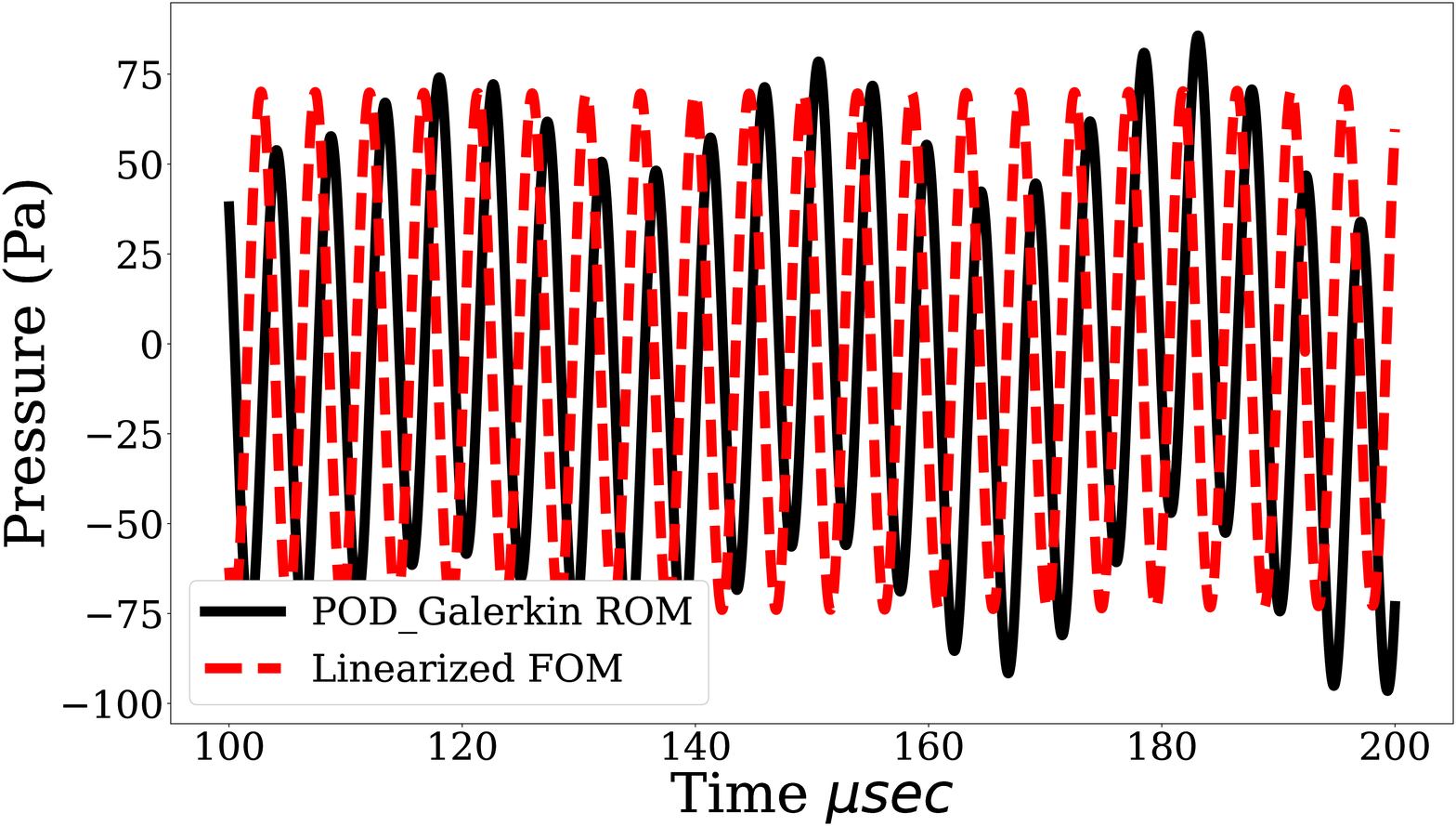}
  \end{minipage}
  \centering
  \begin{minipage}[a]{0.42\textwidth}
    \includegraphics[trim=4 4 4 2cm, clip, width=\textwidth]{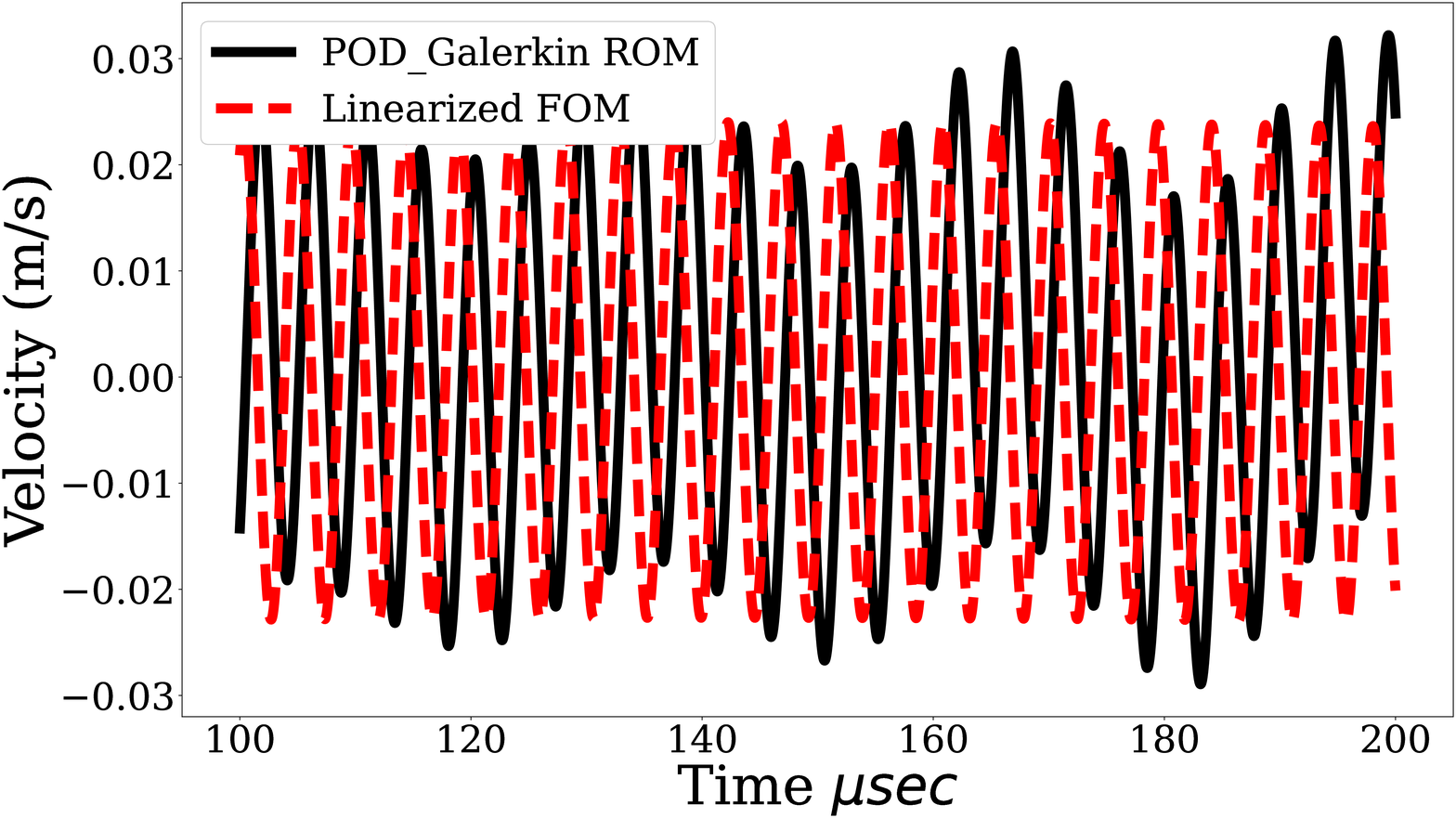}
  \end{minipage}
   \centering
  \begin{minipage}[a]{0.42\textwidth}
    \includegraphics[trim=4 4 4 2cm, clip, width=\textwidth]{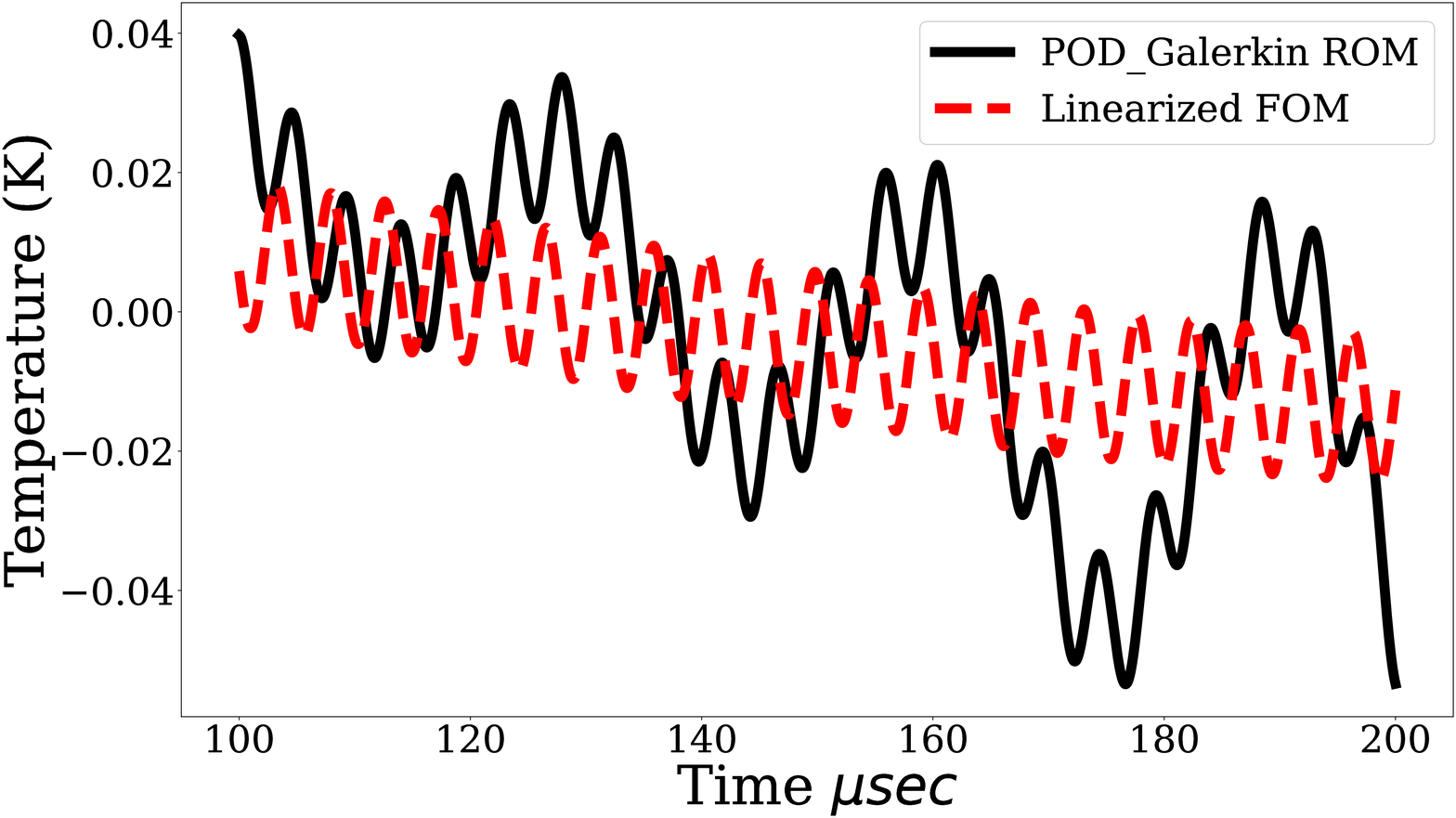}
  \end{minipage}
   \centering
  \begin{minipage}[a]{0.42\textwidth}
    \includegraphics[trim=4 4 4 2cm, clip, width=\textwidth]{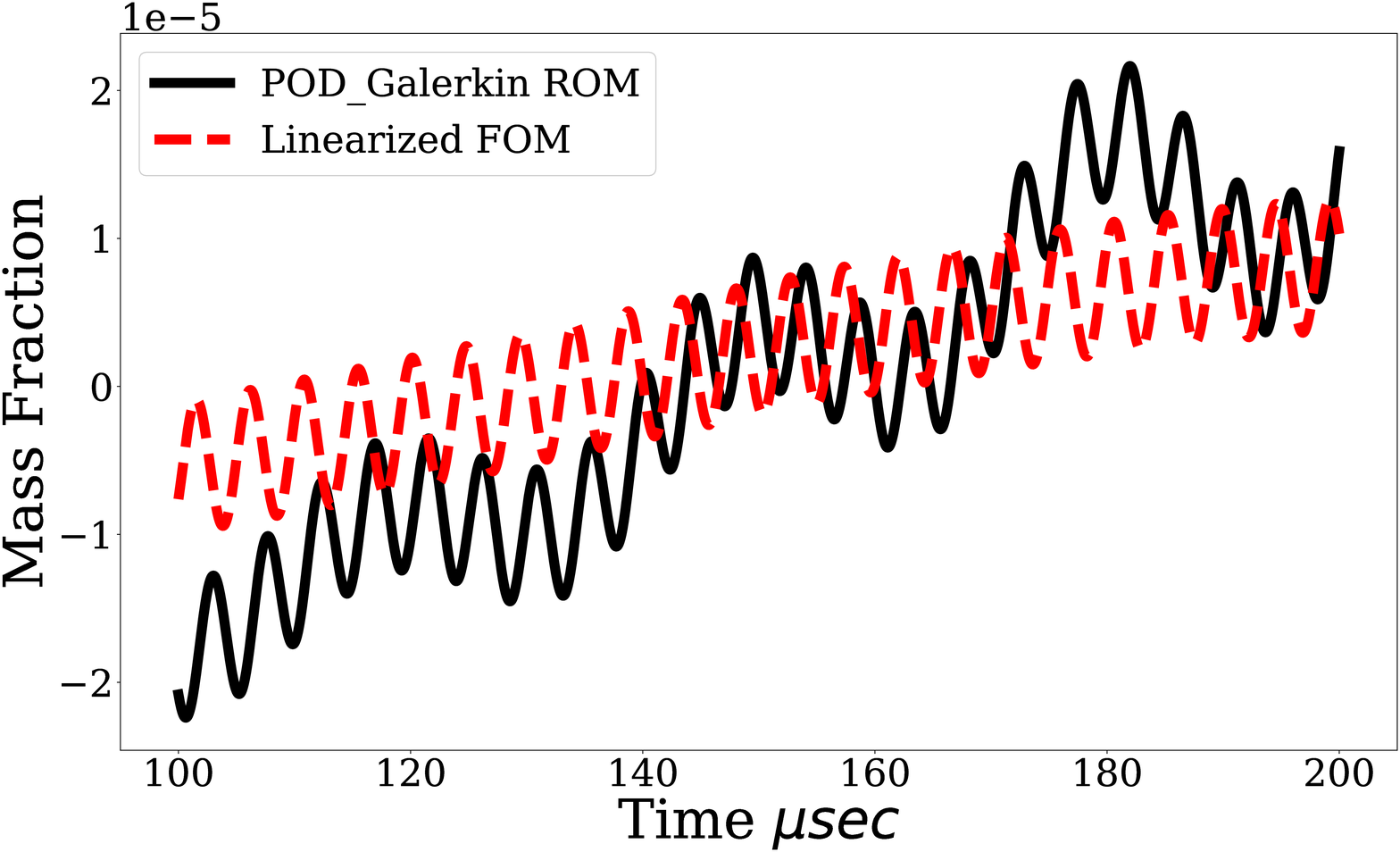}
  \end{minipage}
   \centering
  \caption{POD-Galerkin ROM prediction of the perturbation variables at $x=0.0045 m$ for sinusoidal pressure forcing with amplitude $0.01 \% p_{back}$ and a frequency of 215 kHz. ROM is trained by impulse response.}
   \label{f:PODGImpulse}
\end{figure}

Figure~\ref{f:PODGImpulse} shows the performance of the standard POD-Galerkin ROM, when the ROM is trained with the impulse response snapshots and tested with a sinusoidal input with an amplitude of $0.01 \% p_{back}$ and a frequency of 215 kHz. Numerical errors start to accumulate early in the simulation and ultimately blow up the solution.

\begin{figure}[h!]  
  \centering
  \begin{minipage}[a]{0.42\textwidth}
    \includegraphics[trim=4 4 4 2cm, clip, width=\textwidth]{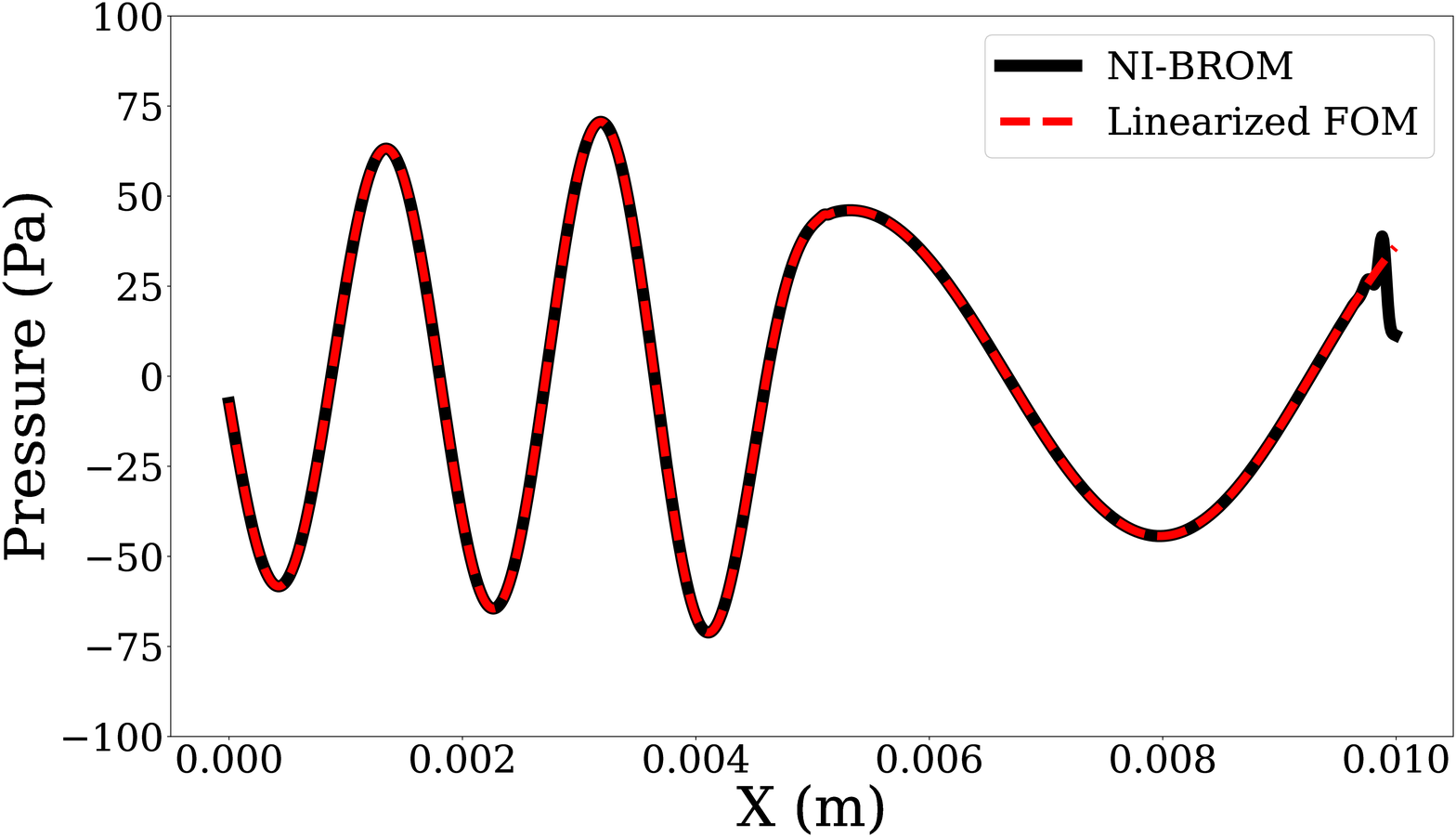}
  \end{minipage}
  \centering
  \begin{minipage}[a]{0.42\textwidth}
    \includegraphics[trim=4 4 4 2cm, clip, width=\textwidth]{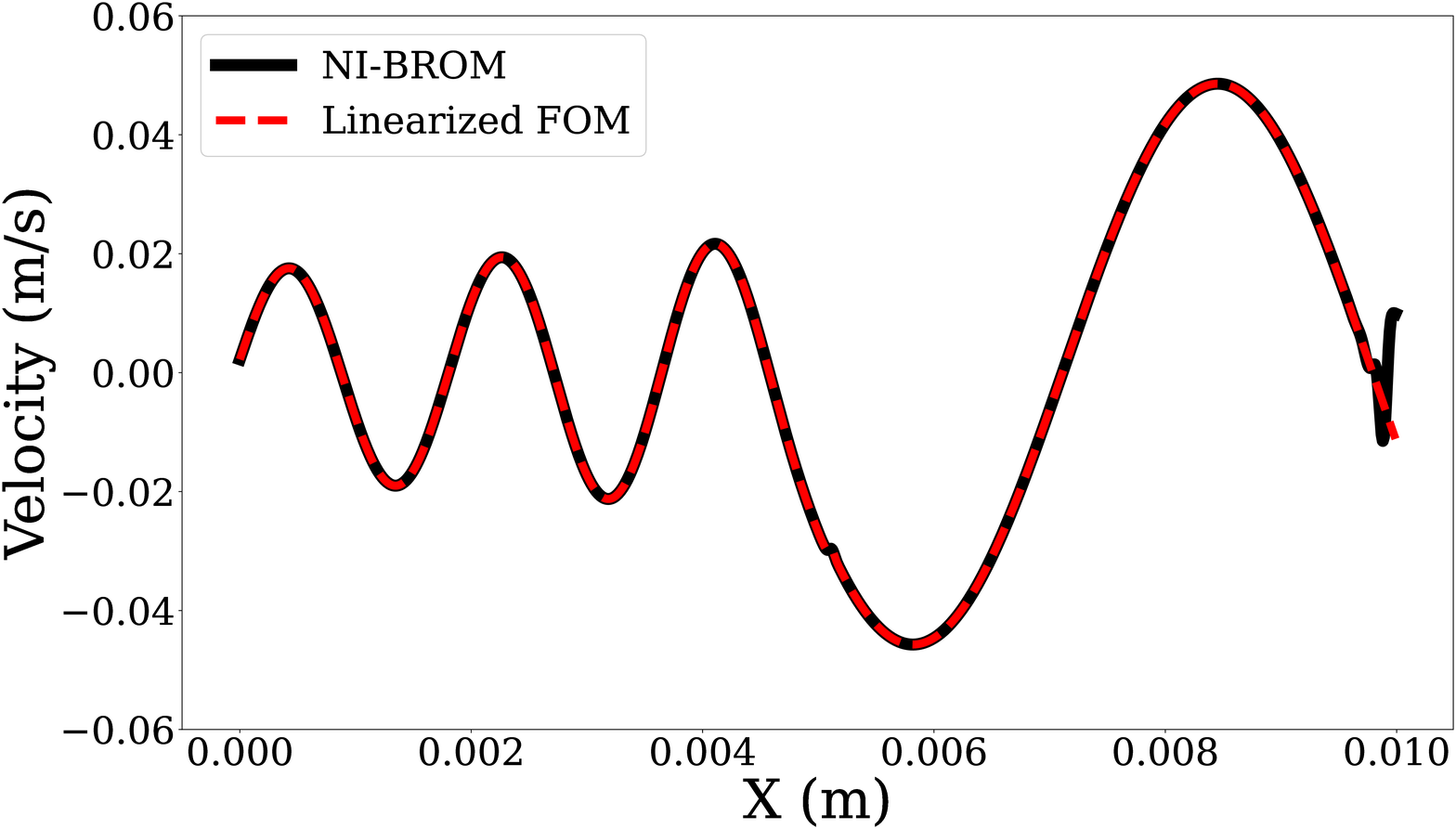}
  \end{minipage}
   \centering
  \begin{minipage}[a]{0.42\textwidth}   
    \includegraphics[trim=4 4 4 2cm, clip, width=\textwidth]{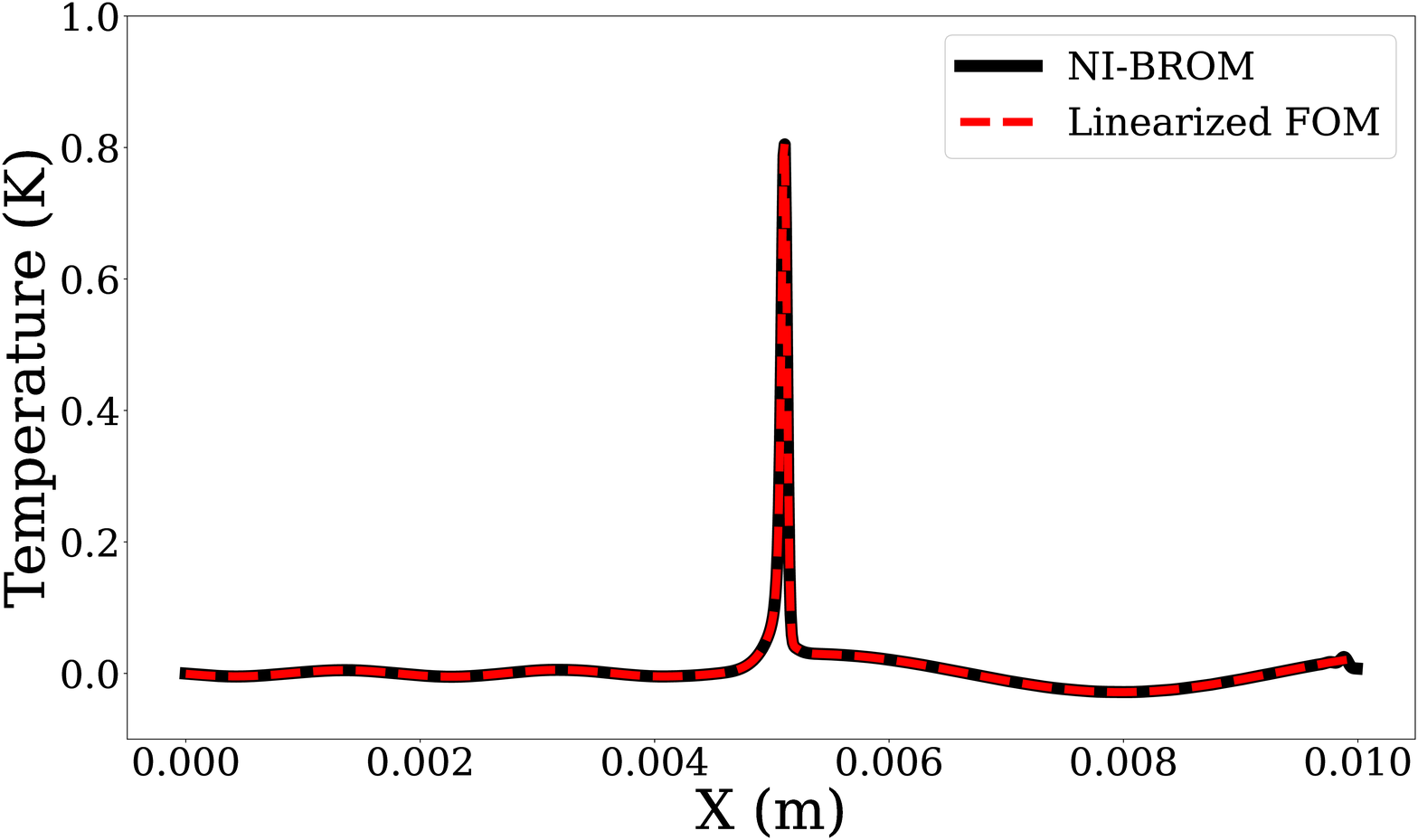}
  \end{minipage}
   \centering
  \begin{minipage}[a]{0.42\textwidth}   
    \includegraphics[trim=4 4 4 2cm, clip, width=\textwidth]{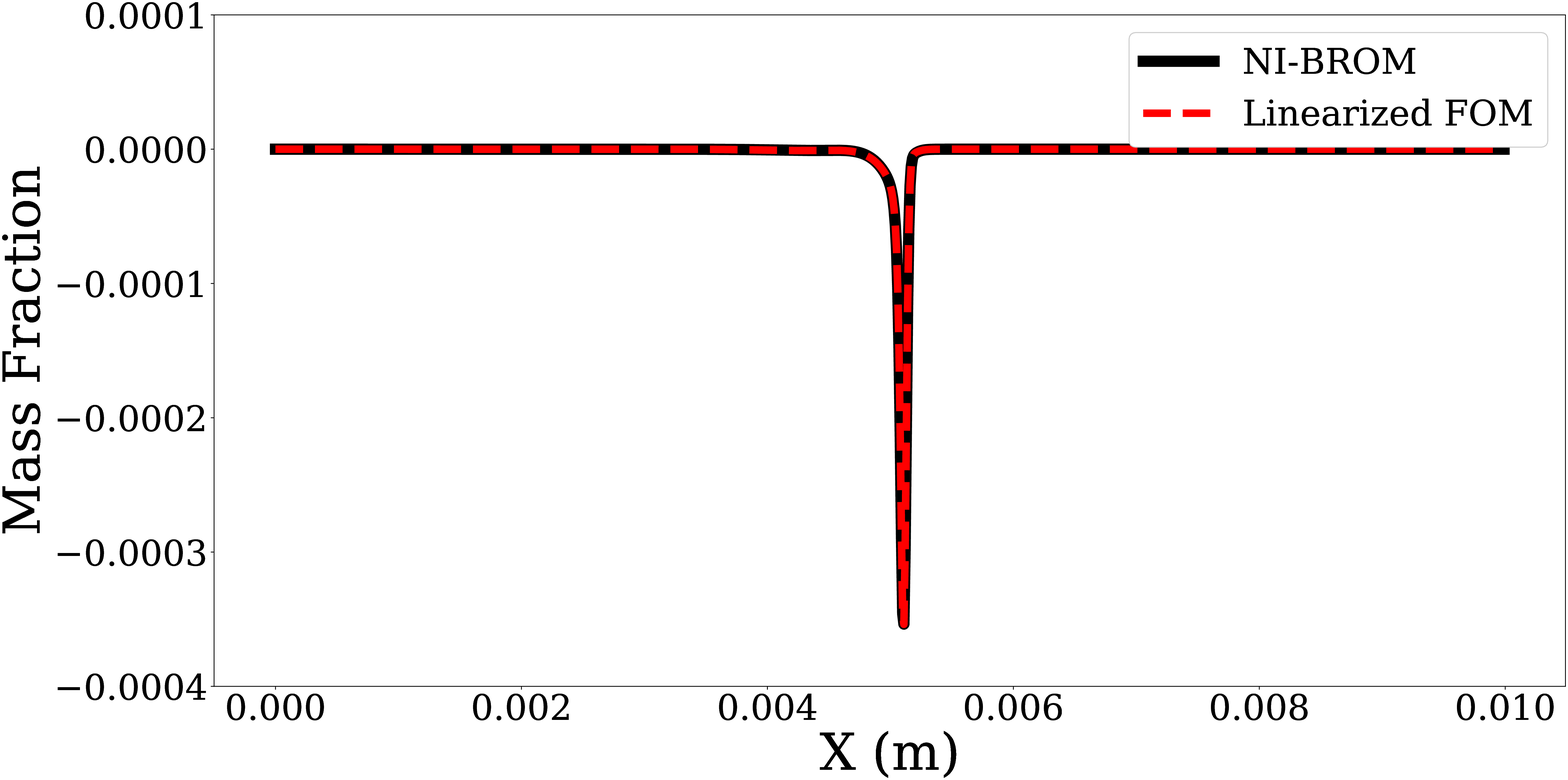}
  \end{minipage}
   \centering
  \caption{Balanced ROM (NI-BROM) predictions of pressure (a), velocity (b), temperature (c), and species mass fraction (d) compared against FOM at $t = 19 \mu sec$. Back pressure is perturbed with sinusoidal input with an amplitude of $0.01 \% p_{back}$ and frequency of 215 kHz}
   \label{f:ERAPred215}
\end{figure}

\begin{figure}[h!]  
  \centering
  \begin{minipage}[a]{0.40\textwidth}
    \includegraphics[trim=4 4 4 2cm, clip, width=\textwidth]{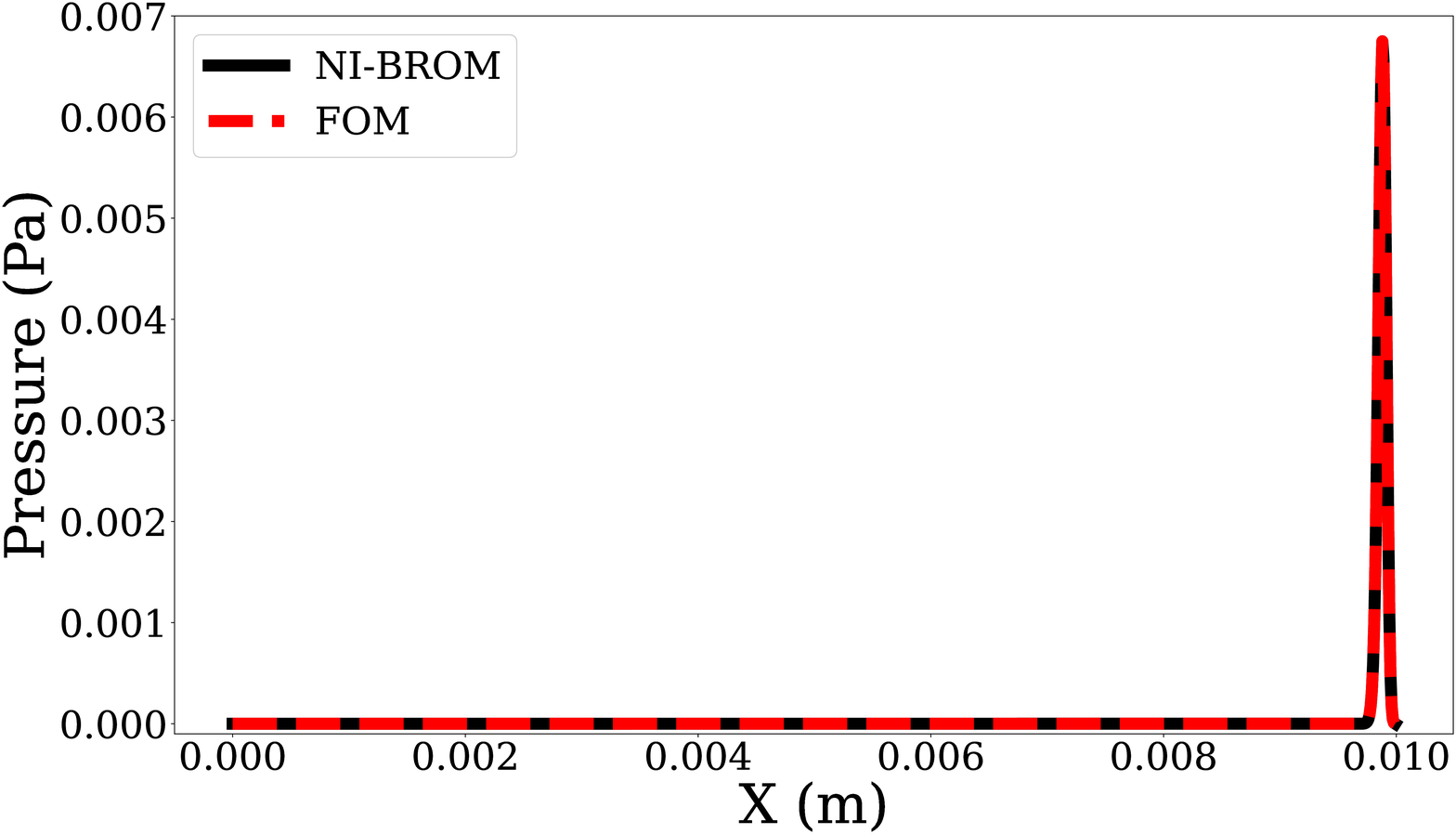}
  \end{minipage}
  \centering
  \begin{minipage}[a]{0.40\textwidth}
    \includegraphics[trim=4 4 4 2cm, clip, width=\textwidth]{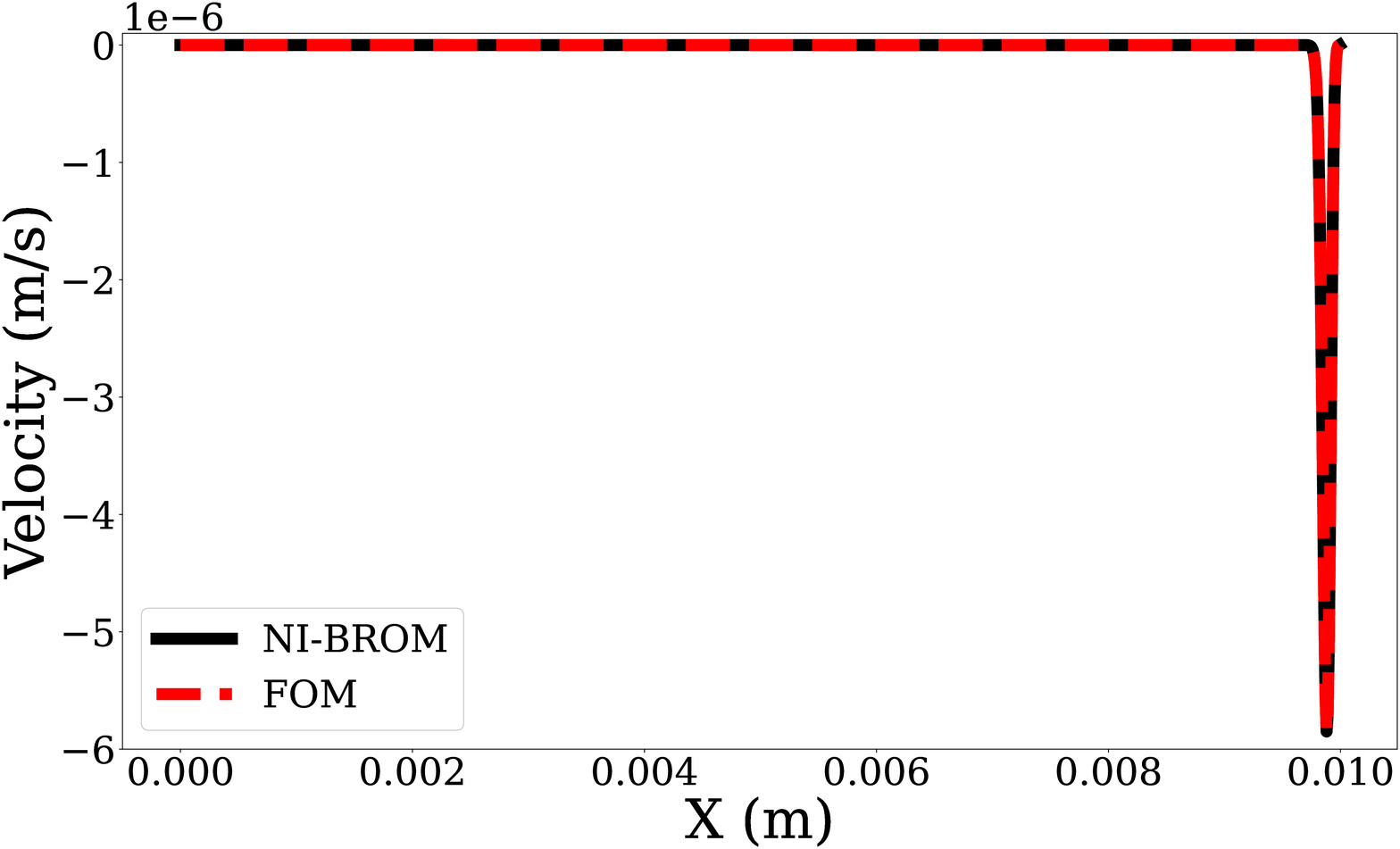}
  \end{minipage}
   \centering
  \begin{minipage}[a]{0.40\textwidth}
    \includegraphics[trim=4 4 4 2cm, clip, width=\textwidth]{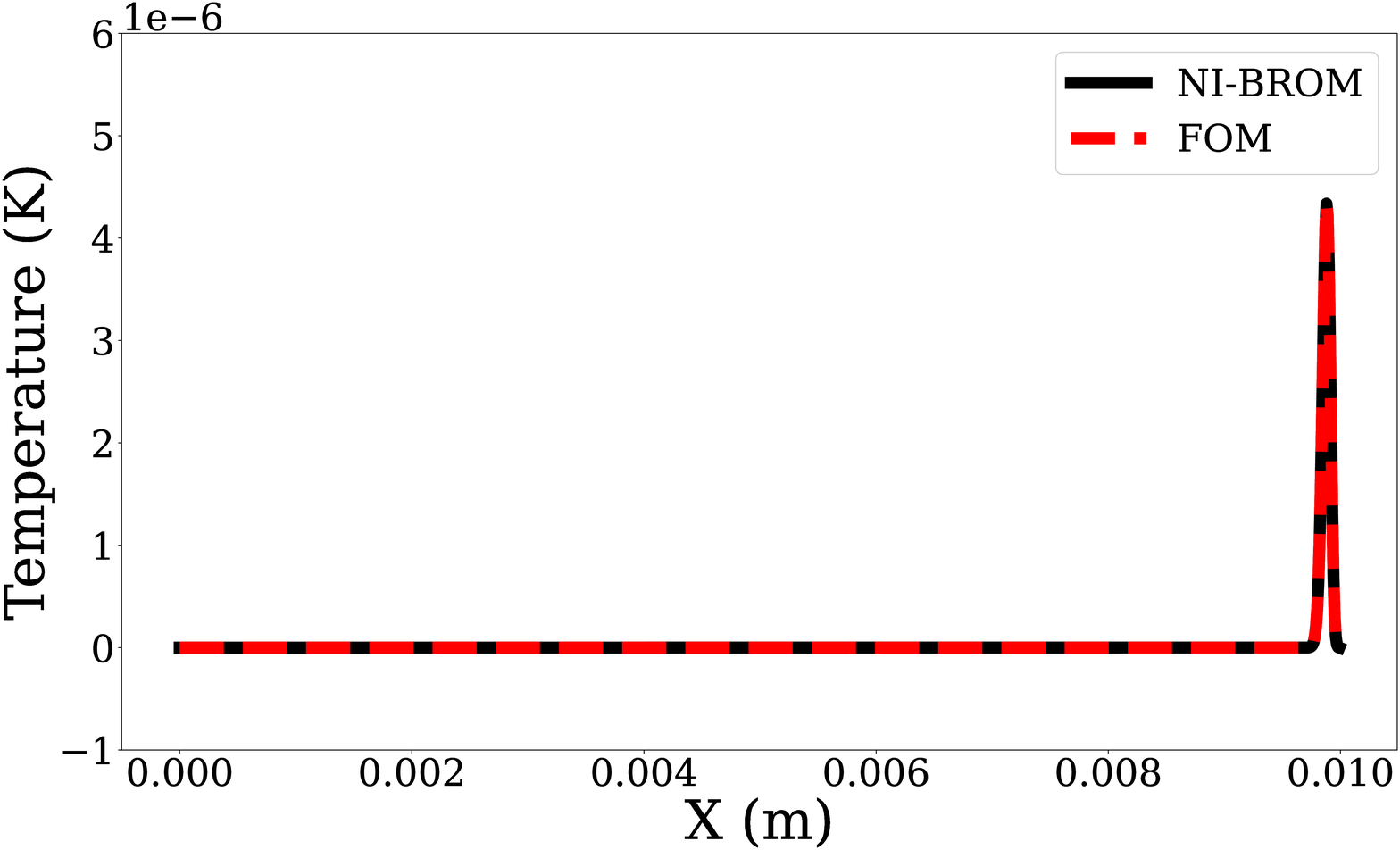}
  \end{minipage}
   \centering
  \begin{minipage}[a]{0.40\textwidth}
    \includegraphics[trim=4 4 4 2cm, clip, width=\textwidth]{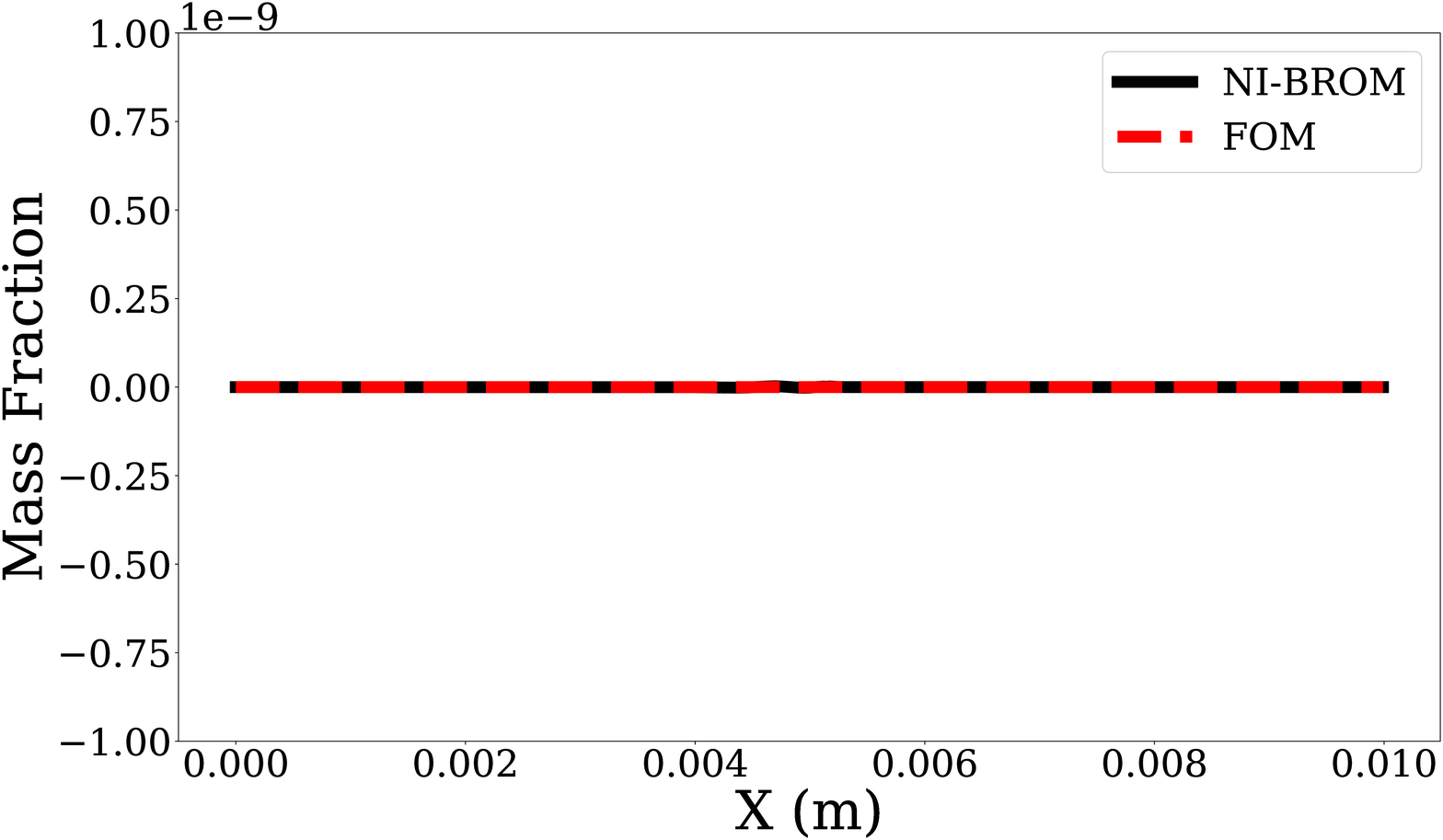}
  \end{minipage}
   \centering
  \caption{Impulse response reconstruction by balanced ROMs (NI-BROM) compared against FOM at $t = 0.1 \mu sec$. Sharp gradients develop in pressure, velocity and temperature close to the right boundary.}
   \label{f:IRespReconst1}
\end{figure}

The performance of ERA is demonstrated in Figure~\ref{f:ERAPred215} in a purely predictive setting. Balanced ROMs are trained with unit impulse response and tested here with sinusoidal input of amplitude $0.01 \% p_{back}$ and frequency of 215 kHz. Unit impulse response is sampled every 100 time steps (equivalent to a sampling period of $T_s = 0.1 \mu sec$) and a total of 1000 snapshots are collected, which covers $100 \mu sec$ of the dynamics. The offline stage of ERA including construction of the Hankel matrix and computing the balanced ROM matrices takes 114.16 seconds of wall-clock time. It is noted that this is a one-time cost, and that the balanced ROMs can be used in a truly predictive setting for unseen conditions, as will be demonstrated. The online stage (time advancement) on the other hand, takes 0.116 seconds, which contributes to an online speedup factor of 138.12.
It is shown in Figure~\ref{f:ERAPred215} that the balanced ROM results for the species mass fraction perfectly matches the FOM. For pressure, velocity and temperature, ROMs overlap with the FOM everywhere in the domain except for at the vicinity of the right boundary. The discrepancy at the right boundary is an artifact of low sampling frequency. Snapshots of the impulse response are collected every 100 time steps to build the Hankel matrix, which is clearly not sufficient to capture the sharp gradients (shown in Figure~\ref{f:IRespReconst1}) that evolve close to the right boundary immediately after the unit impulse enters the computational domain. 
As a result, the predictive performance of ROMs with sinusoidal input, that similarly relies on the balancing modes computed based on the unit impulse response, is affected by the sampling rate. While the unit impulse propagates in the domain sharp gradients are attenuated and lower sampling frequencies absorb the essential dynamics. Therefore, this discrepancy is confined to the right end of the domain. 

\begin{figure}[h!]
  \centering
  \begin{minipage}[a]{0.45\textwidth}
    \includegraphics[trim=4 0.3cm 4 2cm, clip, width=\textwidth]{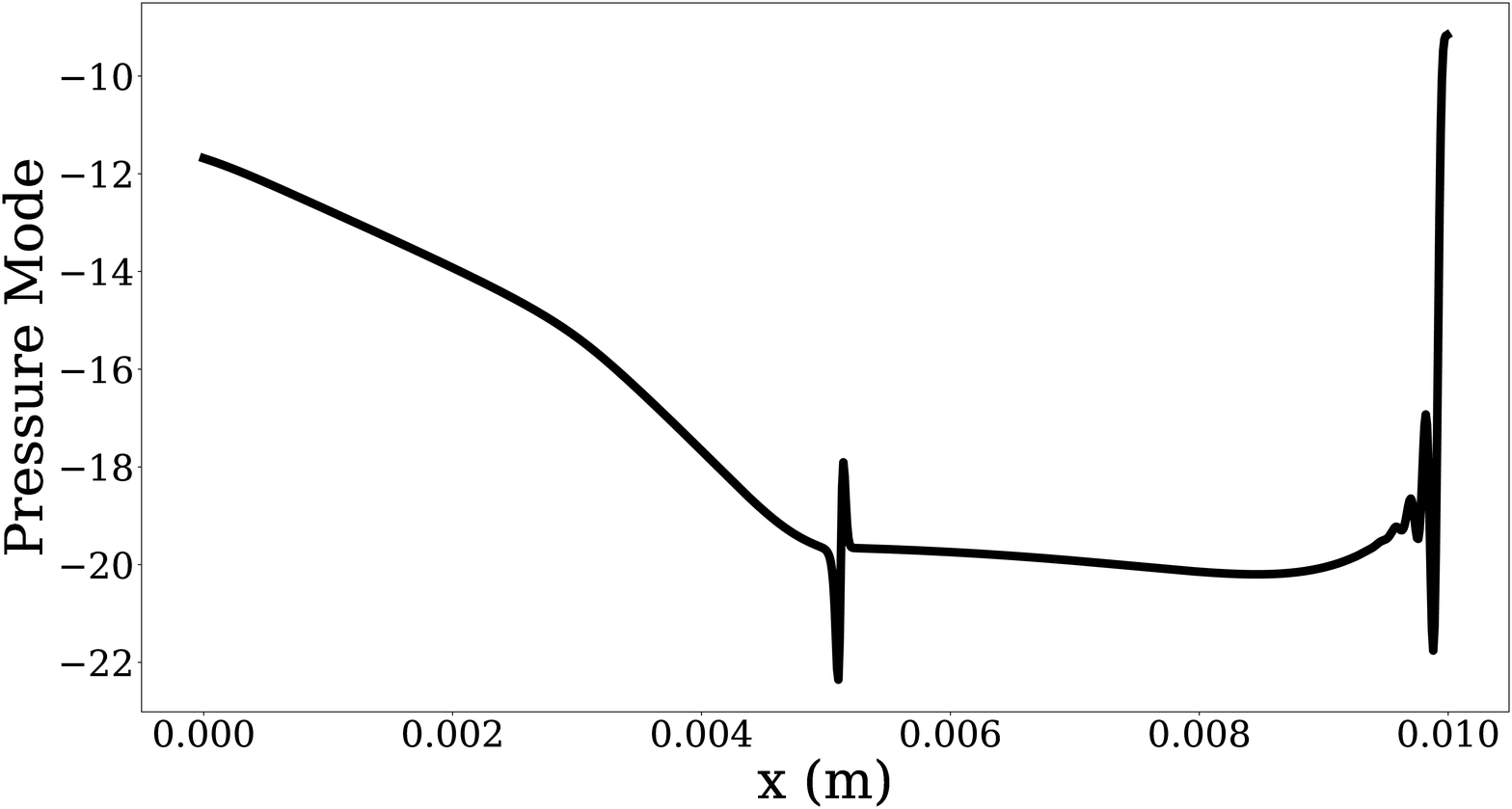}
     \rput(0.9,0.45){\psscalebox{0.9}{\color{black} \textbf{a)}}}
     %\vspace{0.2cm}
  \end{minipage}
  \centering
  \begin{minipage}[a]{0.44\textwidth}
    \includegraphics[trim=4 0.3cm 4 2cm, clip, width=\textwidth]{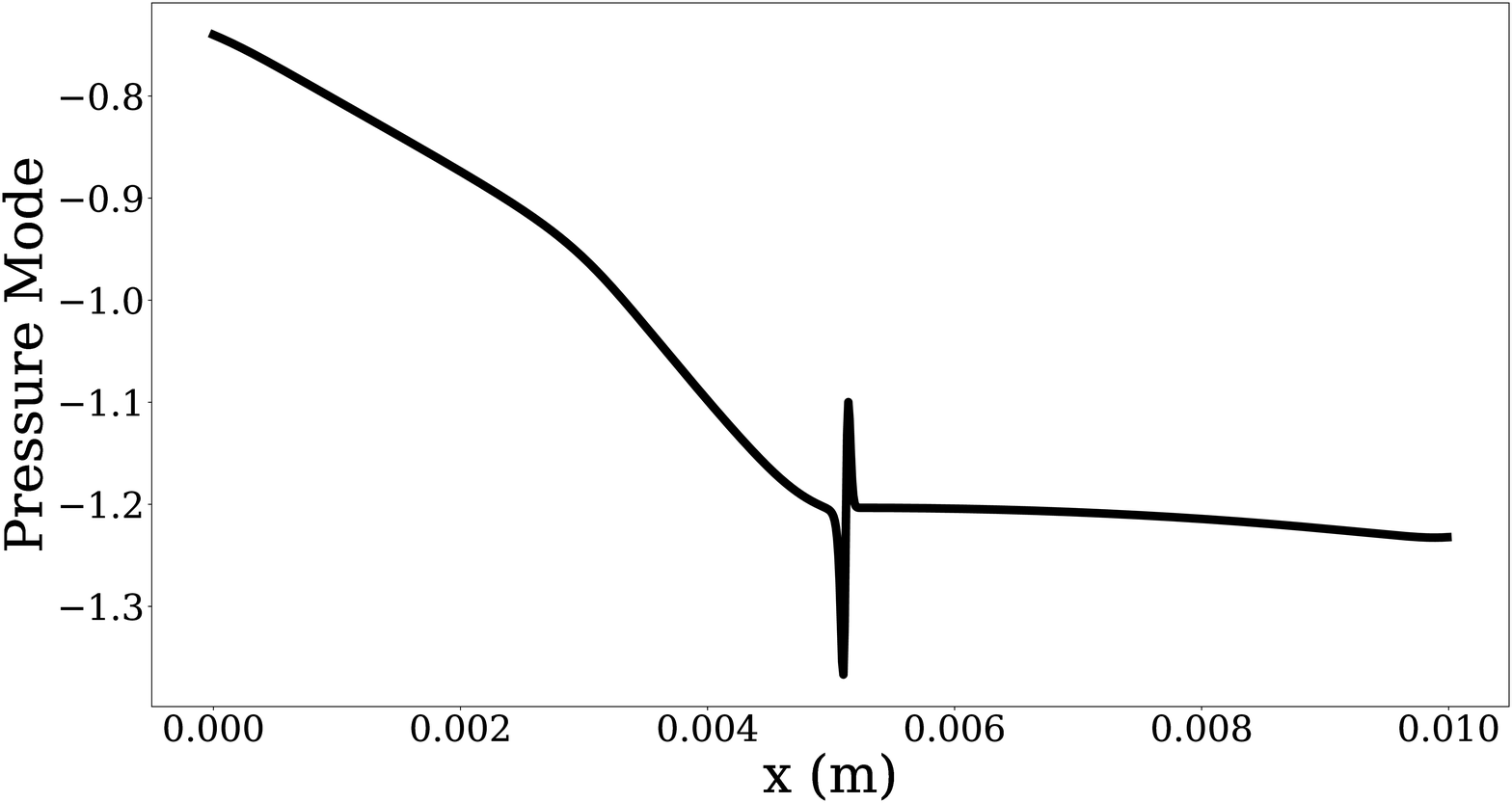}
     \rput(0.9,0.45){\psscalebox{0.9}{\color{black} \textbf{b)}}}
     %\vspace{0.2cm}
  \end{minipage}
   \centering
  \caption{Direct pressure mode 1 obtained with the balanced ROM trained with unit impulse response (a), and Gaussian impulse response (b).}
   \label{f:pressureMode}
\end{figure}

This is more clear in Figure~\ref{f:pressureMode}, which illustrates the first direct mode of pressure. Numerical errors can be seen close to the right boundary. In confirmation of our hypothesis that the errors are caused by insufficient sampling frequency, we also computed the Hankel matrix using the system response to a Gaussian input with zero mean and a standard deviation of 100. Figure~\ref{f:pressureMode} (b) demonstrates the first direct pressure mode with Gaussian input response. The Gramians are not balanced with a Gaussian input and the ROMs identified by ERA are not accurate. However, at the early stages, numerical errors are removed as a result of smoother gradients with a Gaussian-shaped input, which suggests that the low sampling frequency is indeed the source of errors at the vicinity of the right boundary. Similar improvements are observed by reducing the sampling period to $T_s=0.001 \mu sec$, but at a higher computation cost.

\begin{figure}[h!]  
  \centering
  \begin{minipage}[a]{0.40\textwidth}
    \includegraphics[trim=4 0.1cm 4 1.2cm, clip, width=\textwidth]{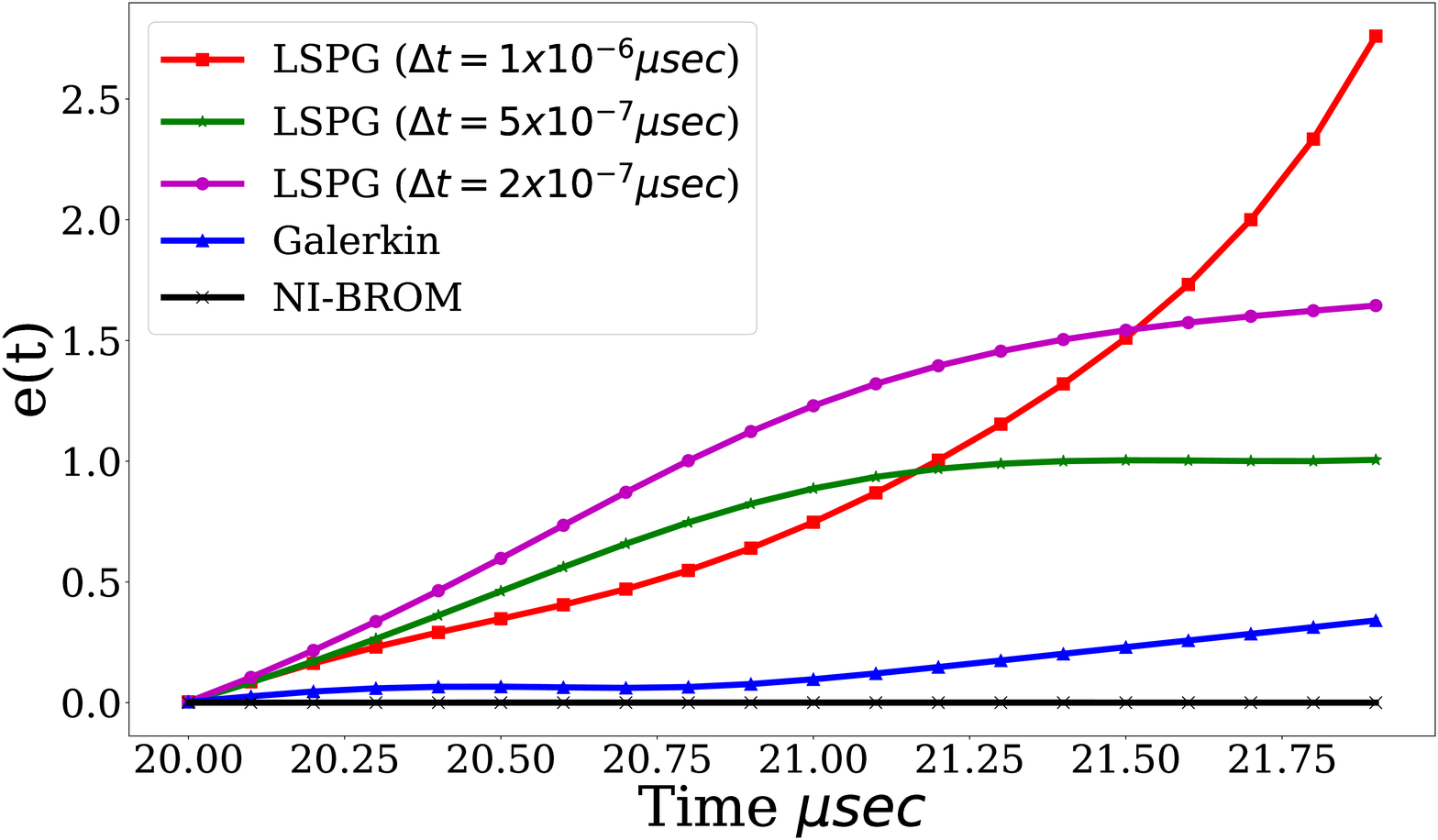}
     \rput(0.9,0.45){\psscalebox{0.9}{\color{black} \textbf{a)}}}
     \vspace{0.1cm}
  \end{minipage}
  \centering
  \begin{minipage}[a]{0.40\textwidth}
    \includegraphics[trim=4 0.1cm 4 1.2cm, clip, width=\textwidth]{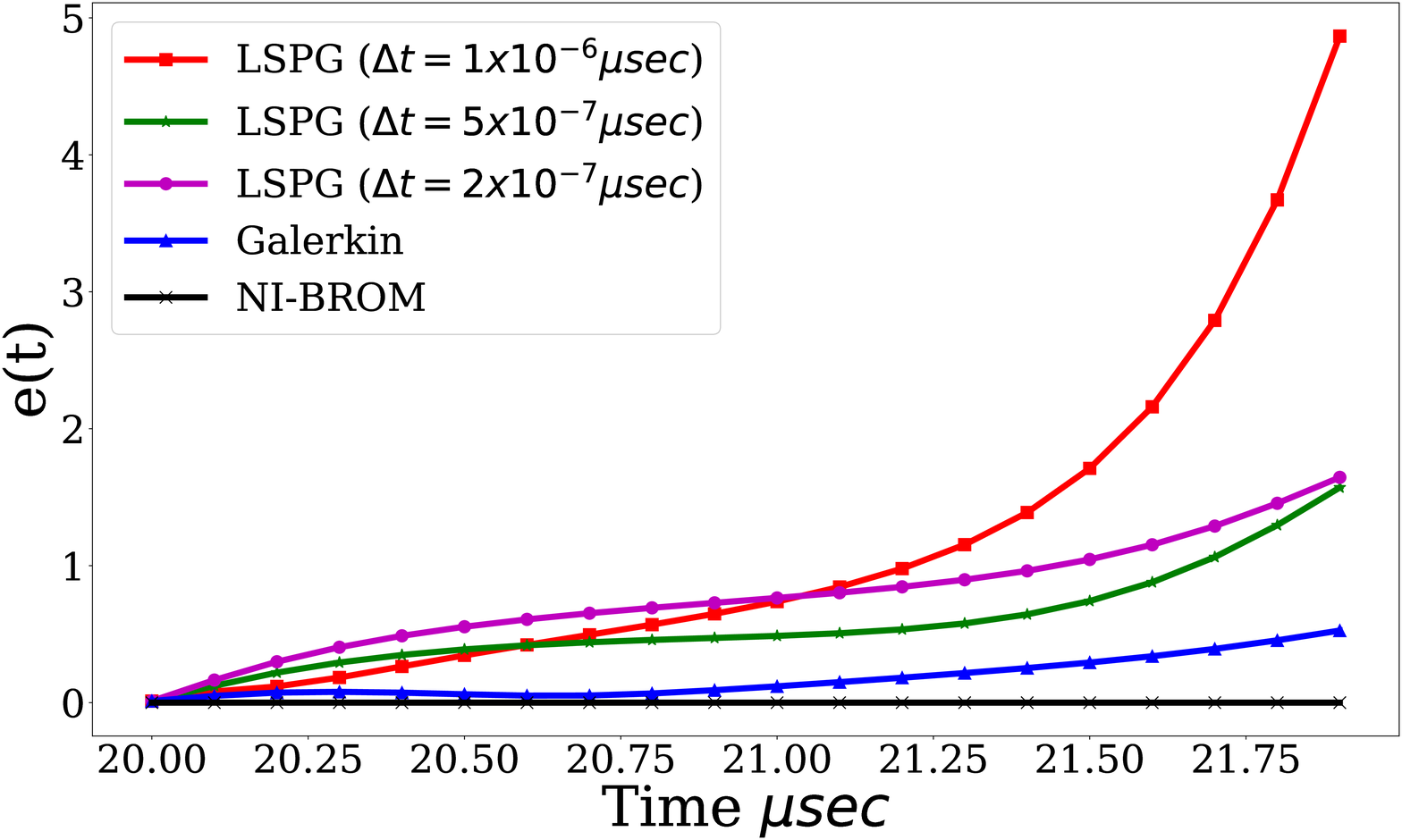}
     \rput(0.9,0.45){\psscalebox{0.9}{\color{black} \textbf{b)}}}
     \vspace{0.1cm}
  \end{minipage}
   \centering
  \begin{minipage}[a]{0.40\textwidth}
    \includegraphics[trim=4 0.1cm 4 1cm, clip, width=\textwidth]{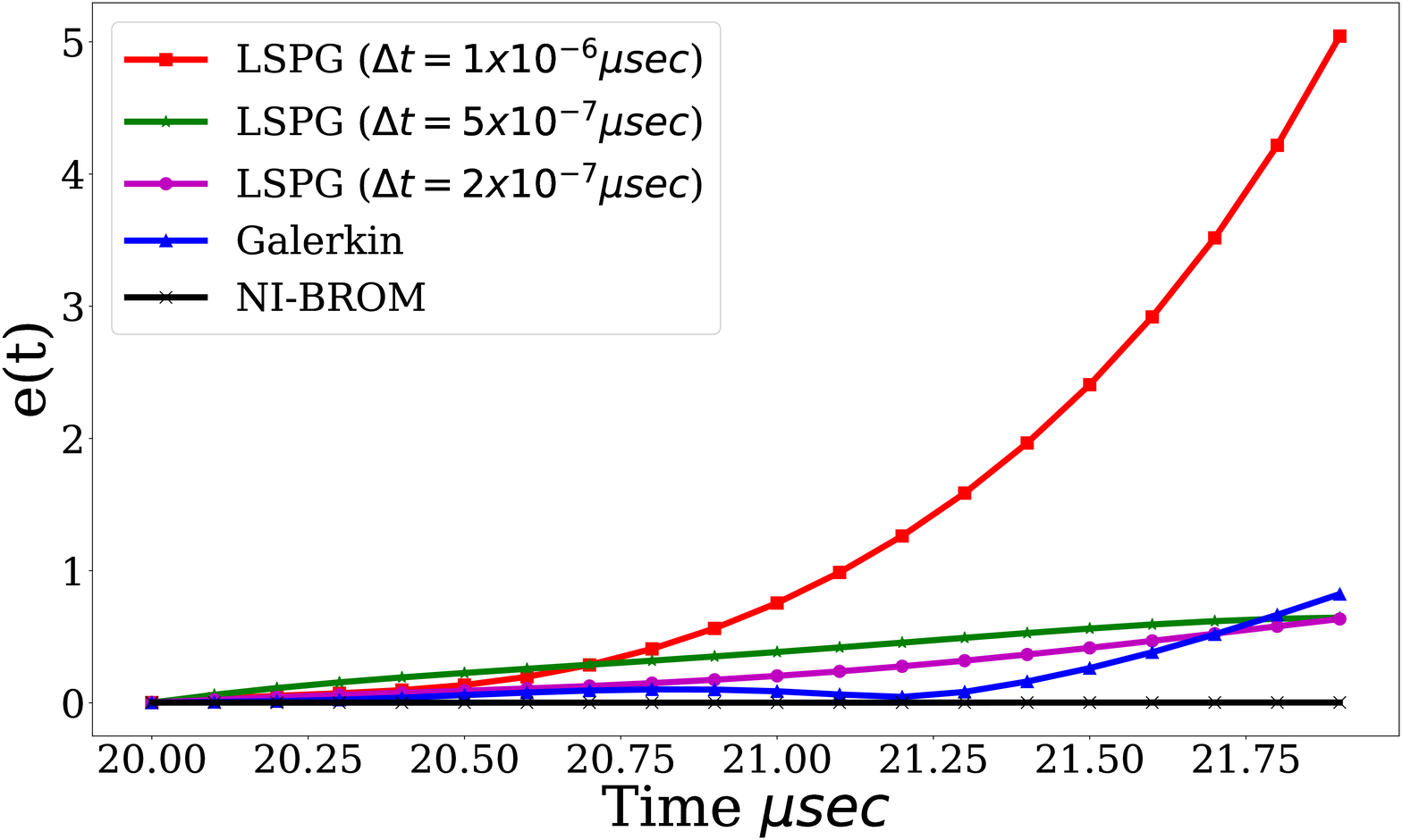}
     \rput(0.9,0.45){\psscalebox{0.9}{\color{black} \textbf{c)}}}
     %\vspace{0.2cm}
  \end{minipage}
   \centering
  \begin{minipage}[a]{0.40\textwidth}
    \includegraphics[trim=4 0.1cm 4 1cm, clip, width=\textwidth]{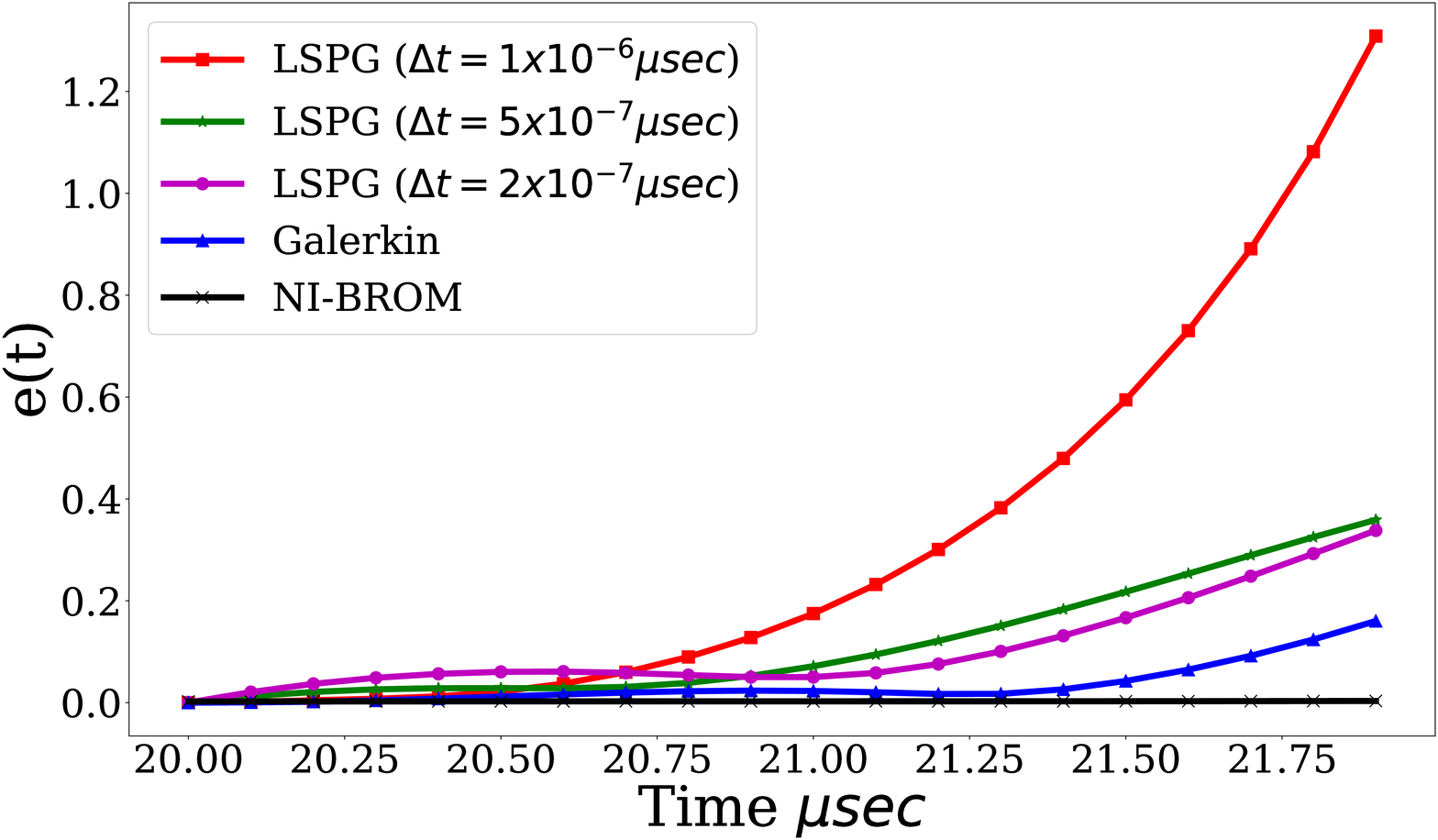}
     \rput(0.9,0.45){\psscalebox{0.9}{\color{black} \textbf{d)}}}
     %\vspace{0.2cm}
  \end{minipage}
   \centering
  \caption{Relative prediction error for forced flame with a perturbation amplitude of $0.01 \% p_{back}$ and a frequency of 215 kHz. Galerkin, LSPG and ERA are compared for pressure (a), velocity (b), temperature (c), and species mass fraction (d). ERA is trained with unit impulse response, and Galerkin and LSPG ROMs are trained with combined forcing frequencies of 200, 210, and 220 kHz. The last 100 cells (at the right end of the domain) are removed when computing the relative errors.}
   \label{f:ROMPred215}
\end{figure}

Demonstrated in Figure~\ref{f:ROMPred215} is the predictive performance of the standard POD-Galerkin and LSPG ROMs compared against balanced ROMs. To train the projection-based ROMs, POD modes are computed by the snapshots obtained with pressure forcing with an amplitude of $0.01 \% p_{back}$ and combined frequencies of 200, 210, and 220 kHz. Similar to ERA, the projection-based ROMs are also tested for perturbation frequency of 215 kHz. The LSPG ROM is sensitive to the temporal step size, therefore, its performance is evaluated with three different step sizes. Unlike the balanced ROM however, relative error values show that both projection-based ROMs exhibit poor predictive performance and unbounded numerical errors.

\begin{figure}[h!]  
  \centering
  \begin{minipage}[a]{0.40\textwidth}
    \includegraphics[trim=4 0.3cm 4 2cm, clip, width=\textwidth]{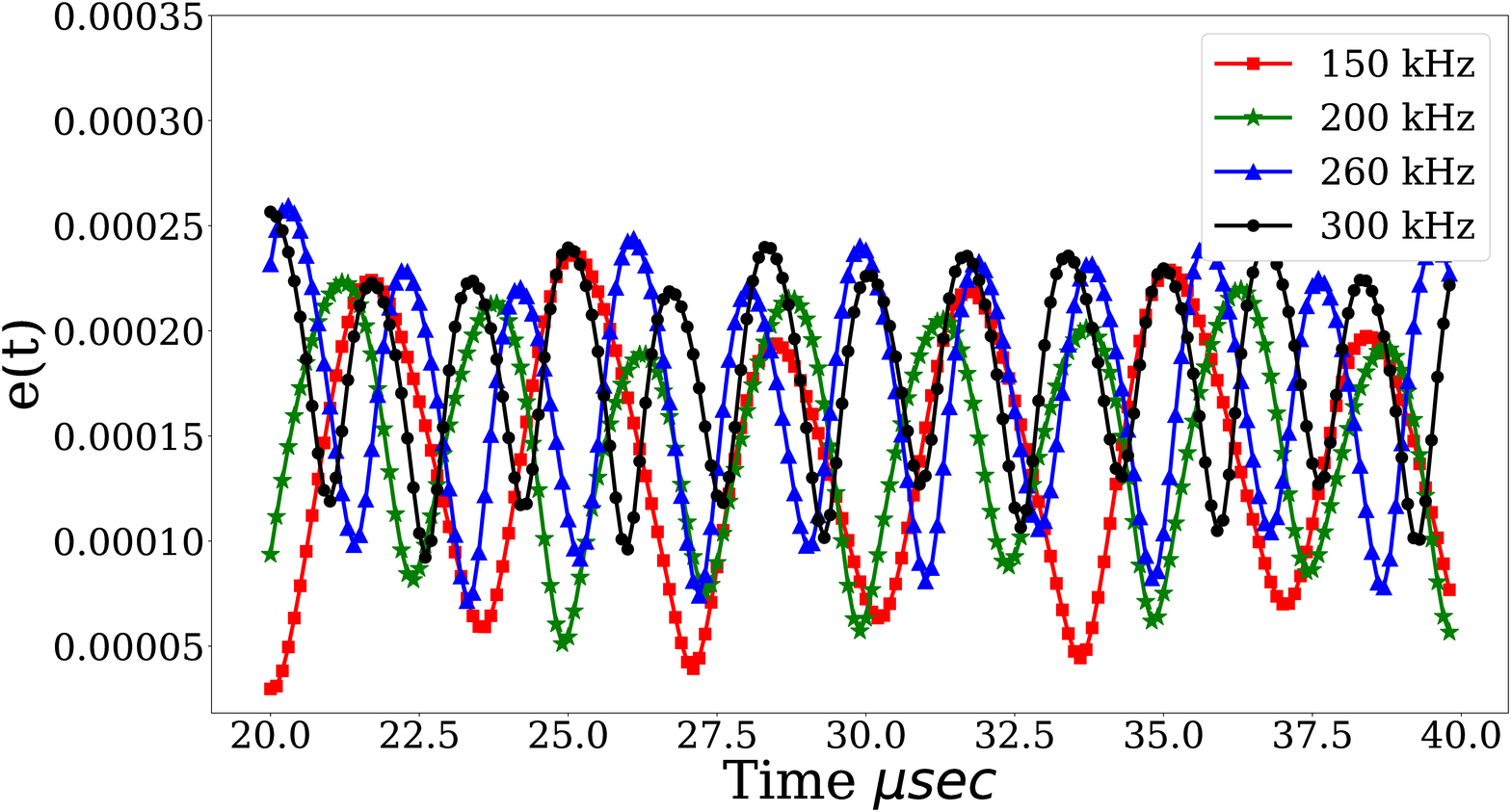}
     \rput(0.9,0.45){\psscalebox{0.9}{\color{black} \textbf{a)}}}
     \vspace{0.1cm}
  \end{minipage}
  \centering
  \begin{minipage}[a]{0.40\textwidth}
    \includegraphics[trim=4 0.3cm 4 2cm, clip, width=\textwidth]{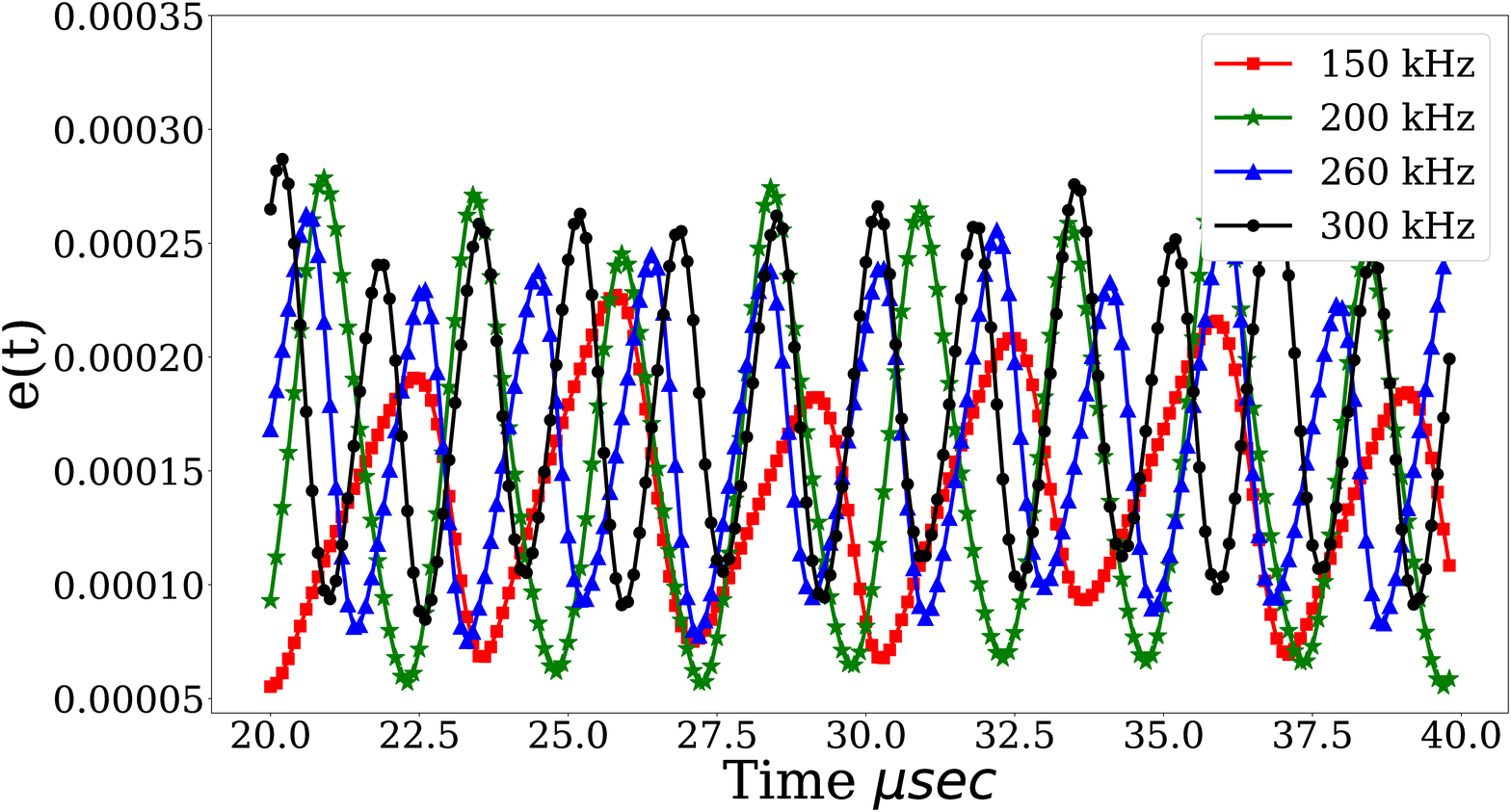}
     \rput(0.9,0.45){\psscalebox{0.9}{\color{black} \textbf{b)}}}
     \vspace{0.1cm}
  \end{minipage}
   \centering
  \begin{minipage}[a]{0.40\textwidth}
    \includegraphics[trim=4 0.3cm 4 2cm, clip, width=\textwidth]{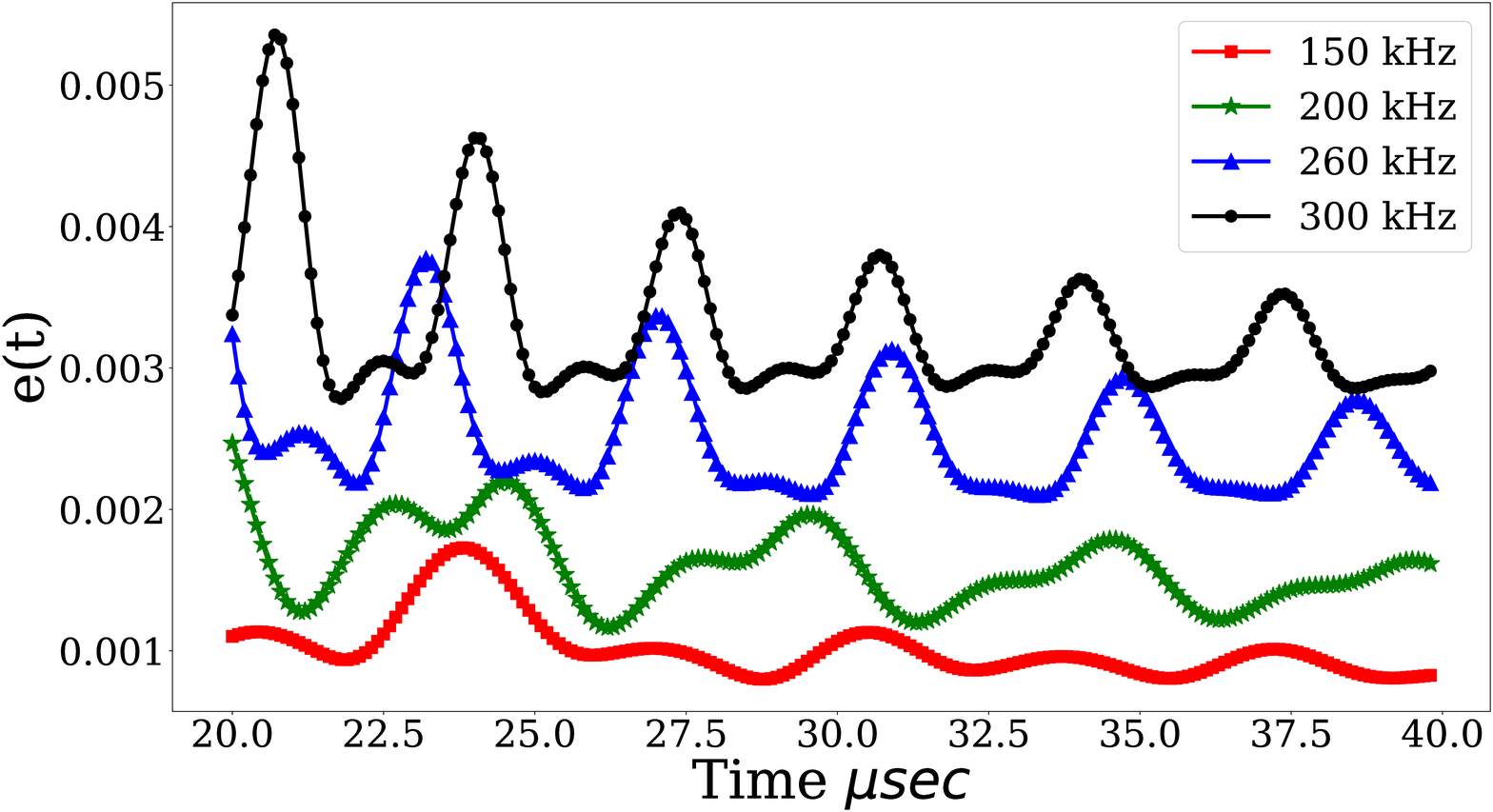}
     \rput(0.9,0.45){\psscalebox{0.9}{\color{black} \textbf{c)}}}
     %\vspace{0.2cm}
  \end{minipage}
   \centering
  \begin{minipage}[a]{0.40\textwidth}
    \includegraphics[trim=4 0.3cm 4 2cm, clip, width=\textwidth]{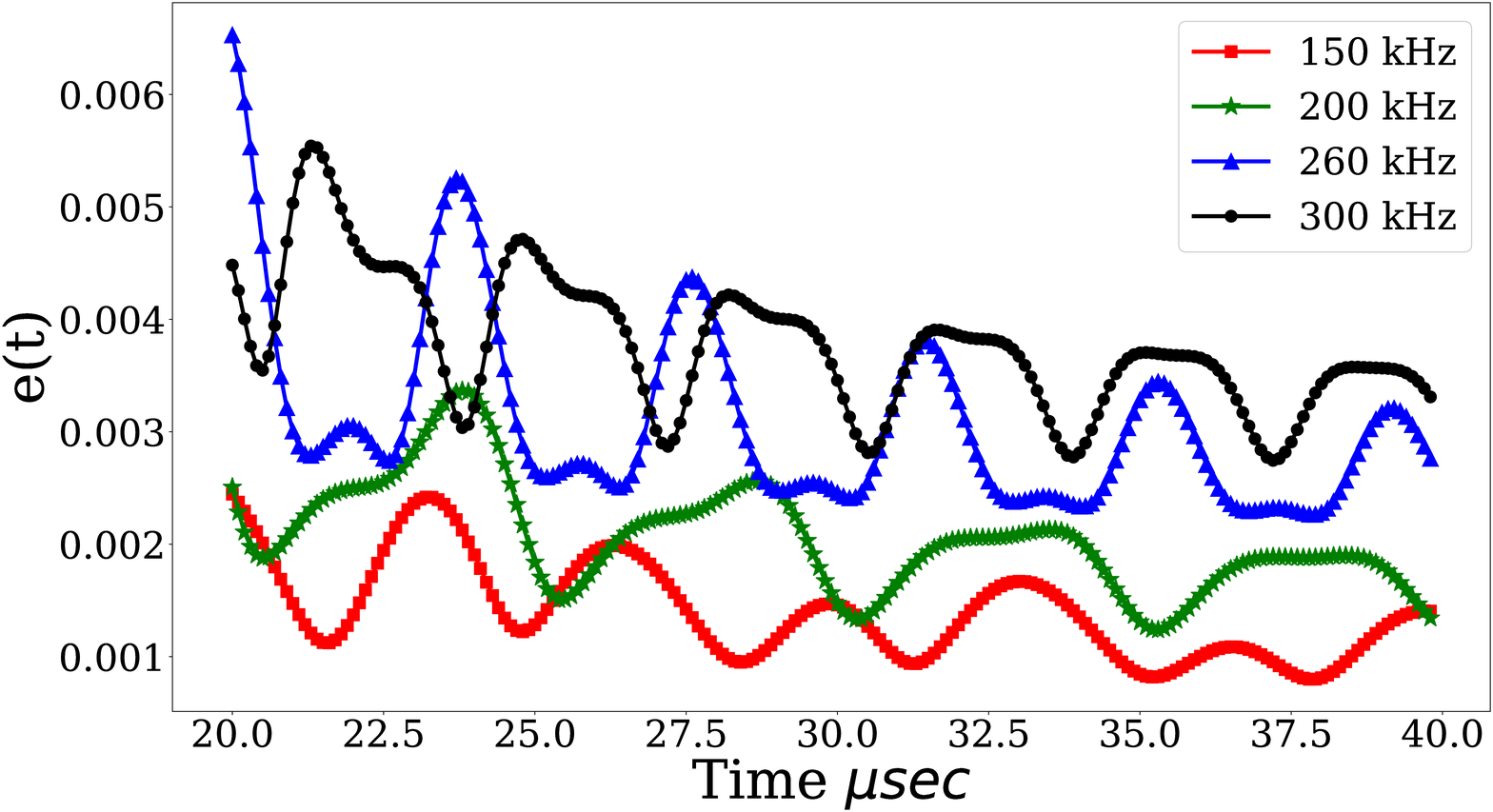}
     \rput(0.9,0.45){\psscalebox{0.9}{\color{black} \textbf{d)}}}
     %\vspace{0.2cm}
  \end{minipage}
   \centering
  \caption{Relative prediction error for forced flame with a perturbation amplitude of $0.01 \% p_{back}$ and different frequencies is demonstrated for pressure (a), velocity (b), temperature (c), and species mass fraction (d). The last 100 cells (at the right end of the domain) are removed before computing the relative errors in all cases.}
   \label{f:ERAPredfpert}
\end{figure}

The analysis so far has been based on a perturbation frequency of 215 kHz. Figure~\ref{f:ERAPredfpert} demonstrates the predictive performance of balanced ROMs for different perturbation frequencies. The perturbation amplitude is retained at $0.01 \% p_{back}$, and the unit impulse response is collected with a sampling period of $T_s=0.1 \mu sec$ and a total of 1000 Markov parameters are used to train the ROMs. The last 100 cells of the domain are removed from the analysis (after computing the solution and before evaluating the relative errors). As shown in Figure~\ref{f:ERAPredfpert} balanced ROM errors decay in time for all frequencies. 

Next, we use the output domain decomposition approach with tangential interpolation to resolve the sharp gradients that develop at the vicinity of the right boundary. We have a single-input multi-output (SIMO) system, therefore, reduction is applied only along the left tangential directions. The full-state output is divided into three subdomains (i.e., $q_m=3$ in equation (\ref{dd})), where $\mathbf{C}_1 \in \mathbb{R}^{950 \times 1000}$ corresponds to ROM 1 that is trained with the largest sampling period of $T_s = 0.1 \mu sec$, $\mathbf{C}_2 \in \mathbb{R}^{45 \times 1000}$ corresponds to ROM 2 trained with a sampling period of $T_s = 0.01 \mu sec$, and $\mathbf{C}_3 \in \mathbb{R}^{5 \times 1000}$ is for ROM 3 trained with the smallest sampling period of $T_s = 0.001 \mu sec$. Tangential interpolation is separately applied to ROM 1 and ROM 2. Since ROM 3 originally has a low-dimensional output (state at 5 cells), further reduction of the number of outputs via projection over the tangential directions affects accuracy, therefore, we maintain the original number of outputs in ROM 3. Table~\ref{t:ERAdd} lists the wall-clock times of subdomain ROMs with and without tangential interpolation. Projection of the impulse response sequence over the dominant tangential directions has clearly reduced the total offline cost of ERA. It is noted that by a more radical truncation along the left tangential directions ROMs remain stable and offline computations are reduced at the cost of accuracy.
\begin{table}[h!]
 \begin{center}
  \caption{Wall-clock time (sec) of offline computations for ERA with output domain decomposition (ERA-DD) and tangential interpolation (TI). ROM 1, ROM 2, and ROM 3 use 1000, 10000, and 18000 Markov parameters, respectively.}
  \label{t:ERAdd}
  \begin{tabular}{llllll}\hline
     &  ROM 1 & ROM 2 & ROM 3 & Total \\\hline
     ERA-DD without TI & 81.89 & 1609.32 & 1102.33 & 2793.54 \\\hline
     ERA-DD with TI & 6.17 & 495.35 & 1102.33 & 1603.85  \\\hline
  \end{tabular}
 \end{center}
\end{table}

\begin{figure}[h!]
  \centering
  \begin{minipage}[a]{0.40\textwidth}
    \includegraphics[trim=4 0.3cm 4 2cm, clip, width=\textwidth]{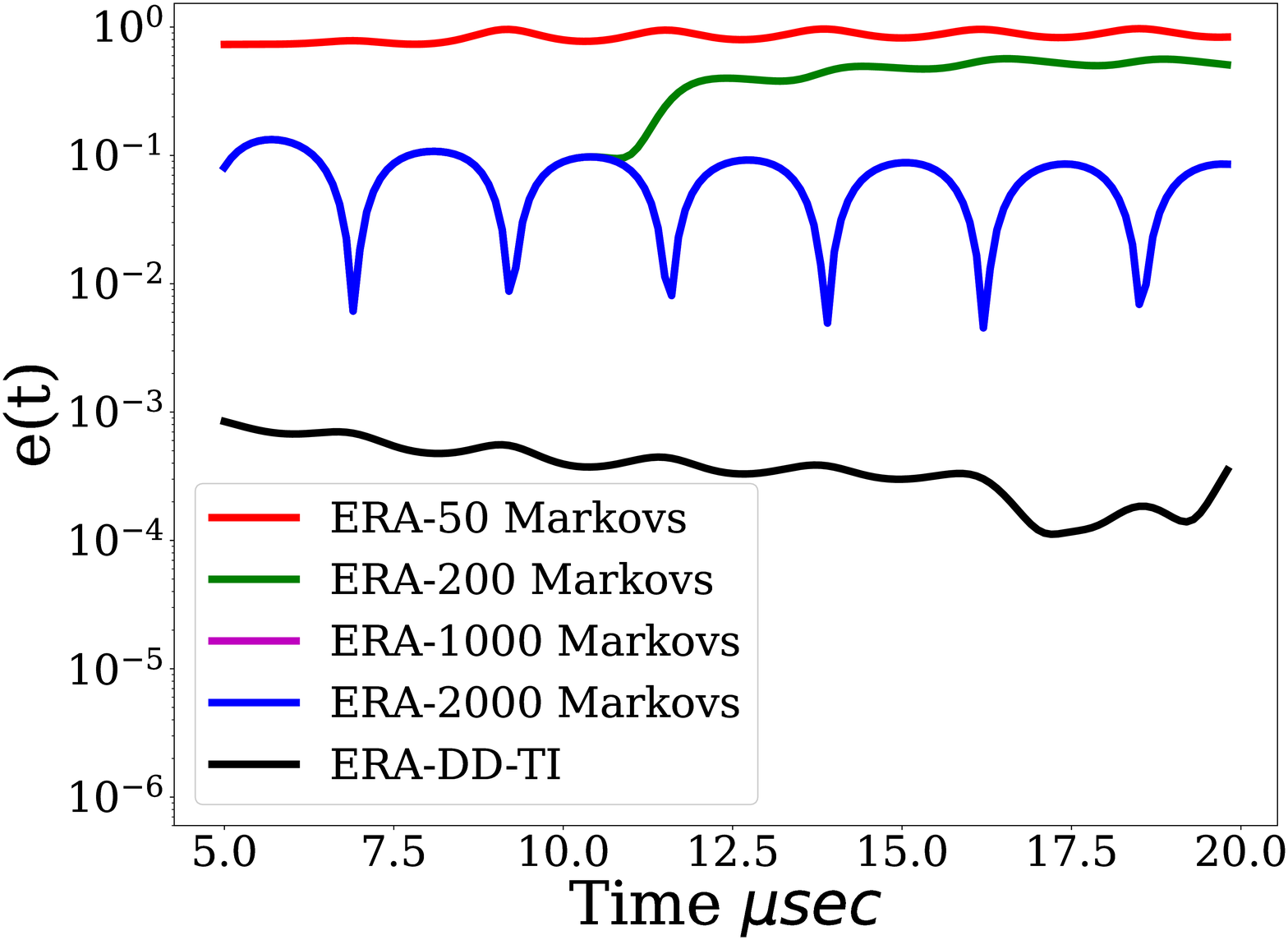}
     \rput(0.9,0.45){\psscalebox{0.9}{\color{black} \textbf{a)}}}
     \vspace{0.1cm}
  \end{minipage}
  \centering
  \begin{minipage}[a]{0.40\textwidth}
    \includegraphics[trim=4 0.3cm 4 2cm, clip, width=\textwidth]{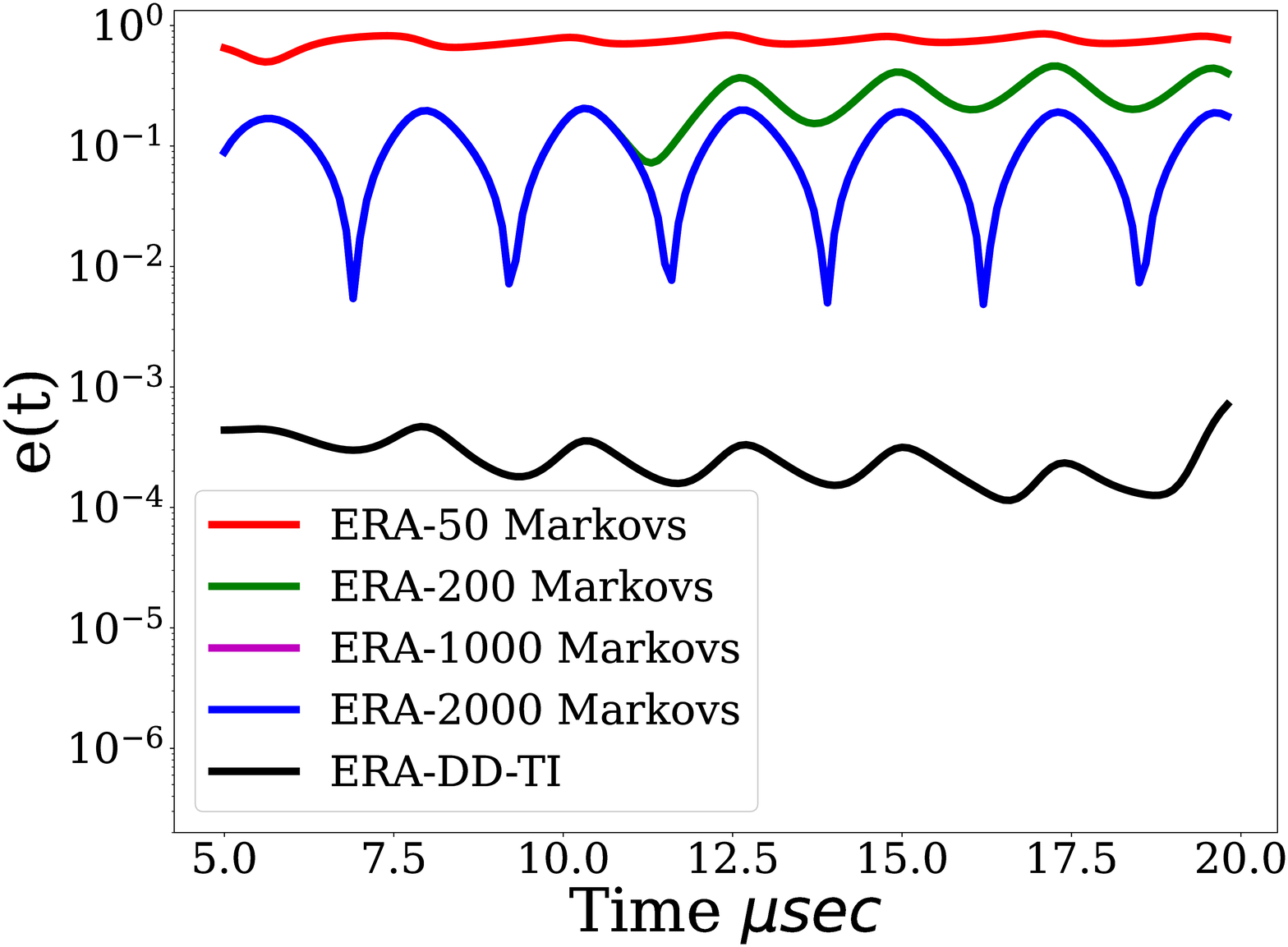}
     \rput(0.9,0.45){\psscalebox{0.9}{\color{black} \textbf{b)}}}
     \vspace{0.1cm}
  \end{minipage}
   \centering
  \begin{minipage}[a]{0.40\textwidth}
    \includegraphics[trim=4 0.3cm 4 2cm, clip, width=\textwidth]{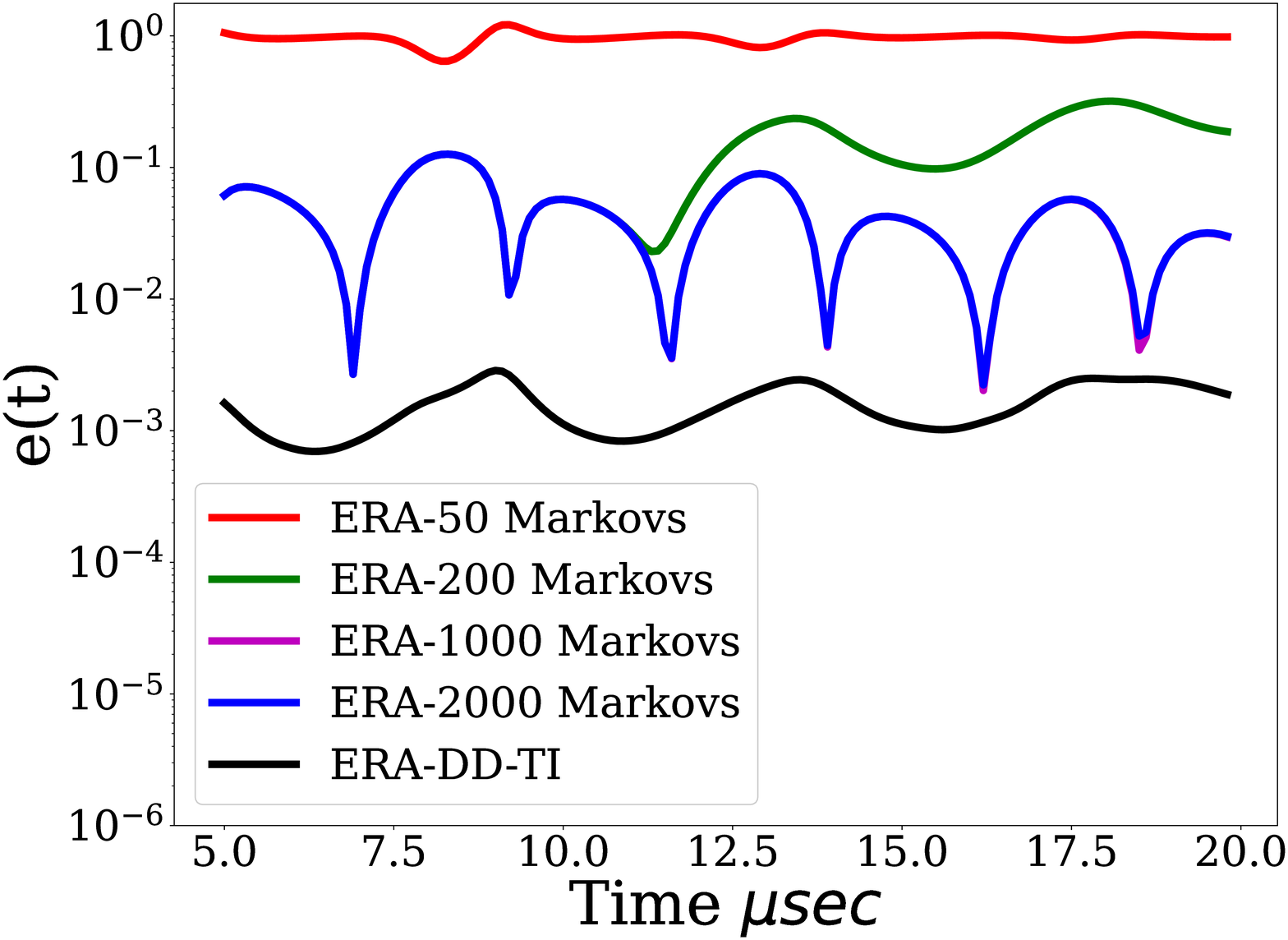}
     \rput(0.9,0.45){\psscalebox{0.9}{\color{black} \textbf{c)}}}
     %\vspace{0.2cm}
  \end{minipage}
   \centering
  \begin{minipage}[a]{0.40\textwidth}
    \includegraphics[trim=4 0.3cm 4 2cm, clip, width=\textwidth]{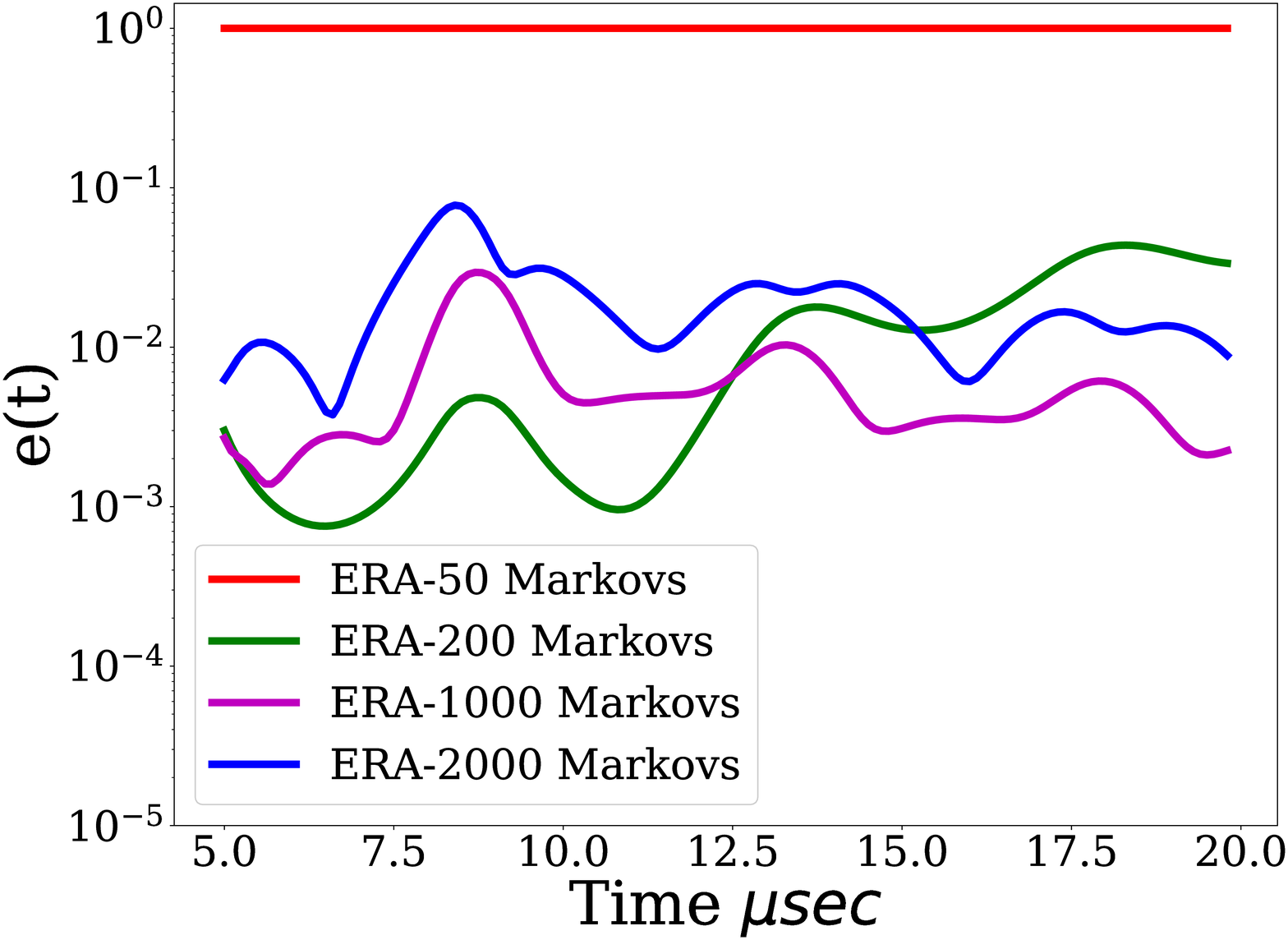}
     \rput(0.9,0.45){\psscalebox{0.9}{\color{black} \textbf{d)}}}
     %\vspace{0.2cm}
  \end{minipage}
   \centering
  \caption{Relative prediction error for forced flame with a perturbation amplitude of $0.01 \% p_{back}$ and a frequency of 215 kHz. Sensitivity of the balanced ROMs with respect to the sampling time (i.e., number of Markov parameters) is demonstrated for pressure (a), velocity (b), temperature (c), and species mass fraction (d). The black line corresponds to ERA with output domain decomposition and tangential interpolation. All other ROMs are trained with a sampling period of  $T_s = 0.1 \mu sec$}
   \label{f:sTimePred215}
\end{figure}

The sensitivity of balanced ROMs constructed with ERA is studied here with respect to the sampling time. As shown in Figure~\ref{f:sTimePred215}, ROMs with 50 and 200 Markov parameters fail to capture the decay of the lightly-damped eigenvector in the impulse response tail and therefore, exhibit a poor predictive performance. Relative prediction error computed for ROMs with 1000 and 2000 Markov parameters overlap. Thus, the impulse response sequence  captures the essential dynamics and further increase in the sampling time does not improve accuracy. The relative error profiles in Figure~\ref{f:ERAPred215} show that the domain decomposition approach has effectively improved accuracy of the balanced ROMs by resolving the numerical errors at the right boundary, despite the impulse response sequence is projected onto a lower-dimensional output space via the tangential interpolation.
We have not used domain decomposition for the species mass fraction, since the sharp gradients are only observed in pressure, velocity, and temperature profiles.

\begin{figure}[h!]
  \centering
  \begin{minipage}[a]{0.40\textwidth}
    \includegraphics[trim=4 0.3cm 4 2cm, clip, width=\textwidth]{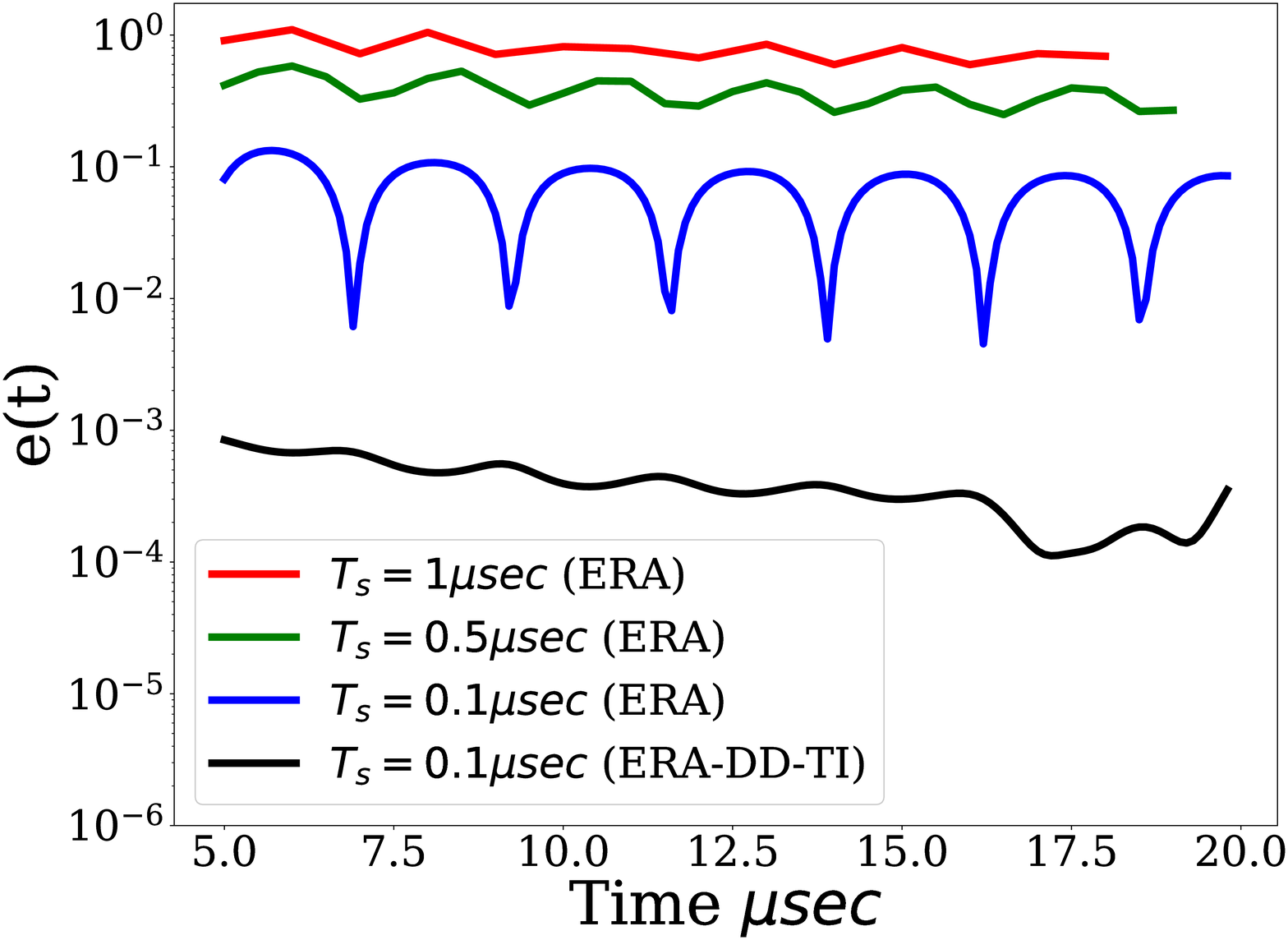}
     \rput(0.9,0.45){\psscalebox{0.9}{\color{black} \textbf{a)}}}
     \vspace{0.1cm}
  \end{minipage}
  \centering
  \begin{minipage}[a]{0.40\textwidth}
    \includegraphics[trim=4 0.3cm 4 2cm, clip, width=\textwidth]{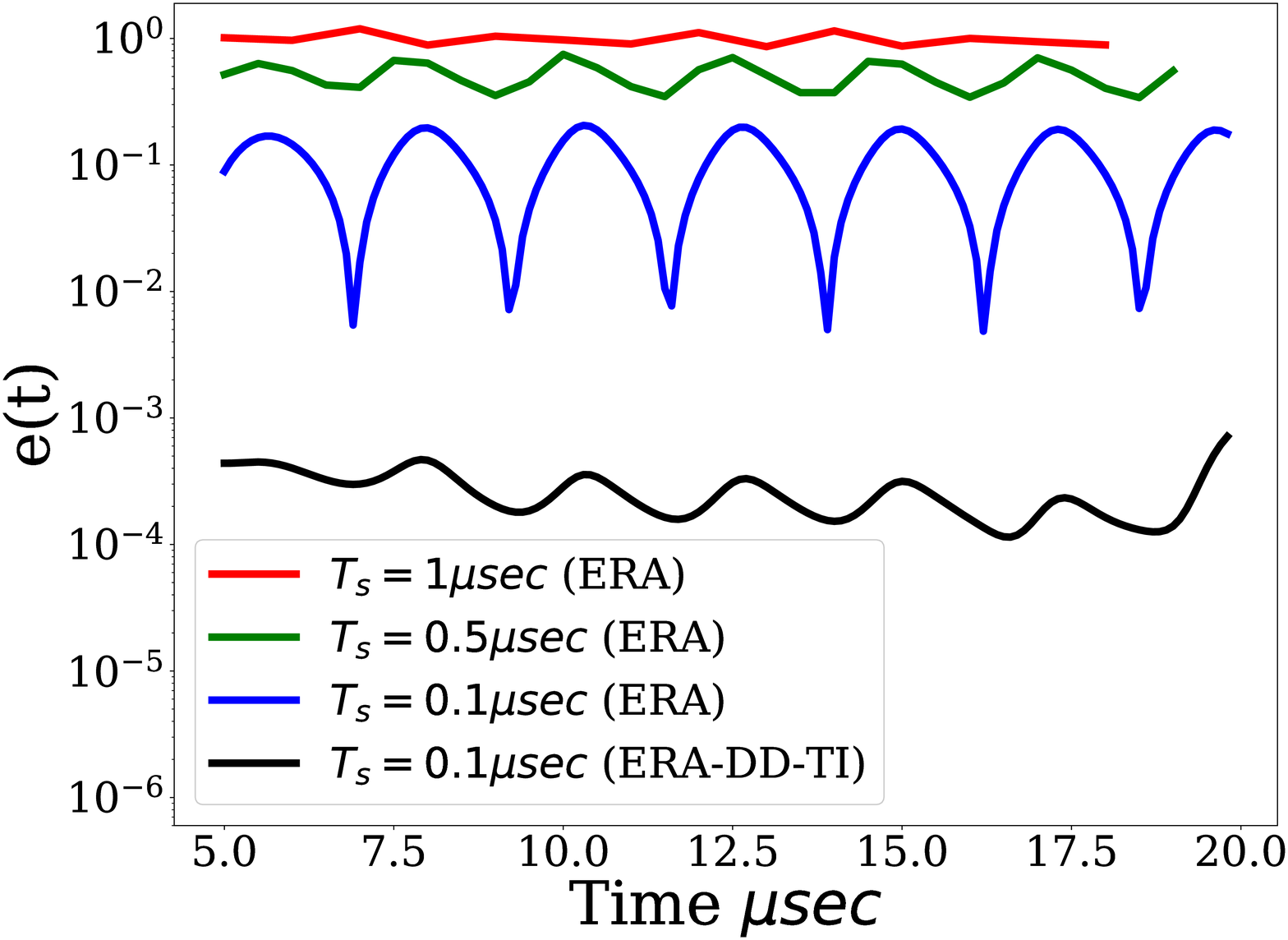}
     \rput(0.9,0.45){\psscalebox{0.9}{\color{black} \textbf{b)}}}
     \vspace{0.1cm}
  \end{minipage}
   \centering
  \begin{minipage}[a]{0.40\textwidth}
    \includegraphics[trim=4 0.3cm 4 2cm, clip, width=\textwidth]{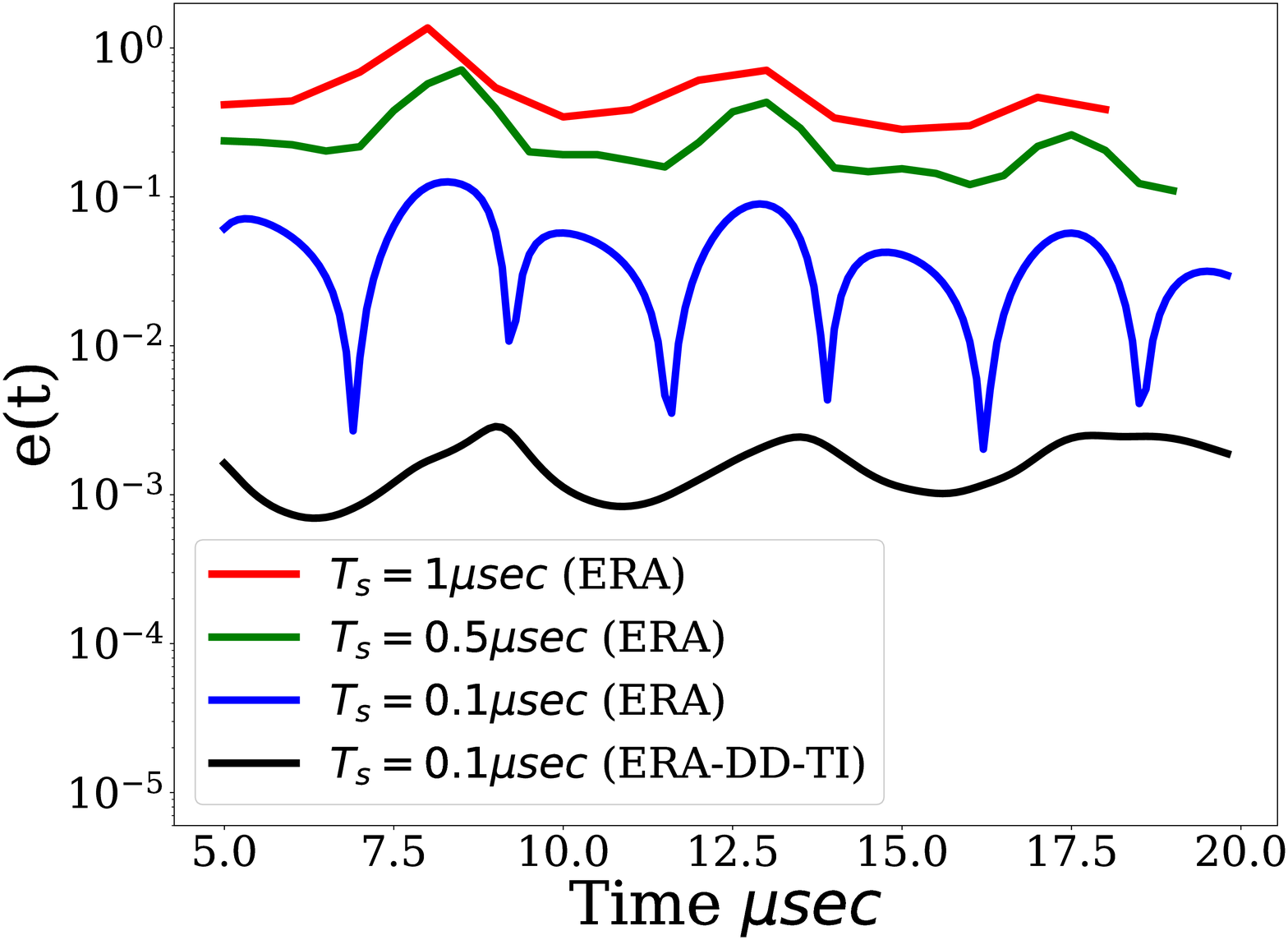}
     \rput(0.9,0.45){\psscalebox{0.9}{\color{black} \textbf{c)}}}
     %\vspace{0.2cm}
  \end{minipage}
   \centering
  \begin{minipage}[a]{0.40\textwidth}
    \includegraphics[trim=4 0.3cm 4 2cm, clip, width=\textwidth]{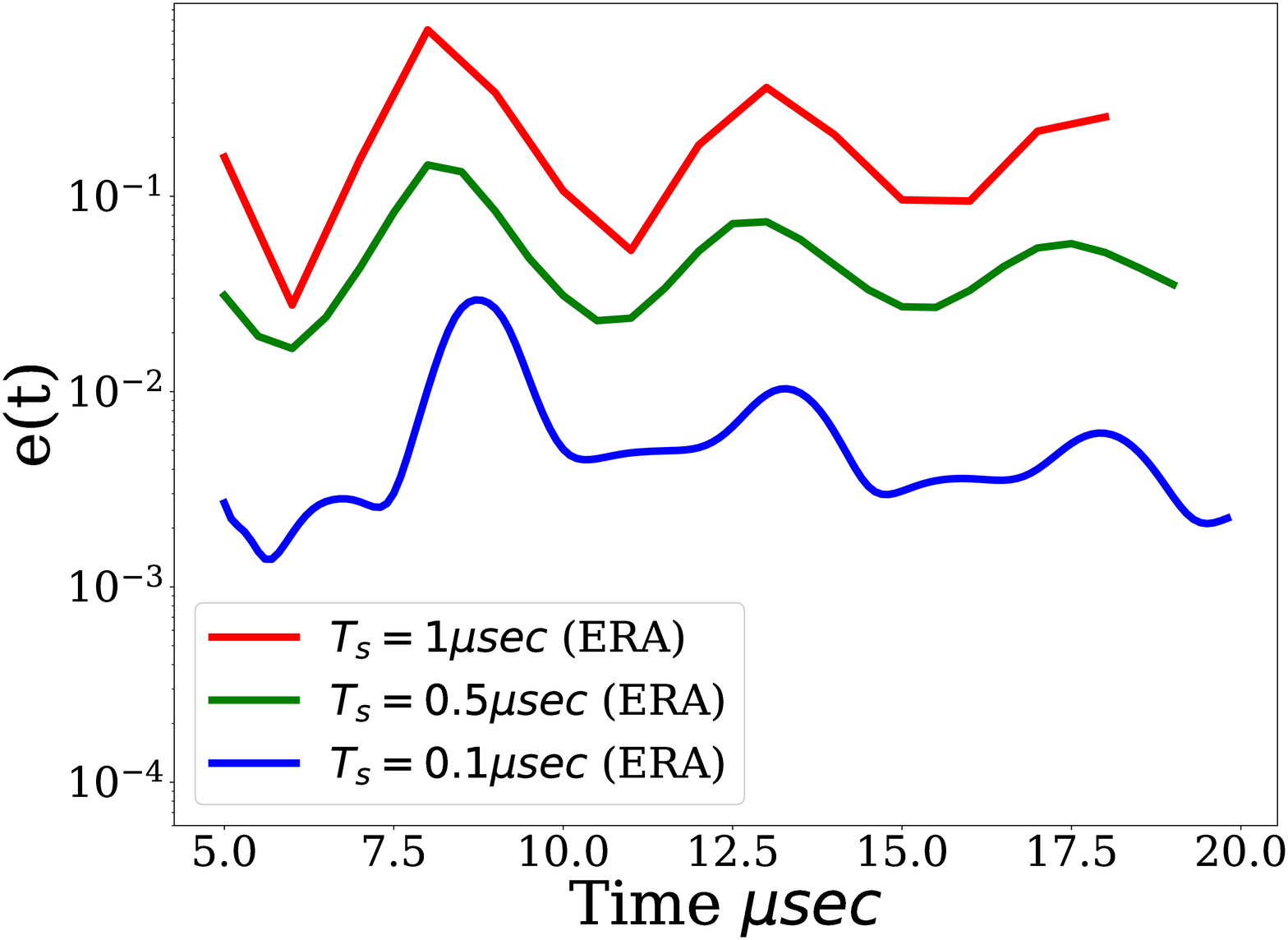}
     \rput(0.9,0.45){\psscalebox{0.9}{\color{black} \textbf{d)}}}
     %\vspace{0.2cm}
  \end{minipage}
   \centering
  \caption{Relative prediction error for forced flame with a perturbation amplitude of $0.01 \% p_{back}$ and a frequency of 215 kHz. Sensitivity of the balanced ROMs with respect to the sampling frequency is demonstrated for pressure (a), velocity (b), temperature (c), and species mass fraction (d). The black line corresponds to ERA with output domain decomposition and tangential interpolation.}
   \label{f:sfPred215}
\end{figure}

We further investigate the sensitivity of the balanced ROMs with respect to sampling properties by evaluating the relative prediction error of ROMs trained with different impulse response sampling frequencies shown in Figure~\ref{f:sfPred215}. Sampling periods of $T_s = 1 \mu sec$ and $T_s = 0.5 \mu sec$ seem to leave out some of the high-frequency structures that are critical to the dynamics of the system. The error reduces as the  sampling period is decreased to $T_s = 0.1 \mu sec$, which showed good agreement with the FOM in Figure~\ref{f:ERAPred215} for unseen forcing input. Training subdomain ROMs with different sampling periods in the output domain decomposition approach has clearly improved accuracy of the state prediction by one to two orders of magnitude.

\section{Conclusion}\label{conclusion}
Balanced truncation (BT) simultaneously accounts for observability and reachability of the dynamical signatures in dimensionality reduction. This is achieved by transformation to a new coordinate system in which the reachability and observability Gramians are equal and diagonal. Therefore, BT addresses some of the fundamental challenges in standard projection techniques such as those based on POD which are only effective for systems in which the most reachable and the most observable directions are aligned. Unfortunately, this is not the case in most transport problems. Therefore, POD-based ROMs typically exhibit poor predictive performance in many practical applications.

Classical Balanced truncation  utilizes the system Gramians to compute the transformation matrix, where the Gramians are obtained by solving a set of Lyapunov equations that are expensive to approach in high-dimensional systems. One way to address this issue is through approximate BT methods by resorting to empirical Gramians or bypassing computation of these matrices altogether. However, they still require adjoint system simulations that are expensive in systems with full-state output and inaccessible in experiments. This motivates non-intrusive BT methods that merely rely on the impulse response of the direct system for computing the balancing transformation. 

Eigensystem realization algorithm (ERA), which is originally a system identification method is demonstrated to be a powerful tool for nonintrusive balancing transformation in discrete-time systems. In this study, we have used ERA to construct balanced ROMs for the linearized Navier-Stokes equations governing a one-dimensional reactive flow with pressure forcing. This system experiences lightly-damped oscillations caused by a few eigenvectors that remain active for a long time after the initial transients and make it challenging to balance computation cost and accuracy in ERA. We study the sensitivity of the balanced ROMs with respect to the sampling time and frequency, and compare the predictive performance of the balanced ROMs with the standard Galerkin and LSPG projections. While Galerkin and LSPG ROMs fail to follow the true dynamics at unseen forcing frequencies, the robust performance of ERA in a truly predictive setting justifies the training cost of BT. 

To resolve sharp gradients in the computational domain while bypassing the prohibitive offline costs associated with a large Hankel matrix as a result of a high sampling frequency and the large number of outputs (full-state), we propose an output domain decomposition approach to confine high sampling frequencies to subdomain ROMs that encounter sharp gradients, and reduce sampling frequency in the rest of the computational domain. We utilize this approach with tangential interpolation to reduce the number of outputs, and consequently the offline computation costs, by projection of the impulse response sequence onto the dominant tangential directions before balancing. As a result, sharp gradients are effectively resolved, and the accuracy of balanced ROMs is improved.

\enlargethispage{20pt}

%\ethics{Insert ethics text here.}

\dataccess{The data and code for the numerical experiments in this manuscript are available at \url{https://github.com/cwentland0/perform.git}.}

\aucontribute{E.R. carried out the numerical analysis and drafted the manuscript. All authors contributed to revising the paper, gave the final approval for publication and agree to be accountable for all aspects of the work.}

\competing{We declare we have no competing interests.}

\funding{Funded by Air Force under grant FA9550-17-1-0195}

\ack{The Air Force is gratefully acknowledged for supporting this research through the Center of Excellence under grant FA9550-17-1-0195. The authors are also thankful to Prof. Karen Willcox for the productive discussions.}

%\disclaimer{Insert disclaimer text here.}

%%%%%%%%%% Insert bibliography here %%%%%%%%%%%%%%

%\section*{References}

\bibliographystyle{RS}
\bibliography{RSPA_Author_tex}

\end{document}